\documentclass[showpacs,aps,prd,preprintnumbers,amsmath,amssymb,superscriptaddress,groupedaddress,nofootinbib]{revtex4}
\usepackage{graphicx,subfigure,multirow}
\usepackage{dcolumn,lipsum}
\usepackage{bm}
\usepackage{natbib}
\usepackage{doi}
\usepackage{soul}
\usepackage[utf8]{inputenc}
\usepackage{slashed} 
\usepackage{dsfont}
\usepackage[toc,page]{appendix}
\makeatletter
\usepackage{color}
\usepackage{url}
\definecolor{myred}{rgb}{0.6,0,0} 
\definecolor{myblue}{rgb}{0,0.2,0.4}
\definecolor{mygreen}{rgb}{0,0.9,0.1}
\definecolor{hc}{rgb}{.9,0.1,0.7}
\definecolor{hcout}{rgb}{.9,0.7,0.9}
\definecolor{Orange}{rgb}{1.,0.65,0.}

\newcommand{\beq}{\begin{equation}}
\newcommand{\eeq}{\end{equation}}
\newcommand{\bea}{\begin{eqnarray}}
\newcommand{\eea}{\end{eqnarray}}
\newcommand{\beas}{\begin{eqnarray*}}
\newcommand{\eeas}{\end{eqnarray*}}
\newcommand{\bi}{\begin{itemize}}
\newcommand{\ei}{\end{itemize}}

\begin{document}
\title{Looking for Minimal Inverse Seesaw scenarios at the LHC\\ with Jet Substructure Techniques}
\author{Akanksha Bhardwaj}
\email{akanksha@prl.res.in}
\affiliation{Physical Research Laboratory (PRL), Ahmedabad - 380009, Gujarat, India}
\affiliation{Indian Institute of Technology, Gandhinagar-382424, Gujarat, India}
\author{Arindam Das}
\email{arindam@kias.re.kr}
\affiliation{School of Physics, KIAS, Seoul 02455, Korea}
\author{Partha Konar}
\email{konar@prl.res.in}
\affiliation{Physical Research Laboratory (PRL), Ahmedabad - 380009, Gujarat, India}
\author{Arun Thalapillil}
\email{thalapillil@iiserpune.ac.in}
\affiliation{Indian Institute of Science Education and Research, Homi Bhabha Rd, Pashan, Pune 411 008, India}
\date{\today}

\begin{abstract}

Simple extensions of the Standard Model (SM) with additional Right Handed Neutrinos (RHNs) can elegantly explain the existence of small neutrino masses and their flavor mixings. Collider searches for sterile neutrinos are being actively pursued currently. Heavy RHNs may dominantly decay into $W^\pm \ell^\mp$ after being produced at the LHC. 
In this paper, we consider collider signatures of heavy pseudo-Dirac neutrinos in the context of inverse seesaw scenario, with a sizable mixing with the SM neutrinos under two different flavor structures, viz., Flavor Diagonal (FD) and Flavor Non-Diagonal (FND) scenarios. For the latter scenario we use a general parametrization for the model parameters by introducing an arbitrary orthogonal matrix and nonzero Dirac and Majorana phases. We then perform a parameter scan to identify allowed parameter regions which satisfy all experimental constraints. 
 As an alternative channel to the traditional trilepton signature, we propose the opposite-sign di-lepton signature in the final state, in association with a fat jet from the hadronic decay of the boosted $W^\pm$. We specifically consider a fat jet topology and explore the required enhancements from exploiting the characteristics of the jet substructure techniques. 
We perform a comprehensive collider analysis to demonstrate the effectiveness of this channel in both of the scenarios, significantly enhancing the bounds on the RHN mass and mixing angles at the 13 TeV LHC. Interestingly the FND scenario can reach up to a $5$-$\sigma$ limit under the presence of the general parametrization at the high luminosity LHC.
\end{abstract}


\maketitle


\section{Introduction}
\label{sec:into}
Neutrino oscillation experiments have unambiguously established the existence of light neutrino masses and lepton flavor mixings \cite{Catanesi:2013fxa,Adamson:2011qu,Abe:2011fz,An:2012eh,Ahn:2012nd}. These are not a priori incorporated in the structure of the SM. Understanding the origin of the fermion mass hierarchy and mixing angles, along with exploring new states and sources of CP violation in the lepton sector are of much current interest.

To satisfy the neutrino oscillation data, a simple extension of the SM in the form of the seesaw mechanism suffices to a large extent \cite{Minkowski:1977sc,Yanagida:1980xy,Schechter:1980gr,Sawada:1979dis,GellMann:1980vs,Glashow:1979nm,Mohapatra:1979ia}. In these frameworks, SM-singlet heavy Majorana RHNs are introduced, which through a dimension five operator \cite{Weinberg:1979sa} subsequently lead to very small Majorana neutrino masses. If the singlet RHNs reside at the electroweak scale, then the RHNs can be produced at the LHC. Being singlets, these RHNs interact with the SM gauge bosons only through mixing with light SM neutrinos. There is another version of the seesaw mechanism \cite{Mohapatra:1986aw,Mohapatra:1986bd,Das:2017hmg,Das:2014jxa,Das:2015toa} wherein the small neutrino mass can be obtained from a naturally small \cite{tHooft:1979rat} lepton number violating parameter, rather than being suppressed by a heavy RHN mass. In this case, the RHN is of a pseudo-Dirac type and their Dirac Yukawa coupling can be large enough to produce RHNs at the LHC.

The Run-II of the LHC has already accumulated significant amounts of data at $\sqrt{s}=13$ TeV. The discovery of a fundamental scalar particle, the Higgs boson in the Standard Model (SM) has laid the foundations for a successful understanding of electroweak symmetry breaking. Nevertheless, the non-appearance of any significant excess, supporting any scenario beyond the SM (BSM), strongly motivates us to develop and apply new strategies. The aim of the latter should be to enhance discovery potentials from existing searches as well as to help efficiently explore difficult corners of signal and phase space. 

 The powerful techniques of jet substructure is one such strategy and in many contexts takes one along untrodden paths. LHC searches have benefitted immensely from developments in jet substructure techniques over the past many years and have enabled investigations of many hitherto challenging signals. Starting from the earlier ideas in jet substructure \cite{Seymour:1993mx,Butterworth:2007ke,Brooijmans:2008zza,Butterworth:2008iy} we now have a large toolkit of methods suited to a diverse array of theoretical and experimental challenges. As mentioned, the lack of any unambiguous indications of new physics at collider and non-collider experiments is perplexing though. This has recently reinvigorated searches for novel methodologies and paradigms, for instance, to categorize anomalous objects \cite{Chakraborty:2017mbz}, or efficiently groom jets either stochastically \cite{Roy:2016qfv} or using the unparalleled power of machine learning \cite{Komiske:2017ubm}. Please see \cite{Adams:2015hiv,Larkoski:2017jix} and references therein for detailed discussions.
 
\begin{figure}[t!]
	\begin{center}
		\includegraphics[scale=0.45]{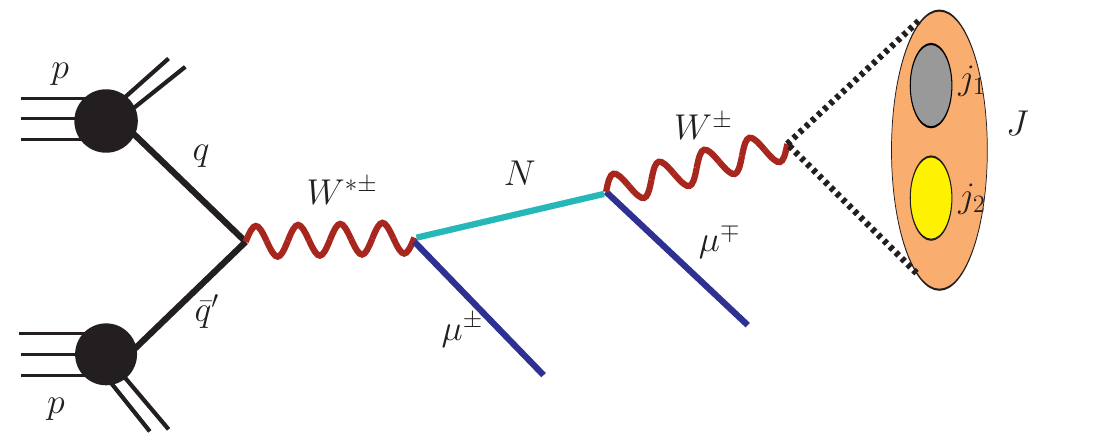}
	\end{center}
	\caption{Representative parton level diagram for production of heavy neutrino at hadron colliders through off-shell $W$ boson and its decay into an opposite sign muon and $W$ boson.  This boosted $W$ originated from a heavier exotic decay results into a fat jet after decaying hadronically. 
	}
	\label{fig:diag_proc}
\end{figure}

In collider searches for sterile neutrinos and related models, there have indeed been studies that have effectively utilized collimated, merged or large-radius objects in the signal topology \cite{Izaguirre:2015pga,Antusch:2016ejd,Mitra:2016kov,Dube:2017jgo,Cox:2017eme,Leonardi:2015qna,Sirunyan:2017xnz,Tang:2017plx,Dev:2015kca}. The use of jet substructure methods, that have proven so powerful in other searches, have nevertheless remained underutilized in sterile neutrino searches; especially in the relevant di-lepton + jet(s) topologies. For example, the importance of jet substructure over and above a merely boosted or collimated topology, in improving significance and mitigating backgrounds, is given by the seminal BDRS paper \cite{Butterworth:2008iy}.
In an earlier work, while considering a generic model with Majorana type heavy neutrinos, we presented a novel strategy, leveraging final states with same-sign di-leptons (SSDL) in association with a fat jet \cite{Das:2017gke}. This was done in tandem with jet substructure tagging that proved crucial. There, we showed that the additional jet substructure can give enough handles to make substantial improvements to exclusion limits.
In the present work, we aim to extend the idea towards probing heavy pseudo-Dirac neutrinos in minimal inverse seesaw scenario, which would produce a more challenging, albeit clean signature, those with opposite-sign di-leptons (OSDL) in association with a fat jet. 
The prototypical signal topology ($l^\pm l^\mp J$) of interest is shown in Fig.~\ref{fig:diag_proc}. In this scenario, the background rejection is much more exacting compared to the SSDL case. More over, as we shall detail in the next section, the OSDL channel can probe several new models where it is the only possible final state in the current context. This broadens the scope of sterile neutrino searches to encompass a larger region of the model space. Due to these reasons pursuit of the much more demanding OSDL+fat jet channel may be considered pertinent.

The paper is organized in the following way -- in Sec.~\ref{sec:model} we discuss the prototypical model of interest for the searches at the LHC.
In Sec.~\ref{sec:coll} we then proceed to our analysis, present details of our simulation, benchmark points and the final results. 
Finally, in Sec.~\ref{sec:sumcon} we summarize our results and conclude.

\section{Inverse Seesaw scenario}
\label{sec:model}

Contrary to our vanilla seesaw mechanism where a mass suppression from heavy neutrinos generates small neutrino mass, the inverse seesaw \cite{Mohapatra:1986aw, Mohapatra:1986bd, Das:2017hmg} rely on the tiny lepton number violating parameters. Here such a dimension-full small parameter generates the active light neutrinos in the presence of heavy right-handed neutrinos as well as sterile fermionic states, still with a natural $\mathcal{O}(1)$ Yukawa couplings ensuring a sizable mixing between active and the sterile states. Hence, contrary to type-I seesaw,  such heavy neutrinos, which turn out to be pseudo-Dirac in nature, can naturally be produced at the high energy colliders and probed.

In the SM one can introduce arbitrary number of SM gauge-singlets particles without affecting the anomalies allowing us to naturally incorporate a low-scale alternative seesaw mechanism. One such low scale seesaw \cite{delAguila:2008cj} can be realized by extending the SM with $n$ SM-singlet RHN and $m$ sterile neutrinos. In this basis we consider the charged leptons in their mass eigenstates. The interaction Lagrangian can be written as
\bea 
-\mathcal{L}_{1} &=& Y_a \overline{\ell_L} H N_R+ Y_b \overline{\ell_L} H S + M_N \overline{N_R^c} S + \frac{1}{2} \mu \overline{S^c} S + \nonumber \\
&+& \frac{1}{2} M_1 \overline{N^c_R} N_R+ h. c.
\label{int11}
\eea
where $\ell_L$ is the SM lepton doublet and $H$ is the SM Higgs doublet. The Yukawa couplings are $Y_a$ and $Y_b$ having dimensions of $(3 \times n)$ and $(3 \times m)$ respectively. $M_1$ and $\mu$ are the Majorana mass matrices which correspond $N_R$ and $S$ of dimensions $(n \times n)$ and $(m \times m)$, respectively. Presence of the mass parameters $\mu$ and $M_1$ explicitly break the lepton number. After the electro-weak symmetry breaking, we get from Eq.~\ref{int11}
\bea
-\mathcal{L}_{mass} &=& M_{D}  \overline{\nu_L} N_R + M_2 \overline{\nu_L} S + M_N \overline{N^c_R} S+ \frac{1}{2} \mu \overline{S^c} S \nonumber \\
&+& \frac{1}{2} M_1 \overline{N^c_R} N_R+ h. c.
\label{Lmass0}
\eea
where $M_D=Y_1 \frac{v}{\sqrt{2}}$, $M_2=Y_2\frac{v}{\sqrt{2}}$ and $<H>=\frac{v}{\sqrt{2}}$. Hence the Eq.~\ref{Lmass0} can be written as 
\bea
-\mathcal{L}_{mass} =\frac{1}{2} \begin{pmatrix} \overline{\nu_L}&\overline{N^c_R}&\overline{S^c} \end{pmatrix} \begin{pmatrix}0&M_D&M_2\\ M_D^T&M_1&M_N\\M_2^T&M_N^T&\mu \end{pmatrix} \begin{pmatrix}\nu_L^c\\ N_R\\S \end{pmatrix}.
\label{Lmass00}
\eea 
A variety of seesaw scenario can be realised from the Eq.~\ref{Lmass00} after choosing some entries as extrermely small or zero. Among these choices the simplest choice might be the inverse seesaw mechanism which can be constructed considering $M_2 \to 0$ and $M_1 \to 0$ \cite{Mohapatra:1986aw, Mohapatra:1986bd}, which has been validated  further in \cite{Malinsky:2009df, Garg:2017iva} using vacuum stability and fitting with the neutrino oscillation data. In this case, the sub-matrices $\mu$ and $M_N$ did not arrive from the $SU(2)_L$ symmetry breaking whereas $\mu$ is the term violating the lepton number. Hence there is a hierarchy $M_N >> M_D>>\mu$. The smallness $\mu$ can be occured due to the 't Hooft's naturalness criteria \cite{tHooft:1979rat} since the expected degree of the violation of the lepton number becomes small naturally. Commonly each of the mass matrices $M_N$, $M_D$ and $\mu$ are $(3 \times 3)$ in dimension. For example, Ref. \cite{Abada:2014vea} has studied such a minimal scenario. In our present study, we consider a minimal scenario where two generations of the RHNs are involved so that the structure can minimally satisfy the neutrino oscillation data. There is an alternative procedure, too. Alternatively  one can assign the lepton numbers to correspond to the singlet RHNs $N_R$ as $+1$and $S$ as $-1$, respectively. If the entries $(13)$, $(31)$ and $(22)$ of the Eq.~\ref{Lmass00} do not arise, then a purely inverse seesaw scenario is obtained. Finally, the effective light neutrino mass matrix can be written as,
\bea
M_{\nu} \sim M_D (M_N^T)^{-1} \mu M_N^{-1} M_D^T.
\label{iseesaw}
\eea

Here, the smallness of $M_\nu^{\rm{light}}$ is naturally acquired both from the tiny $\mu$ term, and smallness of $\frac{M_D}{M_N}$. 
Hence $M_\nu^{\rm{light}}\sim \mathcal{O}(0.1)$ eV can be obtained naturally from $\mu\sim \mathcal{O}(100)$ \rm{eV}, and  a choice of corresponding fraction as $\frac{M_D}{M_N}\sim 0.01$. Hence the scale of seesaw could be lowered using a rather sizable $Y_a\sim \mathcal{O}(0.1)$ considering a TeV choice of scales, such as, $M_D \sim 10$ GeV and $M_N \sim 1$ TeV.
As stated earlier, the small mass term $\mu$ splits the heavy neutrino masses into the three pairs of nearly degenerate pseudo-Dirac neutrinos of masses of order $M_N \mp \mu$.
The inverse seesaw scenario has also been discussed under the general parametrization using Casas-Ibarra conjecture for general $Y_D$ \cite{Das:2012ze}.

For simplicity we consider degenerate RHNs, with $M= M_N \times \mathds{1}$. $\mathds{1}$ is the unit matrix as before and $M_N$ is the RHN mass eigenvalue.
 With these assumptions, the neutrino mass matrix may be simplified as 
\begin{eqnarray}
M_{\nu}=\frac{1}{M_N^2} M_D \mu M_D^T \; .
\end{eqnarray}

Consider a typical flavor structure of the model where $M_D$ and $M_N$ are proportional to the unit matrix such as $M_D \to M_D \times \mathds{1}$ and $M_N \to M_N \times \mathds{1}$ respectively and flavor structure is carried out by the $2\times 2$ matrix $\mu$. We refer to this scenario as Flavor Diagonal (FD). It has been shown that the FD case in the inverse seesaw mechanism is also accommodated by neutrino oscillation data \cite{Das:2012ze}. Another flavor structure possible in the inverse seesaw scenario is where $M_D$ carries flavor structure while $\mu \to \mu \times \mathds{1}$ and $M\to M_N\times \mathds{1}$. This is called the Flavor Non-Diagonal (FND) scenario. This has been studied for different signals in \cite{Das:2012ze,Das:2017nvm}, under general parametrization \cite{Casas:2001sr}.


Assuming $M_D M_N^{-1} \ll 1$, 
 we write down the light Majorana neutrino flavor eigenstates ($\nu$) of 
 as a combination of the 
 the light neutrino ($\nu_{\bar{m}}$) mass eigenstates
  and the heavy Majorana neutrino mass eigenstates ($N_{\bar{m}}$)
\bea 
  \nu \simeq {\cal N} \nu_{\bar{m}}  + {\cal R} N_{\bar{m}},  
  \label{rel-1}
\eea 
where 
\bea
 {\cal R} = M_D M_N^{-1}, \; 
 {\cal N} =  \left(1 - \frac{1}{2} \epsilon \right) U_{\rm PMNS} ,\;
  \epsilon = {\cal R}^* {\cal R}^T, 
 \label{rel-2}
 \eea
where $U_{PMNS}$ is the usual neutrino mixing matrix 
which diagonalizes the mass matrix $m_\nu$ as 
\bea
   U_{PMNS}^T m_\nu U_{PMNS} = {\rm diag}(m_1, m_2, m_3). 
  \label{rel-3} 
\eea
In the presence of $\epsilon$, the mixing matrix ${\cal N}$ is not unitary. The interaction Lagrangian for the charged current (CC) is written in terms of the heavy neutrino mass eigenstates as 
\begin{eqnarray} 
\mathcal{L}_{CC} \supset 
 -\frac{g}{\sqrt{2}} W_{\mu}
  \bar{e} \gamma^{\mu} P_L   \mathcal{R} N_{\bar{m}}  + \rm{H.c}. \; , 
\label{CC}
\end{eqnarray}
where $e$ is the charged lepton and $P_L$ is considered as  $\frac{1}{2} (1- \gamma_5)$. Similarly, the neutral current (NC) interaction in terms of the heavy neutrino mass eigenstates can be written as
\begin{eqnarray} 
\mathcal{L}_{NC} \supset && - \frac{g}{2 c_w}  Z_{\mu}    \Big[
  \overline{N}_{\bar{m}} \gamma^{\mu} P_L (\mathcal{R}^{\dagger} \mathcal{R}) N_{\bar{m}} \nonumber \\
&+&
  \overline{\nu_{\bar{m}}} \gamma^{\mu} P_L (\mathcal{N}^{\dagger} \mathcal{R})  N_{\bar{m}} 
  + \rm{H.c.}
\Big]  \; ,
\label{NC}
\end{eqnarray}
 where $c_w=\cos \theta_w$ with $\theta_w$ being the Weinberg angle. The heavy neutrino production cross section from the Eqs.~(\ref{CC}) and (\ref{NC}) in association with a SM lepton is proportional to $|V_{\ell N}|^2$.

In our analysis, we will consider two degenerate pseudo-Dirac type RHNs separately being coupled to the SM charged leptons $e$ and $\mu$ respectively. Hence in our analysis we consider $M_N \to M_N \times \mathds{1}_{2\times 2}$. In this model framework we will also separately study the case when a RHN is coupled with $\mu$, which we name as the single flavor case. The two flavor case, without considering flavor detection efficiencies, will roughly double the number of signal events relative to the single flavor case.

At this point, for completeness, we must also comment that the seesaw and inverse seesaw mechanisms in the context of Left-Right (LR) models \cite{Das:2017hmg,Das:2016akd} will also produce OSDL + fat jet final states.  In the seesaw framework we have already tested the same-sign di-lepton (SSDL) signature in association with a fat jet \cite{Das:2017gke}; OSDL+ fat jet provides another important channel towards completeness of RHN searches. Also, the neutral charge multiplet in type-III seesaw \cite{Franceschini:2008pz} is a Majorana candidate and may also be studied in the final state of interest. Finally, there is another version of the seesaw mechanism, commonly known as linear seesaw \cite{Bambhaniya:2014kga,Bambhaniya:2014hla} where pseudo-Dirac RHNs are introduced, which too contributes to this channel.

The elements of the $\mathcal{N}$ and $\mathcal{R}$ matrices in the Eqs.~\ref{rel-1}- \ref{rel-3} can be constrained by the experimental results. 
To do this we adopt the current neutrino oscillation data:
$\sin^{2}2{\theta_{13}}=0.092$ \cite{An:2012eh},  
along with the other oscillation data \cite{Patrignani:2016xqp}:
 $\sin^2 2\theta_{12}=0.87$,
 $\sin^2 2\theta_{23}=1.0$, 
 $\Delta m_{12}^2 = m_2^2-m_1^2 = 7.6 \times 10^{-5}$ eV$^2$, 
  $\Delta m_{23}^2= |m_3^2-m_2^2|=2.4 \times 10^{-3}$ eV$^2$. 
The neutrino mixing matrix  is given by 
\begin{widetext}
\begin{eqnarray}
U_{\rm{PMNS}}=
 \begin{pmatrix} C_{12} C_{13}&S_{12}C_{13}&S_{13}e^{i\delta}\\-S_{12}C_{23}-C_{12}S_{23}S_{13}e^{i\delta}&C_{12}C_{23}-S_{12}S_{23}S_{13}e^{i\delta}&S_{23} C_{13}\\ S_{12}C_{23}-C_{12}C_{23}S_{13}e^{i\delta}&-C_{12}S_{23}-S_{12}C_{23}S_{13}e^{i\delta}&C_{23}C_{13} \end{pmatrix} 
 \cal{P}
\label{pmns}
\end{eqnarray}\\
\end{widetext}
where $C_{ij}=\cos\theta_{ij}$, $S_{ij}=\sin\theta_{ij}$ and  the Majorana phase matrix as $ {\cal P} = \text{diag}(1, e^{i\rho}, 1)$.
In this analysis $\delta$ (Dirac CP phase) and $\rho$ (Majorana phase) are considered to be the free parameters.

The elements of ${\cal N}$ (mixing matrix) are extremely constrained by the data obtained from the neutrino oscillation experiments, 
 the precision measurements of SM $W, Z$ decays
 and the SM charged Lepton-Flavor-Violating (LFV) decays \cite{Antusch:2006vwa, Abada:2007ux,Ibarra:2010xw, Ibarra:2011xn, Dinh:2012bp} because of the non-unitarity effects.
 Using the most recent data for the LFV experiments \cite{Adam:2011ch, Aubert:2009ag, OLeary:2010hau, TheMEG:2016wtm}
 we write  
\begin{widetext}
 \bea
|{\cal N}{\cal N}^\dagger| =
\begin{pmatrix} 
 0.994\pm0.00625& <1.288 \times 10^{-5} &  < 8.76356\times 10^{-3}\\
 <1.288 \times 10^{-5} & 0.995\pm 0.00625 & <1.046\times 10^{-2}\\
 < 8.76356 \times 10^{-3}& < 1.046 \times 10^{-2} & 0.995\pm 0.00625
\end{pmatrix}. 
\label{Eq-NU}
\eea
\end{widetext}

The diagonal elements of the Eq.~\ref{Eq-NU} are from the precision measurements of decays of the weak gauge boson
 where the SM predictions are $1$ for the diagonal elements. The off-diagonal elements are the upper bounds from the LFV decays, e.g.,
 the bounds on the $(12)$ and $(21)$ elements come from the $\mu \to e \gamma$, $(23)$ and $(32)$ elements come from the $\tau \to \mu \gamma$
 and $(13)$ and $(31)$ elements come from the $\tau \to e \gamma$ processes respectively. Hence we can estimate $\epsilon$ using ${\cal N}{\cal N}^\dagger \simeq {\bf 1} - \epsilon$.
The strongest limits is coming from the $(12)$ element which is obtained by the $\mu \to e \gamma$ process.
   
In the minimal scenario, one eigenstate can be predicted as massless. 
For the light neutrino mass spectrum, we consider both the normal hierarchy (NH) and the inverted hierarchy (IH).  
In the NH case, the diagonal mass matrix is given by 
\bea 
  D_{\rm{NH}} ={\rm diag}
  \left(0, \sqrt{\Delta m_{12}^2},
           \sqrt{\Delta m_{12}^2 + \Delta m_{23}^2} \right),  
\label{DNH}
\eea 
while in the IH case 
\bea 
  D_{\rm{IH}} ={\rm diag}
\left( \sqrt{\Delta m_{23}^2 - \Delta m_{12}^2}, 
 \sqrt{\Delta m_{23}^2}, 0 \right).  
\label{DIH}
\eea
For the FND case, we describe $\epsilon$ as 
\bea 
 \epsilon = \frac{1}{M^2} m_D m_D^T 
          = \frac{1}{\mu} U_{PMNS} D_{NH/IH} U_{PMNS}^T,  
\eea 
and calculate the minimum $\mu$ value ($\mu_{min}$) 
to get $\epsilon_{1 2}=1.288 \times 10^{-5}$ 
 we use the oscillation data. 
We have found $\mu_{min}=611.4$ keV and $383.2$ keV 
 for the NH and IH cases, respectively. The smallness of the $\mu$ ($\cal{O}$$(100)$ keV) parameter rules out the possibility of generating the lepton number violating signature at the collider under this model set-up. The limit on the light-heavy mixing in this region could be as small as $10^{-10}$-$10^{-8}$ \cite{Adhikari:2016bei} leading to an extremely long-lived scenario. Discussion of such species is beyond the scope of this paper.

In this analysis we consider real parameters and hence  we can calculate the elements of the mixing matrices of
${\cal R}$ and ${\cal N}$ taking
 $\mu = \mu_{\min}$ into consideration. This optimizes 
 the heavy neutrino production cross sections at the LHC. 
Considering a general parametrization of $Y_D$. we study the  for the FND case.
Hence we obtain from the inverse seesaw formula,
 \bea
 m_\nu &=& \mu {\cal R} {\cal R}^T \nonumber \\
   &=&\frac{\mu}{M^2}m_Dm_D^T \nonumber \\
   &=& U_{PMNS}^*D_{NH/IH} U_{PMNS}^\dagger, 
\eea
Finally we write ${\cal R}$ under the general parametrization as 
\bea
{\cal R}(\delta,\rho,X,Y)= 
\frac{1}{\sqrt{\mu}}U_{PMNS}^*\sqrt{D_{NH/IH}}O ,
\label{R}
\eea
with $O$ as a general $2X2$ orthogonal matrix which can be written as 
 \bea 
O &=&
\begin{pmatrix}
    \cos \alpha & \sin \alpha \\
    -\sin \alpha & \cos\alpha\\
\end{pmatrix} 
\eea
with $\alpha$ as a comples number which can be written as $X +i Y$.
Thus in the general parameterization we express 
\bea \nonumber
&&\epsilon(\delta, \rho, Y )  = \mathcal{R}^\ast \mathcal{R}^T \\ 
&&=
 \frac{1}{\mu} \, U_{PMNS}\sqrt{D_{NH/IH}}O^{*}O^{T}\sqrt{D_{NH/IH}}^{T}U_{PMNS}^{\dagger}. \quad  \quad 
\eea
 Note that 
\bea
O^{*}O^{T}=
\begin{pmatrix}
   \cosh^2 Y+ \sinh^2 Y&-2i \cosh Y \sinh Y \\
   2i \cosh Y \sinh Y & \cosh^2 Y + \sinh^2 Y
\end{pmatrix}.
\eea
Finally we get that $O^{*}O^{T}$ is not depending up on $X$ and from that the $\epsilon$-matrix is a function of $\delta$, $\rho$ and $Y$\footnote{In this context we point out that we have used the parametrization prescribed in \cite{Abada:2007ux} for $\cal{N}$, a different parametrization in \cite{Fernandez-Martinez:2016lgt} uses $\cal{N}\sim$ $1-\epsilon$ which over constrains the parameter space by a factor of $2$ when calculating $\cal{N}\cal{N}^{\dagger}\sim$ $1-2\epsilon$, which is $1-\epsilon$ in our case.}.

\section{Collider Analysis and Results}
\label{sec:coll}

We are interested in a very specific decay topology arising from the production and decay of heavy sterile neutrinos. The schematic of the prototypical parton level process, at the leading order, is shown in Fig.~\ref{fig:diag_proc}.
\begin{eqnarray}
q~\bar{q}\,'\rightarrow W^{\pm *}\rightarrow \mu^\pm N,   \,\,  N \rightarrow \mu^\mp W^\pm,   \,\,  W^\pm \rightarrow J
\end{eqnarray}
We focus on opposite-sign (OS) muon pair final states, in association with a reconstructed fat jet, at $\sqrt{s}=13$ TeV LHC. For simplicity, we demonstrate explicitly our analysis assuming a simple, single flavor scenario where the light-heavy mixing is non-zero only for the muon flavor. This is also motivated by the fact that muons provide a clear detection at the LHC with high efficiency and hence is of primary interest. We will however also include the electron channel while discerning the final exclusion results. \footnote{In this article we have studied the $eejj$ and $\mu\mu jj$ signals. We have also used the general parametrization for the detailed analysis. Later we found an article dealing with the OSDL signature with different flavors of the leptons, such as $e\mu jj$ signal \cite{Antusch:2018bgr}, studying the lepton flavor violating scenario at the collider.}.

As motivated earlier, the OSDL signature, unlike our previously studied SSDL signature, is prone to much larger SM backgrounds -- coming from $t \bar{t}$, mono-boson, di-boson and tri-boson productions. This makes the analysis challenging and interesting. Here we will argue and demonstrate that the additional W-like fat jet can be identified effectively by looking at different jet substructure parameters and that this consequently will lead to clear OSDL signatures, emerging over and above the humongous backgrounds. 

We generate signal and all the background events using {\tt Madgraph5} (v2.5.4) \cite{Alwall:2011uj,Alwall:2014hca} followed by {\tt Pythia} (v8) \cite{Sjostrand:2006za} for showering and hadronization. We use {\tt NN23LO1} \cite{Ball:2014uwa} parton distribution function and with the default dynamic renormalisation and factorisation scales \cite{madgraph_scale} in {\tt MadGraph5 aMC@NLO}. MLM matching \cite{Mangano:2006rw,Hoche:2006ph} of the shower jets and the matrix element jets. Matched background is generated using the default kt-MLM algorithm with {\tt Xqcut} $= 30 $ GeV and the corresponding  jet matching parameter (QCUT) is 1.5 times the {\tt Xqcut} \cite{Alwall:2007fs}. $p_t^j$ and $\Delta R_{jj}$ are set to zero for {\tt kt-MLM} matching. MLM matching reduces double counting of jets coming from the showers and the matrix element partons. Subsequent to this, the detector simulation is implemented using {\tt Delphes-v3.4.1} \cite{deFavereau:2013fsa}. We use {\tt Fastjet}-v3.3.2 \cite{Cacciari:2011ma} to identify fat jets using the Cambridge-Aachen (CA) algorithm \cite{Dokshitzer:1997in}. 
  The signal process, generated through the production of intermediate to heavy mass Dirac neutrinos and their boosted decay, is expected to have minor correction from such jet matching. We however included them for completeness to obtain our results. We adopt an {\tt Xqcut} $= 70$ GeV, or above, for the different mass scenarios we consider.
Jet parameters corresponding to $R = 0.8$ and  $p_T^{min} = 10$ GeV are adopted.

\begin{figure*}[t!]
\includegraphics[scale=0.8]{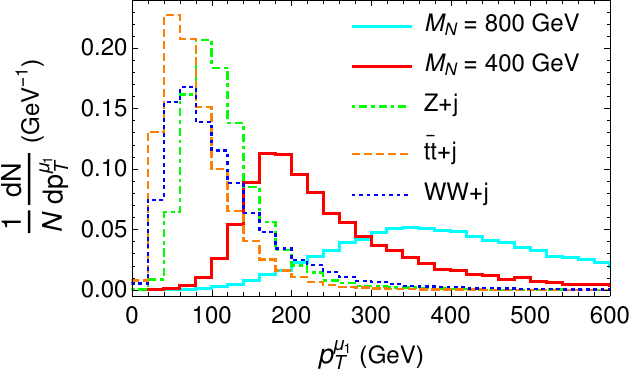}
\includegraphics[scale=0.8]{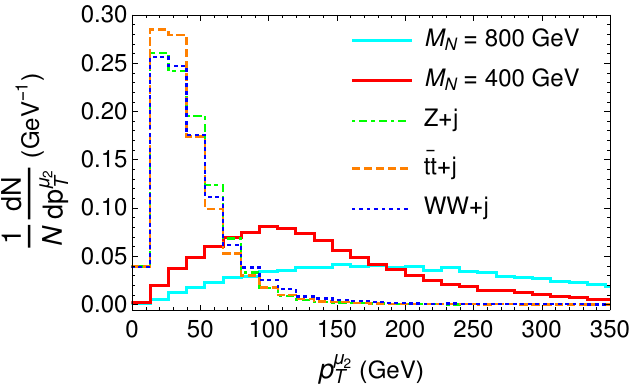}\\
\caption{Normalized differential distributions of transverse momentum $p_T$ of leading muon (left) and sub-leading muon (right). These differential distributions are after the $baseline~selection~cuts$ 
The distribution of heavy neutrino benchmark points with $M_N = 400$ and $800$ GeV is shown along with three dominating background processes.
}
\label{fig:dist_var1}
\end{figure*}

Opposite-sign di-leptons can arise from different production channels with gauge boson decays. Leptonic decays from $t \bar{t}$ can also give a substantial contribution. Our signal characteristic of a W-like fat jet can be faked by all such channels in association with additional QCD jets. Hence, to be consistent and thorough, all background production channels were produced with additional partons; with proper matching to showers. Moreover, associated $W^\pm$ bosons decaying hadronically may also generate irreducible backgrounds. We considered all the relevant dominant SM backgrounds which can mimic the OS di-muon and fat jet signal. 

Significantly large contribution can come from $Z$ + jets when the $Z$ boson decays leptonically. This is a large background and can be effectively controlled by applying much stronger cuts on the invariant mass of opposite-sign di-leptons ($M_{ll}$). QCD jets in these process can be controlled in addition through jet substructure. A significant background is also expected from $t\bar{t}+$ jets, where top decays leptonically. Vetoing b-jets and proper implementation of fat jet variables can again control this background. The efficiency of b-tagging is approximately $70\%$ while misidentification of a light parton jet as a b-tagged jet is $1.5 \%$ \cite{Chatrchyan:2012jua}. Additional modes that may contribute include $VV$ + jets and $VVV$ + jets, where either of the vector bosons ($V=W^\pm, Z$) decay leptonically to generate di-lepton pairs. Note that a number of these backgrounds subsequently produce missing neutrino(s) and/or missing charged leptons that can substantially add to the missing transverse momentum. In the signal process of interest whereas this is not the case, since we are considering hadronic decays of the $W^\pm$. The only dominant source of missing energy in the signal arises from possible jet energy mis-measurements. We use next-to-next-leading order estimate in QCD perturbation theory for the production cross section for Z boson as 2089 pb  \cite{Catani:2009sm} and $ W^\pm Z  = 51.11$ pb  \cite{Grazzini:2016swo}.  Furthermore, $W^+W^-$ and $W^+W^-Z$ the production cross section is computed at $NLO$ to be $112.64$ pb \cite {Campbell:2011bn} and $103.4$ fb  \cite{Nhung:2013jta} respectively. For $t\bar{t}$ we use production cross section as $ 835.61$ pb computed at $N^3LO$ \cite{Muselli:2015kba}.The next-to-leading order QCD correction for heavy neutrino production and scale uncertainties are studied in  \cite{Das:2016hof}, see also \cite{Degrande:2016aje}. For signals, we use the NLO cross section as in \cite{Das:2016hof} for different benchmark mass.
Before moving for our analysis we list our basic selection criteria as following.

{\bf Primary selection criteria -} To identify the leptons as well as the fat jet, we implement the following {\sf baseline~selection} \textbf{(C1)} of the events.

\begin{itemize}
\item  Two opposite sign muons are selected with $p_T > 10$ GeV within the detector rapidity range $|\eta_{\mu}| < 2.4$, assuming a muon detection efficiency of 95\%. We veto the event if any additional reconstructed lepton with $p_T > 10$ GeV  is present. The choice of leading (sub-leading) opposite sign muons are identified as $\mu_1$ ($\mu_2$) based on their $P_T$ and is irrespective of their charge.

\item We demand at least one fat jet, reconstructed adopting the CA algorithm with radius parameter $R = 0.8$ and $|\eta^{J_0}| < 2.4$. We select events with the hardest reconstructed fat jet ($J_0$) having minimum transverse momentum $p_T^{J_0} > 100$ GeV.
\end{itemize}

Let us now discuss the main kinematic characteristics that may be important in differentiating  signal events from the various large backgrounds. Several such features were already identified during the description of the background processes and they were suggestive in their effectiveness in controlling specific background channels. Before moving further, we identify our signal benchmark points -- labelled in terms of the sterile neutrino mass $M_N$ and mixing angle $|V_{\mu N}|^2$, they are $M_N = 400$ GeV, $800$ GeV and $|V_{\mu N}|^2 = 0.01$. Kinematic distributions are independent of the mixing angle and they are presented as normalized distributions, with differences between signal benchmark points and background processes highlighted. Two extreme mass points are chosen to establish the significantly different kinematic characteristics,  which could be leveraged to identify the optimized selection cuts for various masses.

\begin{figure*}[t!]
	\includegraphics[scale=0.8]{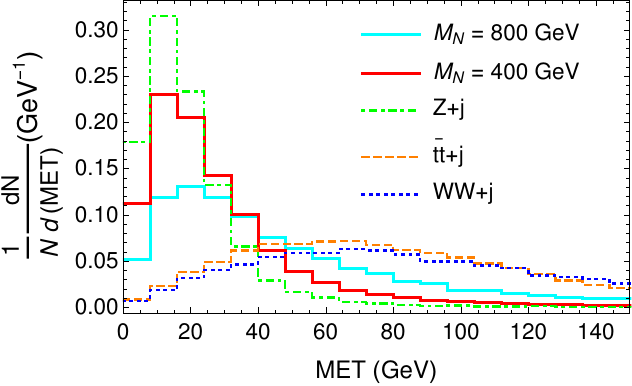} 
	\hspace{4pt} 
	\includegraphics[scale=0.8]{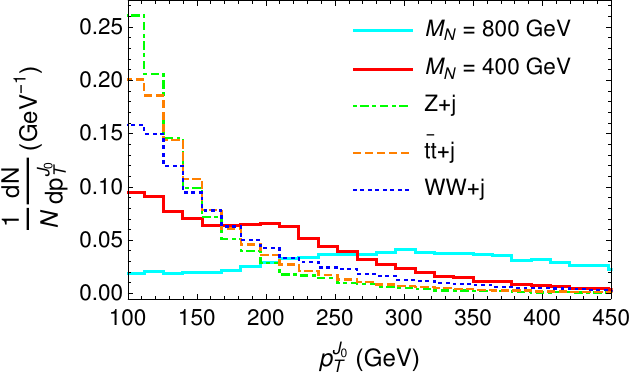}
	\caption{Normalized differential distributions of missing transverse energy (MET) (left) and the transverse momentum of the leading fat jet $p_T^{J_0}$  (right). These differential distributions are after the $baseline~selection~cuts$. The distribution of heavy neutrino benchmark points with $M_N = 400$ and $800$ GeV is shown along with three dominating background processes. }
	\label{fig:dist_var2}
\end{figure*}

As the two leptons in the signal process are produced at two different stages of decay, they carry distinctly different transverse momentum profiles. The second lepton originating from the heavy neutrino decay is expected to be significantly boosted, since the relevant $M_N$ are large. The hardest lepton in the signal event is hence generally expected from this stage and is expected to peak around $(M_N^2 - M_W^2)/(2 M_N)$. This may be noted in Fig.~\ref{fig:dist_var1} (left). All the SM backgrounds display milder hard-lepton transverse momentum profiles in comparison.  Distributions for next leading muons is also presented in Fig.~\ref{fig:dist_var1} (right). All these differential distributions are normalized and are shown after applying the above mentioned {\sf baseline~selection} criteria. 

Now, let us consider the typical missing transverse momenta distributions for signal and backgrounds. In Fig.~\ref{fig:dist_var2} (left) we show the missing transverse momentum (MET) distributions for the two benchmark signals and various backgrounds. MET is calculated from the transverse momentum imbalance of all the isolated objects such as leptons, photons and jets, as well as any unclustered deposits. MET for our signal process is expected to be relatively small, affected only by mismeasurements in clustering and jet reconstructions; no missing particles are involved per se. On the contrary, a large fraction of the background processes come with leptons from $W^\pm$ decays which are always associated with corresponding neutrinos. These thereby produce substantial MET contribution over and above contributions from jet mismeasurements. This trend is discernible in the plots.

The next three distributions we discuss primarily define the characteristics of the highest transverse momentum fat jet ($J_0$), which we rely upon heavily to mitigate backgrounds further. We will primarily utilize fat jet transverse momentum ($p_T^{J_0}$), jet mass ($M^{J_0}$) and N-subjettiness ($\tau_{21}^{J_0}$) for signal background discrimination and tagging. 

Boosted fat jet topologies and their associated jet substructures have proven crucial in various supersymmetric and non-supersymmetric LHC searches \cite{Adams:2015hiv}. In the $l^\pm l^\mp J$ topology of present interest, the fat jet evolves from the boosted, hadronically decaying $W^\pm$; the right handed sterile neutrinos $N_R$ are heavier than $W^\pm$ giving the latter large boosts. In the analysis, the importance of jet substructure therefore primarily manifests as a means to efficiently tag boosted, hadronically decaying $W^\pm$. As mentioned, we will utilize two well-known jet substructure variables towards this requirement --  N-subjettiness \cite{Thaler:2010tr,Thaler:2011gf} and jet-mass. 

The fat jet appearing from $W^\pm \rightarrow q \bar{q}'$ potentially retains some information of its two-prong structure. We would like to leverage this aspect to help tag it.  N-subjettiness \cite{Thaler:2010tr,Thaler:2011gf} is defined as 
\begin{eqnarray}
 \tau_N^{(\beta)} = \frac{1}{\mathcal{N}_0} \sum\limits_i p_{i,T} \min \left\lbrace \Delta R _{i1}^\beta, \Delta R _{i2}^\beta, \cdots, \Delta R _{iN}^\beta \right\rbrace.
 \label{eq:nsub_N}
\end{eqnarray}
Here, $\mathcal{N}_0=\sum\limits_i p_{i,T} R_0$ for a jet radius $R_0$, with $i$ running over the constituent particles, and $p_{i,T}$ is the respective transverse momentum. We compute N-subjettiness with the thrust measure $\beta = 2$.  The $\eta-\phi$ distance between a candidate $\alpha$-subjet and constituent particle $i$ is defined as $ \Delta R_{i\alpha} = \sqrt{(\Delta \eta)^2_{i\alpha}+(\Delta \phi)^2_{i\alpha}}$. N-subjettiness tries to quantify how much the original jet seems to be composed of N daughter subjets. A small value of $\tau_N$ suggests that the original jet may consist of $N$ or fewer subjets. It has been demonstrated that a good discriminant to tag an N-subjet object is to consider ratios of adjacent N-subjettiness values \cite{Thaler:2010tr,Thaler:2011gf}. For W-tagging, since the $W^\pm$ yields two subjets that are collimated, the variable of interest would therefore be $\tau_{21}=\tau_2/\tau_1$. The mass of the fat jet ($M_J$), is another discriminant that may be leveraged to identify the jet as originating from a hadronically decaying $W^\pm$. The fat jet four momenta is the vector sum of all the constituent four momenta, in the E-scheme. From this reconstructed fat jet four momenta ($P_T^J$) the invariant fat jet mass ($M_J^2$) may be computed.

\begin{figure*}[t!]
    \includegraphics[scale=0.8]{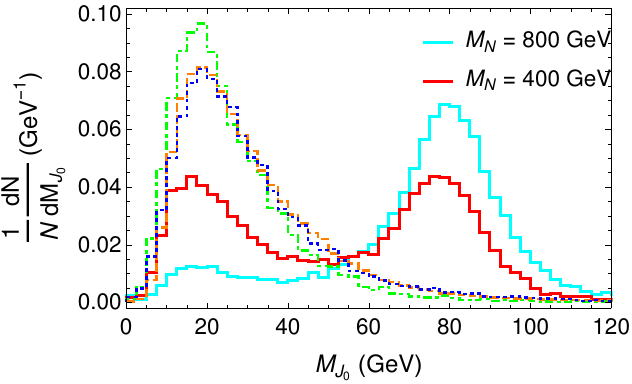}
    	\hspace{4pt} 
	\includegraphics[scale=0.8]{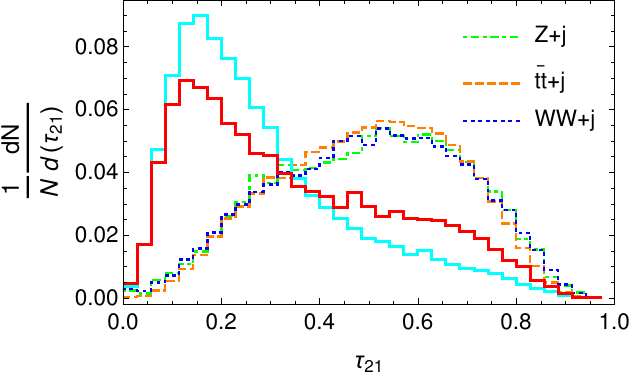}
    \caption{Normalized differential distributions of invariant mass $M^{J_0}$ (left) and N-subjettiness ratio $\tau_{21}^{J_0}$ (right) of the leading fat jet. The selection criteria are same as Fig.~\ref{fig:dist_var1}. The distribution of heavy neutrino benchmark points with $M_N = 400$ and $800$ GeV is shown along with three dominating background processes.    }
	\label{fig:dist_var5}
\end{figure*}

\texttt{Delphes 3.3.2} \cite{deFavereau:2013fsa} hadron calorimeter outputs are clustered using \texttt{FastJet 3.1.3} \cite{Cacciari:2005hq,Cacciari:2011ma} to reconstruct the candidate fat jet. The N-subjettiness extension, available through \texttt{FastJet-contrib}, is used to compute $\tau_{21}$. For tagging the hadronically decaying $W^\pm$ we adopt parameter choices from a CMS analysis \cite{Khachatryan:2014vla}, as a starting point. We choose Cambridge-Achen \cite{Dokshitzer:1997in,Wobisch:1998wt} for the recombination algorithm, with a jet-cone radius $R=0.8$. Further refinements for W-tagging are then made by requiring specific cuts on $\tau_{21}$ and $M_J$. 

The $P_T$ of the boosted $W^\pm$ scales as $P_T^W \sim (M_N^2-M_{W}^2)/(2 M_N)$. Fig.~\ref{fig:dist_var2} (right) presents the distributions for fat jet transverse momenta $P_T^{J_0}$. With the minimum  transverse momentum of 100 GeV already implemented during primary selection, one notices the spread and second peak (towards higher values) for the signal distributions suggestive of its origin from the decay of the heavy $N$. This second peak in comparison to one at the lower value becomes more and more prominent as expected for larger $M_N$. The $P_T$ of candidate fat jets from all background processes monotonously fall. Evidently, larger values for the transverse momentum cut helps us in selecting relatively more signal-like  fat jets, in comparison to background events. This may probably be at the cost of some signal events but would nevertheless also help mitigate backgrounds, and potentially result in a net significance gain.

The two plots in Fig.~\ref{fig:dist_var5} highlight the internal characteristics of the identified fat jets, in the form of the invariant jet mass $M^{J_0}$ (left)  and the N-subjettiness $\tau_{21}^{J_0}$ (right). These jet substructure variables help correctly tag the candidate fat jet as W-like or not. Construction of these variables are as defined earlier in this section and they provide a powerful tool to discriminate the QCD jet contaminations. 

Signal distributions for $M^{J_0}$ clearly peak at $M_W$ reflecting their origin as W-like jets. For low $M_N$, the $W^\pm$ boosts are smaller and with the $P_T >100\,\mathrm{GeV}$ cut and $R=0.8$ jet radius some of the $W^\pm$ hadronic decay products are not captured inside the cone. This is evident as a secondary, spurious peak at a lower mass value in the plots. However, for heavier $M_N$ or with a choice of a larger transverse momentum cut, only the peak around $80\,\mathrm{GeV}$ survives. We retain the $P_T$ cut at $100\,\mathrm{GeV}$, as this gives an overall higher signal significance across the $M_N$ mass ranges under consideration. The most SM backgrounds peak at low $M^{J_0}$, except those where fat jets are indeed W-like {\it e.g.} backgrounds from $Z^l W^h$ + jets or $Z^l W^l W^h$ (superscript $l/h$ for leptonic/hadronic decay modes). These particular backgrounds are not shown in the plots for readability and for the reason that their final contributions in the present channel will be minuscule after applying all selection criteria. 

The N-subjettiness ratio $\tau_{21}=\tau_2/\tau_1$ is the other jet substructure quantity of interest. It quantifies the two-pronged nature of the fat jet arising from boosted-$W^\pm$ hadronic decays and discriminates it from the structureless jets coming from QCD. The distribution of $\tau_{21}$ for signal and backgrounds is shown in Fig.~\ref{fig:dist_var5} (right). By construction $\tau_{21}^{J_0}$ for W-like fat jets is expected to peak at low values. The separation between the hadronic decay products of $W^\pm$ scale as $M_W/P_T^W$. It is observed that the W-like fat jets from the signal benchmark points peak around $0.15$, whereas most backgrounds with QCD jets peak at much higher values, around $0.6$.

\begin{figure}[t!]
	\includegraphics[scale=0.80,angle=0,keepaspectratio=true]{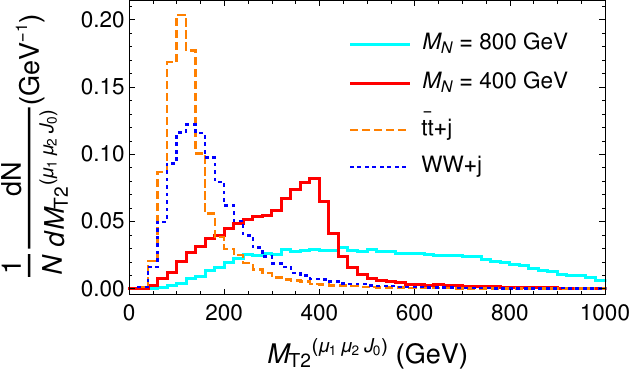}
	\caption{
	Normalized differential distributions of $M_{T2}$ for the two muon and the leading fat-jet for an asymmetric subsystem (${\mu_1\mu_2 J_0}$) is  shown for $M_N = 400$ and $800$ GeV  along with dominant backgrounds from top-pair and $W$-pair.}                                                                               
	\label{fig:dist_M_T2}
\end{figure}

With a detailed understanding of the above kinematic and jet substructure variable distributions we are now in a position to make appropriate choices for the final selection criteria. Choice for the final {\sf event selection criteria} are optimized towards the lower mass regions with the benchmark point at  $M_N=400$ GeV. This is chosen for simplicity of demonstration and the fact that one gets a large cross-section here with a reasonable efficiency from jet characteristics. It nevertheless also provide good signal significance across the full mass range of interest. Various kinematic variables along with fat jet observables are constrained in the following way :

\begin{itemize}

\item  \textbf{C2} The highest $p_T$ muon is selected with $p_T > 100$ GeV and the next $p_T$ ordered muon is selected with $p_T > 60$ GeV. These relatively harder selection criteria are effective in mitigating most of the backgrounds, as motivated from Fig.~\ref{fig:dist_var1}. The large $t \bar{t}$ background is reduced without affecting the signal substantially.

\item  \textbf{C3} To control the huge backgrounds coming from leptonic decays of $Z$ bosons, we veto events if the opposite-sign di-muon invariant mass ($ M_{\mu^+\mu^-}$) is less than 200 GeV. The harder cut on $M_{\mu^+\mu^-}$ also reduces parts of the $t\bar{t}$ background further. 

\item \textbf{C4} We apply a b-veto to reduce the $t\bar{t}$ background without affecting signal acceptance.

\item \textbf{C5} As mentioned earlier, it is evident that our signal does not have any missing particle per se, hence should have relatively low MET. The final $\slashed{P}_T$ would of course get contributions from measurements and uncertainties. Taking into account the unclustered towers, we consider only events with a maximum MET of $60$ GeV. 

\item \textbf{C6} Events with the leading fat jet (${J_0}$) having transverse momentum $p_T^{J_0} > 150$ GeV are selected. This is done in order to increase the purity of the boosted jets further.

\item \textbf{C7} For signal events, the fat jet is reconstructed from the boosted $W$ boson. Hence, we demand for the corresponding mass, $M^{J0} > 50$ GeV. 
  
\item \textbf{C8} We choose events with N-subjettiness $\tau^{J_0}_{21} < 0.4$.

\item \textbf{C9} After identifying $t\bar{t}$ as a major source of irreducible background, one needs to engage some new event constraining variables, beyond just the jet variables. If b-jets, as well as the opposite sign di-leptons are identified, the transverse mass variable  $M_{T2}$~\cite{Lester:1999tx, Cho:2007qv, Barr:2011xt} works exceptionally well---providing a distribution with an upper limit at the top mass. In our present study only one fat-jet is identified which can originate from either of the b-jets. In such a scenario, an asymmetric $M_{T2}$, considering the $b \mu^+ \mu^-$ subsystem~\cite{Burns:2008va, Konar:2009qr} may be shown to follow the same inequality as before; and thus useful in disentangling the signal, where all decay products are produced from a single prong. We choose $M_{T2}^{(\mu_1\mu_2J_0)}$  $\geq$ 250 GeV to reduce $t\bar{t}$ background by a significant amount as shown in Fig.~\ref{fig:dist_M_T2}.

\end{itemize}

 \begin{table*}[t!]
 \tiny	
 \centering
 \renewcommand{\arraystretch}{2.5}
 \begin{tabular}{|c|c||c|c||c|c|c|c|c|c|}
 \hline
 \multicolumn{2}{|c||}{Cuts}    & \multicolumn{2}{c||}{Signal}     & \multicolumn{6}{c|}{Background}        \\ \hline\hline
   &    & $M_N = 300$(GeV)  & $M_N = 400$(GeV) & $ \,Z^l+j $&$t \bar t+j$&$ \,W^l \,W^l+j $& $\,Z^l\,W^h+j$ &$\,Z \,W^l+j $  & $Z^l \,W^l\,W^h +j$   
                                                                               \\ \hline    
\textbf{C1} & Pre-selection +       
 & {1026.85} & {486.05} & {$5.1 \times 10^7$} & {$5.6 \times 10^6$} & {$2.2 \times 10^5$} &  {$5.7 \times 10^5$} & {$1.4 \times 10^4$}  & {120.8}\\ 
& $\mu^+\mu^- + 1 J$              &  [$100\%$]    & [$100\%$]                           & [$100\%$]   &[$100\%$]&[$100\%$]&[$100\%$]&[$100\%$]&   [$100\%$]

\\\hline
 \textbf{C2}&$p_T (l_1) >$ 100 GeV +  & 787.40  & 416.05           & $8.9 \times 10^6$
                          & $6 \times 10^5$          & $3.6 \times 10^4$ 
                          & $10.2 \times 10^4$         & 2737.3
                          & 33.64
 \\& $p_T (l_2) >$ 60 GeV & [$76.68\%$]  & [$85.59\%$]     &[$16.88\%$] 
                          & [$10.68\%$]      &[$16.86\%$]
                          &[$18.27\%$]     &[$19.48\%$]
                          & [$27.85\%$]        
                          
 \\\hline
\textbf{C3}&$M_{\mu^+ \mu^-} >$ 200 GeV  & 588.50 & 349.80         & 583.7
                          & $4.2 \times 10^5$        & $2.2 \times 10^4$
                          & 9.4           & 724.76 
                          & 10.42
\\     &              &[$57.31\%$]      &[$71.96\%$]       &[$0.0010\%$]
                          &[$7.37\%$]       &[$9.87\%$]
                          &[$0.0016\%$]      &[$5.15\%$]
                          & [$8.62\%$]

 \\\hline
\textbf{C4}&b-veto          & 501.12          & 295.7         & 530.6
                           & $6.2 \times 10^4$         & $2.0 \times 10^4$
                           & 8.54         & 647.7
                           & 8.5
 \\  &          &[$48.80\%$]            &[$60.84\%$]       &[$9.9\times10^{-4}\%$]
                           &[$1.1\%$]       &[$9.09\%$]
                           &[$0.0014\%$]      &[$4.6\%$]
                           & [$7.06\%$]

\\\hline
\textbf{C5}& MET $< 60$       & 459.29         & 263.97           & 371.4
                          & $2.1 \times 10^4$          & 7796
                          & 6.0           & 353.04
                          & 3.8
\\     &       &[$44.72 \%$]            &[$54.31\%$]            &[$6.9\times10^{-4}\%$]
                 &[$0.37\%$]                &[$3.54\%$]
                          &[$10.3\times10^{-4}\%$]        &[$2.5\%$] &[$3.13\%$]

\\\hline
\textbf{C6}& $p_T^{J_0}>150$ GeV   & 242.86   & 175.75      & 265.32 
                          & 7635.9        & 3898
                          & 4.27         & 195.96
                          & 2.5
\\ &   &[$23.65\%$]                    &[$36.15\%$]                     &[$4.9\times10^{-4}\%$]               &[$0.13\%$]      &[$1.77\%$]
                          &[$7.3\times10^{-4}\%$]      &[$1.39\%$]
                          & [$2.12\%$]   

\\\hline
\textbf{C7}&$M^{J_0}>50$ GeV    & 161.43      & 132.09        & 42.4 
                          & 1754.2         & 843.08
                          & 0.68          & 48.21 
                          & 1.3
\\ &             &[$15.72\%$]                &[$27.17\%$]   &[$7\times10^{-5}\%$]                 &[$0.03\%$]    &[$0.38\%$]     
                          &[$1.2\times10^{-4}\%$]    & [$0.34\%$] 
                          &[$1.11\%$]  
    
\\\hline
\textbf{C8}& $\tau_{21} < 0.4$    & 141.59    & 117.75        & 21.18  
                          & 928.6         & 396.7
                          & 0.34          & 32.66
                          & 1.0
\\   &    &[$13.78\%$]                  &[$24.22\%$]     &[$3.9\times10^{-5}\%$]
                          &[$0.016\%$]      &[$0.18\%$]
                          &[$5.8\times10^{-5}\%$]      &[$0.23\%$]
                          &[$0.85\%$]             
 \\\hline
\textbf{C9}& $M_{T2}^{(\mu_1\mu_2J_0)}$  $\geq$ 250 GeV  & 135.55    & 113.465      & 14.72 
 & 412.6     & 280.0 
 & 0.24    & 23.22 
 & 0.89 
 \\ & & [$\pm$12.5 $\pm$ 6.7 ] & [$\pm$ 9.04 $\pm$  5.6] & [$\pm$ 1.32 $\pm$ 0.73] & [$\pm$ 82.5 $\pm$  20.63 ] &[$\pm$ 86.8$\pm$ 14.0] & [$\pm$ 0.05 $\pm$   0.01  ] & [$\pm$ 5.5 $\pm$ 1.1]& $[\pm$ 4.45$\pm$ 0.04]
 
 \\  &     &[$13.20\%$]                  &[$23.43\%$]     &[$2.7\times10^{-5}\%$]
 &[$0.007\%$]      &[$0.13\%$]
 &[$4.1\times10^{-5}\%$]      &[$0.16\%$]
 &[$0.74\%$]                          
                          
 \\\hline
 \end{tabular}
 \caption{Expected number of events in $\mu^+ \mu^- + J$ channel after implementation of the corresponding event selection criteria for an integrated luminosity of $3000$ fb$^{-1}$ at the 13 TeV LHC. We choose the value of mixing angle $|V_{\mu N}|^2$ to be 0.01. The signal events are shown for Dirac neutrino mass $M_N = 300$ and $400$ GeV in the case of single flavor. The statistical and systematic uncertainties are reported respectively after the C9 cut. 
 }
 \label{tab:cut-flow}
 \end{table*}

\begin{figure}[h]
\centering
\includegraphics[scale=0.5]{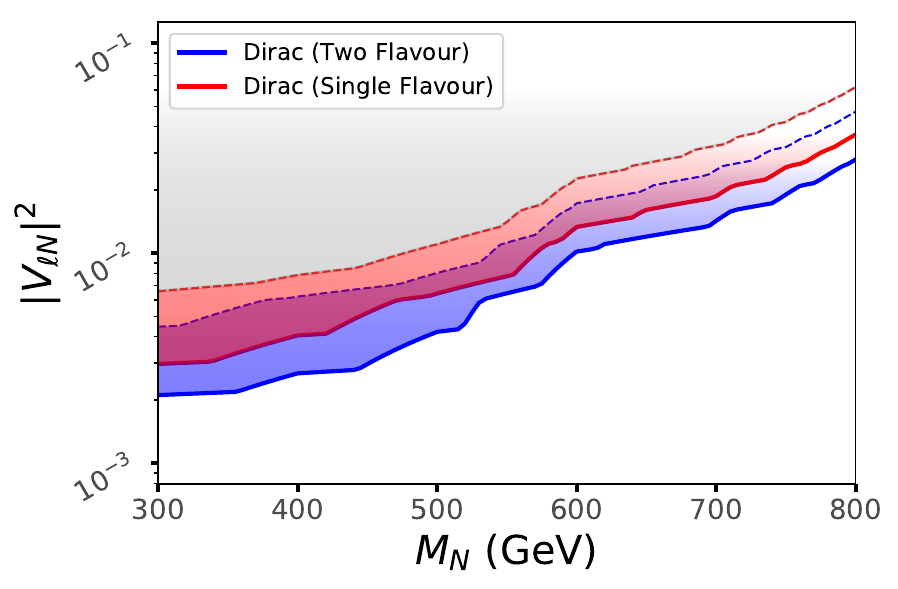}

\caption {The figure shows the 2-$\sigma$ exclusion limits, in terms of heavy neutrino mass $M_N$ and $|V_{\ell N}|^2$, at $3000$ fb$^{-1}$ of integrated luminosity at the 13 TeV LHC. Where dotted line corresponding to each colour represents the $5\%$ systematics uncertainty in the total background estimation. The results for the $2$-flavor ($1$-flavor) case up to $M_N=550$ GeV ( $500$ GeV) are better than the optimistic scenario mentioned in \cite{Atre:2009rg} from the electroweak precision measurement in \cite{Nardi:1994iv,Nardi:1994nw}. The optimistic limit on $|V_{\mu N}|^2$ has been mentioned as $6.0\times 10^{-3}$ in \cite{Atre:2009rg}.}                                                                                      
	\label{fig:dist_exp}
\end{figure}

We present the analysis and describe the results explicitly for a few example benchmark signal points -- $M_N = 300$ GeV and $400$ GeV -- for single flavor Dirac neutrino, together with the main backgrounds. In Table~\ref{tab:cut-flow} we summarize the effect of each selection cut in the order presented before. The analysis presented here is not a template based. The cuts presented in Table~\ref{tab:cut-flow} are same for all the masses of $M_N$ considered in the paper. Hence the
background yield is independent of our benchmark choices.  Expected number of events after baseline selection and the number of surviving events after each subsequent cuts (also in terms of percentages) are presented in the first and successive rows, assuming an integrated luminosity of $3000\,\rm{fb}^{-1}$. Here, one can follow sequentially the cut efficiency for the signal and background events as per our previous discussions. It is seen that harder cuts for leptons indeed reduce all the backgrounds, without affecting the signal significantly. One can have an even harder choice for the highest-PT lepton, when probing larger $M_N$. Veto on b-jets shrinks events from $t \bar{t}$ and missing transverse energy (MET) is effective for all backgrounds possessing additional MET contributions from neutrinos. The other three selections in the form of $p_T^{J_0}$, $M^{J0}$ and $\tau^{J_0}_{21}$ rely on the fat jet substructure and reduce all dominant backgrounds where fat jets are mimicked by QCD jets. We further use $M_{T2}^{\mu_1\mu_2J_0}$ to reduce the dominant $t\bar{t}$ background. Overall efficiency for the $400$ GeV signal can be observed to be around $23 \, \%$, whereas different backgrounds are reduced to between $2.7\times10^{-5}\,\%$ and $0.74\, \%$. Note that the signal cross-section for heavier mass falls significantly, due to production s-channel suppression. However, better substructure efficiencies partially mitigate that reduction. This is evident from the $M_N = 300$ GeV and $400$ GeV results. Statistical significances for the observed signal events ($S$) over the total irreducible standard model backgrounds ($B$) are calculated adopting the familiar expression ${\cal S} = \sqrt{2 \times \left( (S+B) \ln (1+S/B) - S\right)}$.

\subsection{Flavor democratic case}
\label{FD1}
In this section we study the FD scenario where the two degenerate RHNs are equally mixed with the $e$ and $\mu$ leptons. After the signal and SM background analyses we have displayed the exclusion limits on the $|V_{\ell N}|^2$ as a function of the $M_N$ in Fig.~\ref{fig:dist_exp}. To account for the effects coming from systematics, we consider a characteristic $5\%$ systematics uncertainty from \cite{Kim:2018cxf}in the background estimation; which is represented by dotted line, corresponding to all three cases in Fig.~\ref{fig:dist_exp} Furthermore, note that in any actual analyses, data driven techniques, for instance utilising ant b-jet vetoes or ABCD type methods, will help significantly reduce systematic uncertainties in background estimations \cite{ABCD}. We have assumed $3000$ fb$^{-1}$ integrated luminosity at $13$ TeV LHC. There is no direct search result for the RHNs at this mass range for the inverse seesaw scenario at the colliders. In this same figure limits are also indicated if the OSDL along with a fat-jet is searched for the Majorana neutrino. Heavy Majorana neutrino, if exits in nature, should also show up in equal strength producing lepton number violating same sign di-leptons, where backgrounds are immensely suppressed. Evidently SSDL bounds are extremely strong and studied extensively \cite{Das:2017gke} in see-saw framework along with fat-jet. 
Corresponding efficiency for selecting muon signals is at $70\%$ while it is reduced to $50\%$ for electron events. For different seesaw models, the exclusion limits for lower $M_N$ values can be as low as $2 \times 10^{-3}$. 
The heavy neutrino production, especially at heavier mass, can get $(10\%-60\%)$ additional contribution for the mass limit under consideration from $\gamma- W^\pm$ fusion \cite{Dev:2013wba,Alva:2014gxa}, and thus can potentially improve the exclusion limits further.
%

\begin{figure}[t!]
	\includegraphics[scale=0.60,angle=0,keepaspectratio=true]{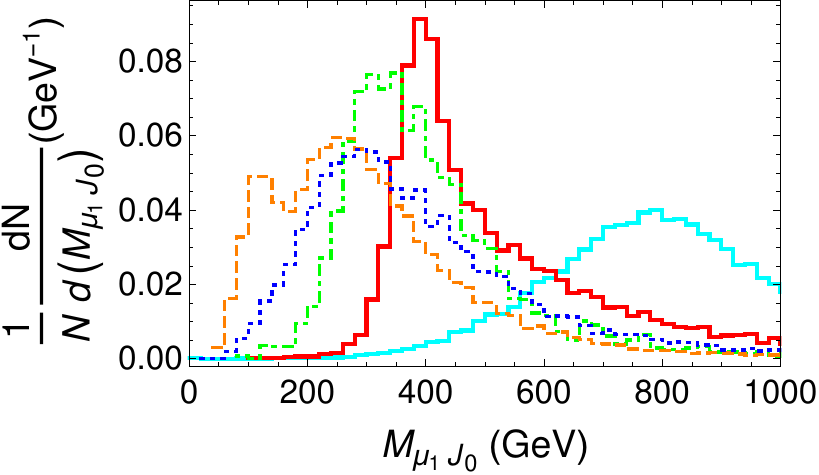}
	\caption{
		Normalized differential distributions of invariant mass for the hardest lepton and the leading fat-jet system ($M_{\mu_1 J_0}$) 
		is  shown for $M_N = 400$ and $800$ GeV  with dominant backgrounds. Choice of line colors and types are similar to previous Fig.~\ref{fig:dist_var5} }                                                                               
	\label{fig:dist_M_mu1J}
\end{figure}

\begin{table}[h]
	\tiny	
	\centering
	\renewcommand{\arraystretch}{2.5}
	\begin{tabular}{|c||c|c||c|c|c|c|c|c|}
		\hline
		Cut    & \multicolumn{2}{c||}{Signal}     & \multicolumn{2}{c|}{Background}        \\ \hline\hline
		& $M_N =600 (GeV)$ & $M_N =800 (GeV)$& $t \bar t+j$&$ \,W^l \,W^l+j $  \\ \hline
		
		Table I + & & & & \\
		$p_T (l_1) >$ 200 GeV  & {$40.01 $ $\pm$8.2 $\pm$ 2.0} & {$11.65$ } $\pm$ 1.4 $\pm$ 0.6 & {$180.5$} $\pm$ 36.0 $\pm$ 9.0 & {$154$} $\pm$ 46.2 $\pm$ 7.7\\

		$M_{\l_1 J_0}>$ 500 GeV    &  [$32.3\%$]    & [$33.62\%$]      &[$3.2\times10^{-3}\%$] & $[0.07\%$]  \\
		\hline
	\end{tabular}
	\caption{Expected number of events  after implementing additional cuts (together with cuts described in Table~\ref{tab:cut-flow}) suited for higher mass probe, {\it i.e.} $M_N > 600 $GeV. The signal events are shown for Dirac neutrino in the case of single flavor. Only two dominant backgrounds are presented here. The statistical and systematic uncertainties are reported respectively.}
	\label{tab:cut-flow2}
\end{table}
We reiterate that the given limits are based on simple criteria, optimized at $M_N = 400$ GeV. There is ample scope for improvements at higher masses. One can readily recognize the quantities which may crucially factor in for higher masses. For instance, RHN possessing mass of several hundreds of GeV would often produce both boosted leptons as well as collimated jets from boosted W bosons. $P_T$ of hardest lepton will evidently shift towards higher values in Fig.~\ref{fig:dist_var1} for these heavier masses, and the peak position will be around half of relevant heavy neutrino mass. We also illustrated the invariant mass of this hardest lepton and the fat-jet system, in Fig.~\ref{fig:dist_M_mu1J}, which peaks around the benchmark heavy neutrino mass.  Effective use of these two variables, as shown in Table~\ref{tab:cut-flow2}, can provide an improvement by a factor of slightly more than two, for the $|V_{lN}|^2$ limits; for the heavy-neutrino masses greater than $M_N = 600$ GeV.
\subsection{General parametrization: Flavor non democratic case}
\label{FND1}
In this section we study the flavor non democratic (FND) scenario where the flavor structure is carried out by the Dirac Yukawa coupling. According to our formalism the mixing between the light and heavy neutrinos $(\mathcal{R}=V_{\ell N})$
 is a function of the Dirac phase $(\delta)$ and the Majorana phase $(\rho)$. $\mathcal{R}^{\ast} \mathcal{R}^T$ is a function of the general parameter $Y$ coming from the general orthogonal matrix $O$. 
We perform a parameter scan by varying these parameters between $-\pi \leq \delta, \rho \leq \pi $ with an interval of $\frac{\pi}{20}$ and $0\leq Y \leq 1$ with an interval of 0.02.
The elements of the Dirac mass matrix grow exponentially with $|Y|$. For a value $Y >$  1, the neutrino oscillation data are realized under the fine-tuning between the large elements.
 Although the neutrino oscillation data are correctly reproduced for any values of Y in the general parametrization, we only consider  $Y \leq 1$ to avoid the fine-tuning.
 The ranges of the independent parameters like $\delta, \rho$ and $Y$ satisfy the constraints on $\epsilon$. Hence we calculate the cross section for the $i-$th generation RHN
 at the LHC through the $W$ boson exchange process  $u\overline{d} \to \ell_{\alpha}^{+} N_i$ and $d\overline{u} \to \ell_{\alpha}^{-} \overline{N_i}$. Hence the production cross section at the 
 LHC can be written as 
 \bea
 \sigma(q\overline{q^\prime} \to \ell_\alpha N_i) =\sigma_{\rm{LHC}} |\mathcal{R}_{\alpha i} (\delta, \rho, Y)|^2
 \eea
 where $\sigma_{\rm{LHC}}$ is the production cross section the RHNs at the LHC.
 The partial decay widths of the RHN $(N_i \to \ell_{\alpha} W^+/ \nu_\alpha Z/ \nu_\alpha h)$ can be found by multiplying the corresponding decay widths by 
 $|\mathcal{R}_{\alpha i}(\delta, \rho, Y)|^2$. As a result the corresponding branching ratios can be expressed in terms of $\delta, \rho$ and $Y$ through the elements of the mixing matrix.
 Running the parameters within the allowed ranges and satisfying the constraints obtained from $\mathcal{N N}^{\dagger}\simeq 1-\epsilon$ we obtain the upper limits on the mixing angles 
 for the NH and IH cases with two electron $(|V_{eN}|^2)$ and two muon $(|V_{\mu N}|^2)$ final states, respectively in Tab.~\ref{tabmix}. We compare our results with the bounds obtained from 
 the EWPD \cite{deBlas:2013gla, delAguila:2008pw, Akhmedov:2013hec} on $|V_{eN}|^2~(1.68 \times 10^{-3})$ and $|V_{\mu N}|^2~(9.0 \times 10^{-4})$ respectively.
 \begin{table}[t]	
 \setlength{\tabcolsep}{3pt}
 \renewcommand{\arraystretch}{1.1}
\setlength\extrarowheight{2pt}
	\begin{tabular}{|c|c|c|}
		\hline
		Mixing Angles   & Calculated upper limits  &  EWPD\cite{deBlas:2013gla, delAguila:2008pw, Akhmedov:2013hec} \\
		\hline \hline
		$|V_{eN}|^2$ (NH)  &  $6.908\times 10^{-4}$  & \multirow{2}{*}{$1.68 \times 10^{-3}$}  \\ 
		$|V_{eN}|^2$ (IH)  &    $1.884\times 10^{-4}$ & \\ \hline
		$|V_{\mu N}|^2$ (NH)  &  $8.963\times 10^{-4}$  & \multirow{2}{*}{$9.0 \times 10^{-4}$}  \\ 
		$|V_{\mu N}|^2$ (IH)  &  $1.923 \times 10^{-4}$  &\\ \hline
	\end{tabular}
	\caption{Calculated upper limits on the mixing angles for the NH and IH cases and comparison with the EWPD}
	\label{tabmix}
\end{table} 
We notice that the allowed upper limit on $|V_{\mu N}|^2$ and $|V_{eN}|^2$ in the NH and IH cases are below the corresponding EWPD limits.

\begin{figure*}[t]
\includegraphics[scale=0.22]{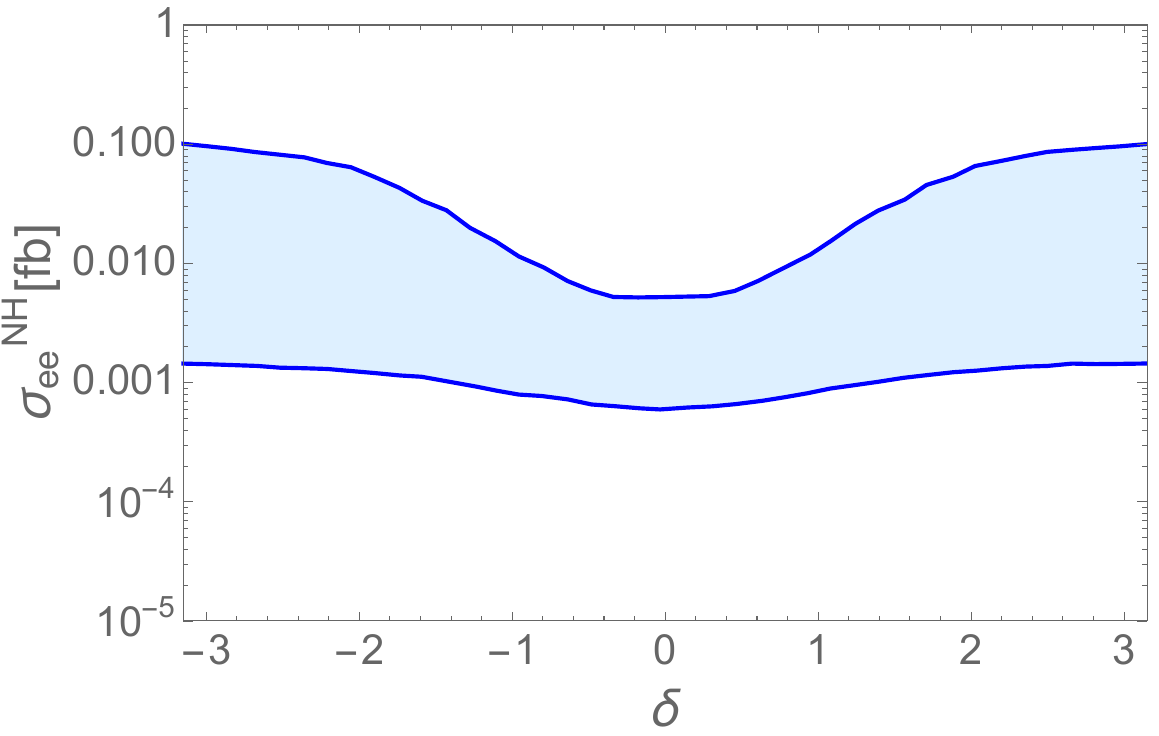}
\includegraphics[scale=0.22]{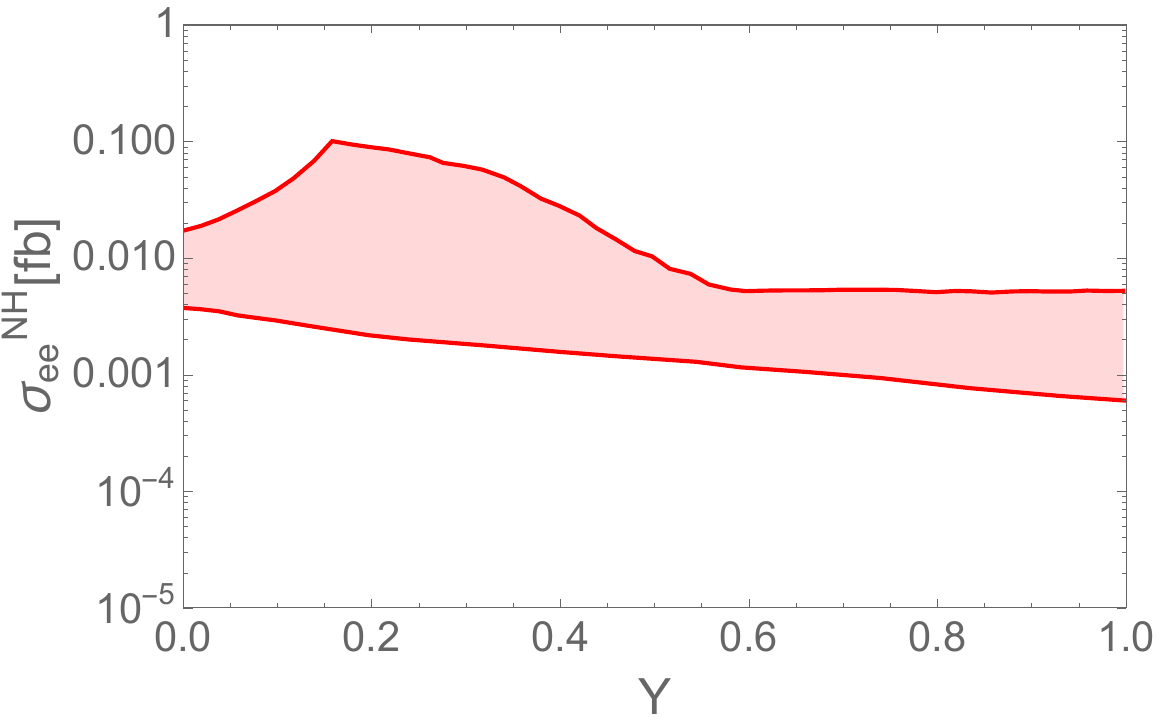}
\includegraphics[scale=0.22]{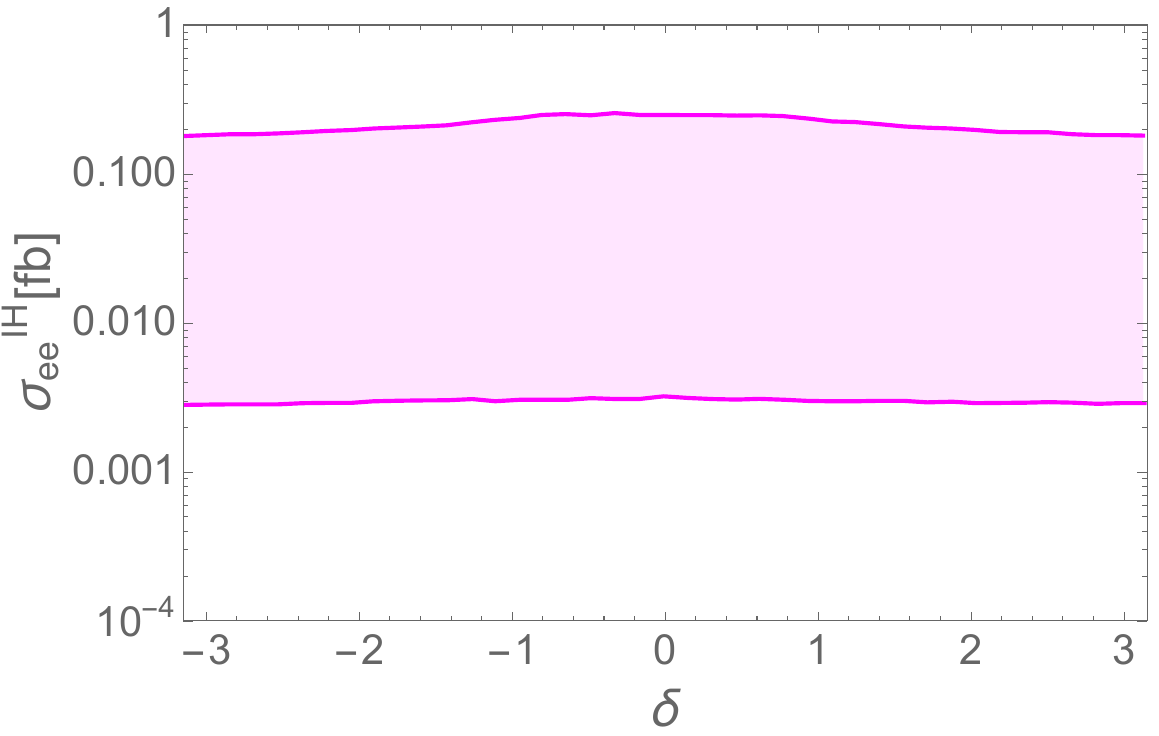}
\includegraphics[scale=0.22]{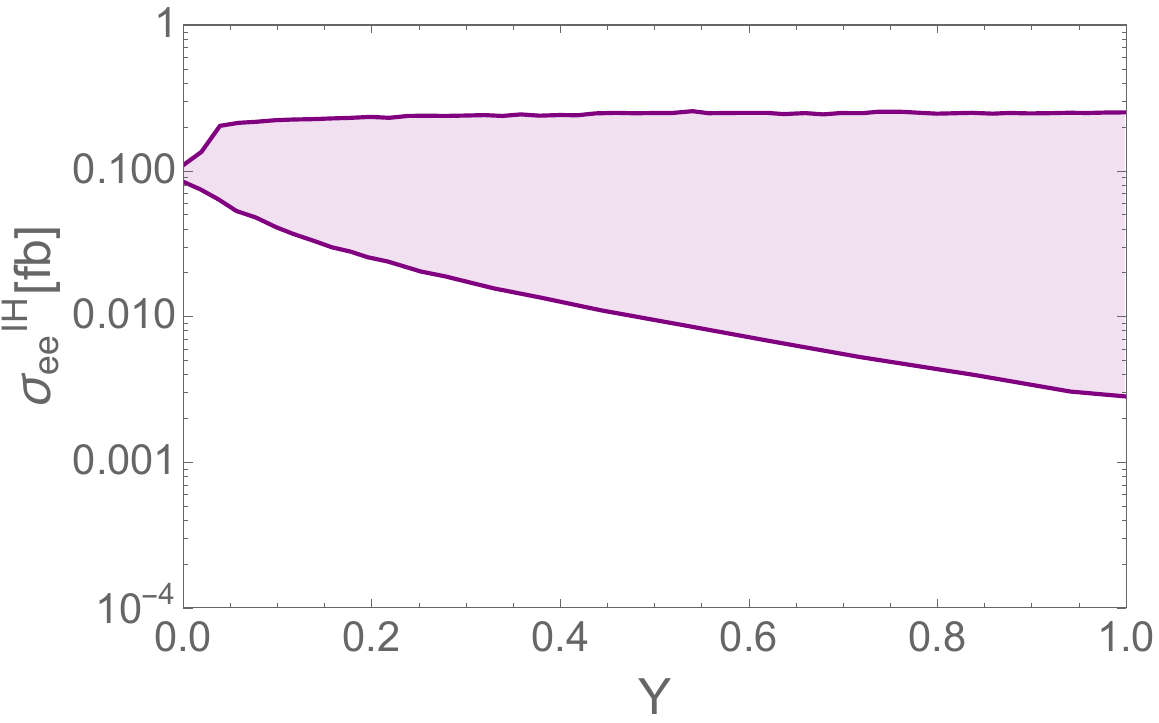}\\
 \includegraphics[scale=0.22]{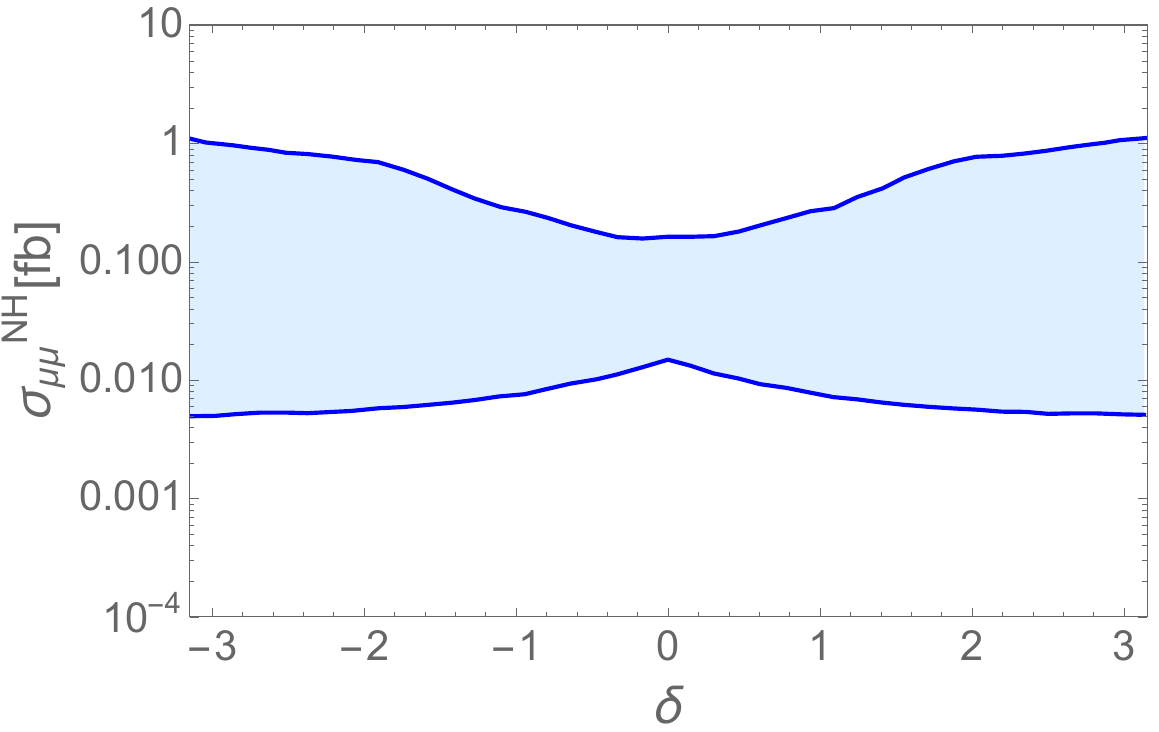}
 \includegraphics[scale=0.22]{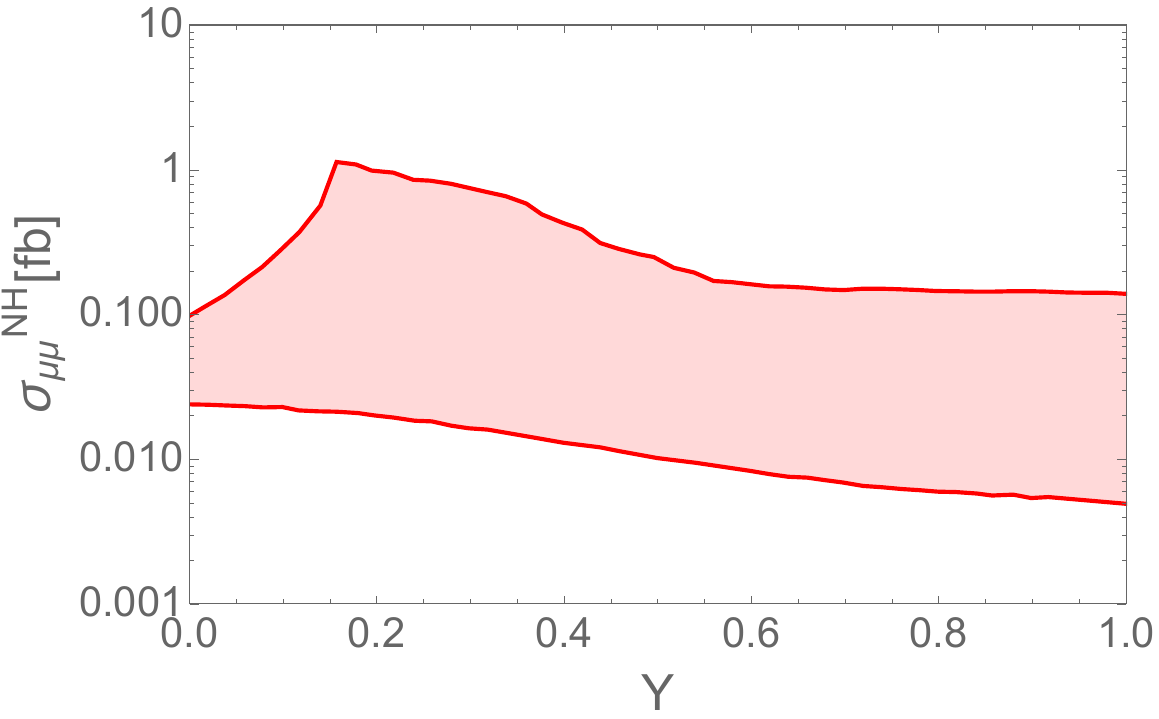}
\includegraphics[scale=0.22]{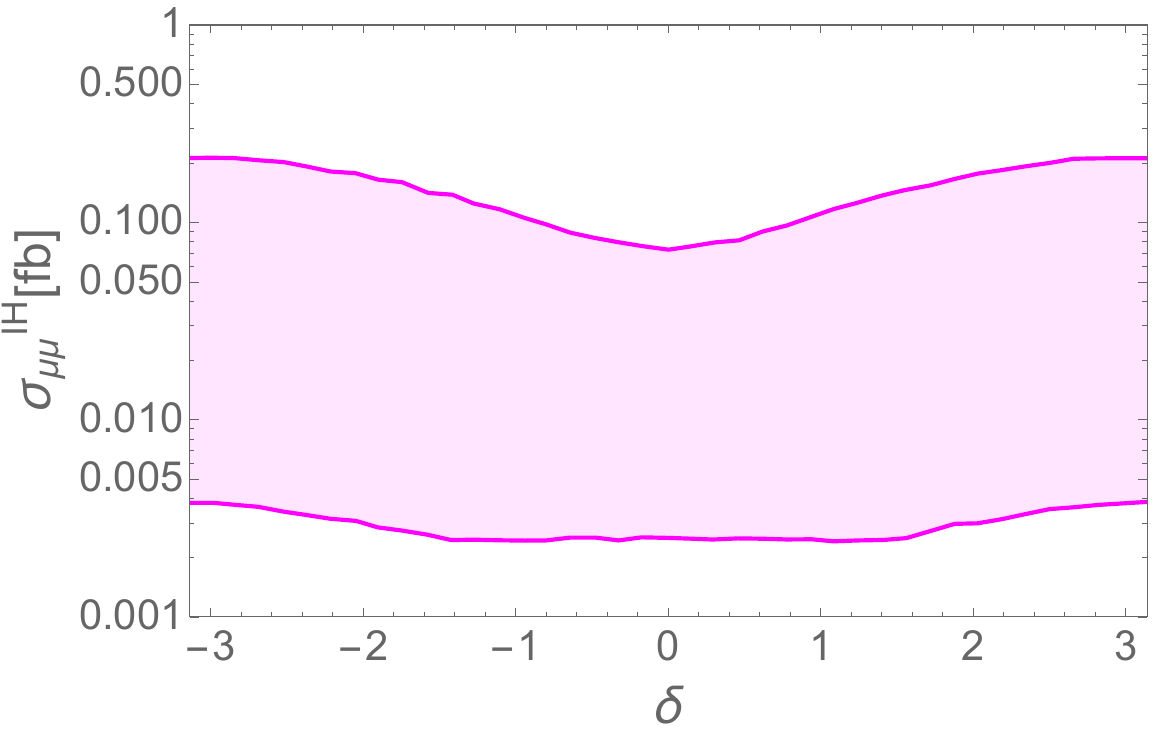}
\includegraphics[scale=0.22]{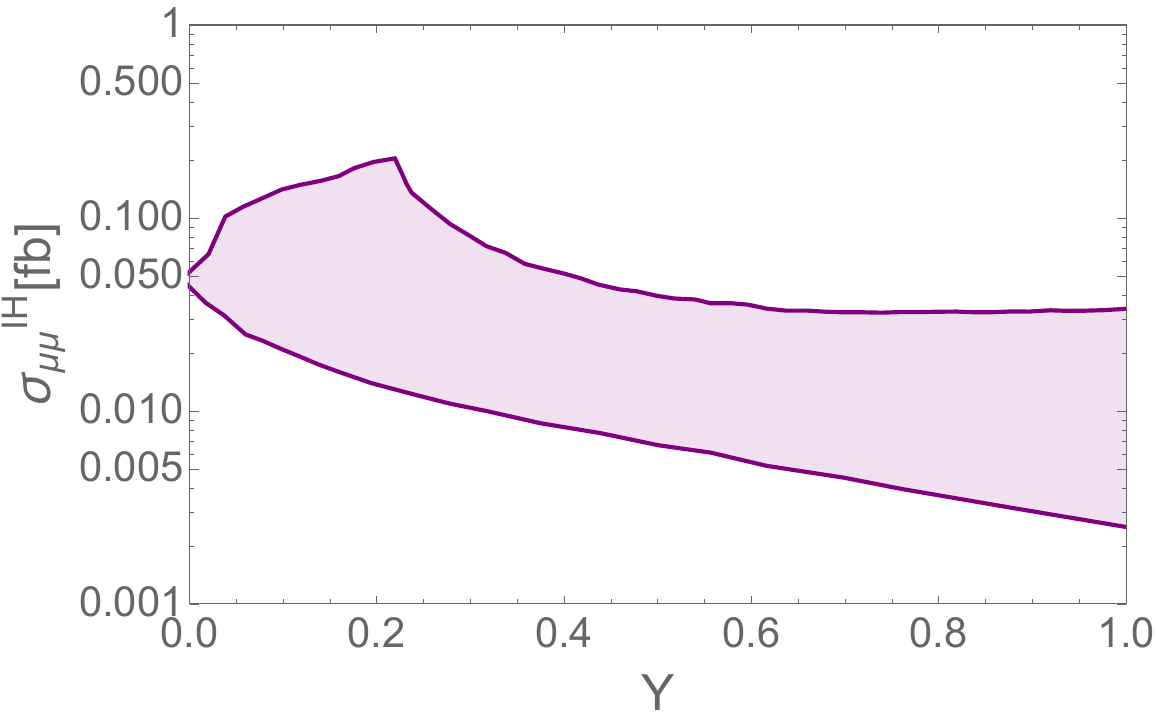}
\caption{LO cross section at the $13$ TeV LHC for the (upper row) $e^{\pm}e^{\mp} jj$ and (lower row) $\mu^{\pm}\mu^{\mp} jj$ final state in the general parametrization applying the constraints on the $\epsilon$-matrix.  $Y$ and $\delta$ parameter dependance were shown, where the left two columns stand for the NH cases whereas the right two columns for the IH case.
}
\label{fig:f1f2}
\end{figure*}

Fig.~\ref{fig:f1f2} shows the results of the parameter scan of the pseudo-Dirac RHN production cross section in the same flavor OSDL final state with a pair of jets coming 
from leading RHN production followed by its decay into a leading mode, $pp\to N\ell, N\to W\ell, W \to jj$ for $\ell=e$ or $\mu$ flavors at the LO for a benchmark value $M_N= 175$ GeV at the $13$ TeV LHC. 
We have three other benchmark points for $M_N$ such as $200$ GeV, $250$ GeV and $300$ GeV.  

Each point in the shaded region of the Figs.~\ref{fig:f1f2}, based on $Y$ or $\delta$ parameter dependance, satisfy all the experimental constraints imposed on $\epsilon$-matrix. 
The left two columns in upper (lower) row of Fig.~\ref{fig:f1f2} show the NH cases whereas the right two columns for the IH cases considering the production of $e^{\pm}e^{\mp}jj~(\mu^{\pm} \mu^{\mp}jj)$ events respectively. 
The estimated NLO cross sections $(\sigma_{\rm{LHC}}^{NLO})$ have been listed in  Tab.~\ref{tab0} matching \cite{Das:2016hof} satisfying all the constraints imposed on the $\epsilon$-matrix.

\begin{figure}
 \includegraphics[scale=0.25]{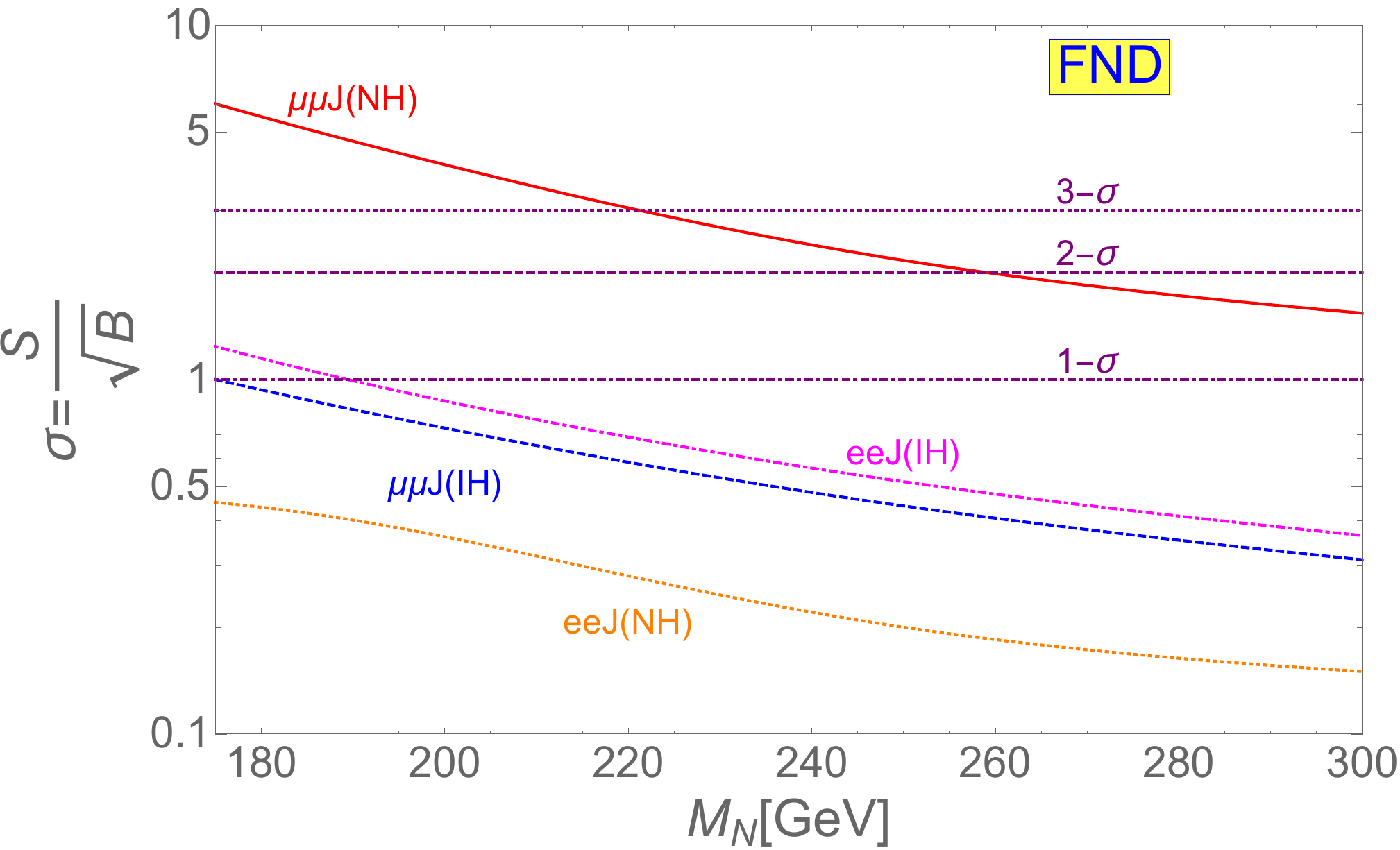}
 \caption{Significance reach as a function of $M_N$ for the FND case at the $13$ TeV LHC with $3000$ fb$^{-1}$ luminosity considering the constraints on the $\epsilon$-matrix. }
\label{fig:fx2}
\end{figure}

The efficiencies have been estimated using the cuts flow between C1-C8 used for the events as shown in Tab.~\ref{tab:cut-flow} for the different benchmark values of $M_N$. The efficiencies are $3.64\%$, $4.93\%$, $7.07\%$ and $9.84\%$ for $M_N= 175, 200, 250$ and $300~\rm{GeV}$ respectively for the $e^{\pm}e^{\mp}J$ signal. On the other hand the cut efficiencies are $5.10\%$, $6.90\%$, $9.9\%$ and $13.78\%$ for the corresponding muon signals. We use the upper limits on $|V_{eN}|^2$ and $|V_{\mu N}|^2$ for the NH and IH cases for the $ee$ and $\mu\mu$ signals from Tab.~\ref{tabmix}. The total SM backgrounds $(B)$ have been estimated in the Tab.~\ref{tab:cut-flow} as $986.06~(1380.50)$ for the $e^{\pm}e^{\mp} J(\mu^{\pm}\mu^{\mp} J)$. Hence we estimate the maximum signal events $(S^{\rm{NH/IH}})$ for $\ell^{\pm}\ell^{\mp} J$ $(\ell = e$ or $\mu)$ for the different benchmark values of $M_N$ using the luminosity of $3000$ fb$^{-1}$ at the $13$ TeV LHC for the NH and IH cases. Using such signal and background events, we estimate the significance of the signal events $\sigma^{\rm{NH/IH}}=\frac{S^{\rm{NH/IH}}}{\sqrt{B}}$ at the different benchmark values of $M_N$ for the NH and IH cases. Significances reach as a function of heavy neutrino mass $M_N$ are plotted in Fig.~\ref{fig:fx2}. 
While all other cases are expected to remain unconstrained, 
normal hierarchy in this general parametrization can be interesting in this OSDL muon search channel, especially at the lower mass region. Even with a relatively small signal efficiency, $M_N=175$ GeV RHN with this flavor structure can be probed more than $5$-$\sigma$ significance using the muon channel, where as $M_N \leq 222$ GeV can be probed up to $\geq 3$-$\sigma$.

 \begin{table}[t]	
 \centering
 \setlength{\tabcolsep}{10pt}
 \renewcommand{\arraystretch}{1.3}
\begin{tabular}{|c||c|c||c|c|}
	\hline
	$M_N$ (GeV) &  \multicolumn{2}{c||}{$e^{\pm}e^{\mp}jj$ (fb)}   &    \multicolumn{2}{c|}{$\mu^{\pm} \mu^{\mp}jj$ (fb)} \\
	                        &       IH                &          NH              &       IH             &          NH  \\

	\hline
	175&0.3232&0.1281&0.233&1.43\\
	200&0.1837&0.0762&0.131&0.7254\\
	250&0.07623&0.0294&0.055&0.271\\
	300&0.0386&0.01562&0.0276&0.138\\
	\hline	
	\end{tabular}
	\caption{Estimated signal cross sections at the NLO level $\sigma_{LHC}^{NLO}$ for a $13$ TeV LHC for different benchmark values of $M_N$ matching values from \cite{Das:2016hof}.The second (third) column represents the $e^{\pm}e^{\mp}jj$ final state for the IH (NH) case. 
	The fourth (fifth) column represents the $\mu^{\pm}\mu^{\mp}jj$ final state for IH (NH) case under the FND scenario.}
	\label{tab0}
\end{table}

\section{Summary and Conclusion}
\label{sec:sumcon}
The Seesaw framework gives an elegant but simple mechanism for tiny neutrino masses and flavor mixings. If the sterile neutrinos in these models appear close to the electroweak scale, they may be probed at the 13 TeV LHC. Conventionally such searches for heavy neutrinos, Majorana or pseudo-Dirac, are made in the di-lepton+jets or trilepton channels. Jet substructure methods have been relatively underutilized in these contexts. In this work we extended the same-sign di-lepton + fat jet channel investigated earlier, to the more challenging opposite-sign di-lepton + fat jet final state. The opposite-sign di-lepton state is expected to encounter a huge standard model background. This channel is nevertheless very important as it may be the only final state, with jets, for a class of models with Dirac or pseudo-Dirac type neutrinos. Hence, strategies to effectively investigate the opposite-sign di-lepton along with a fat jet would greatly broaden the scope of collider sterile neutrino searches -- both in terms of probing model aspects as well as uncovering the nature of the heavy sterile neutrinos. 

In the present analysis we propose a new strategy to search for intermediate to heavy mass sterile neutrinos, when their decays lead to boosted fat jets arising from $W$ boson hadronic decays. By looking into the jet substructure characteristics, boosted jets reveal useful information on their origin and topology. We leveraged the same to achieve good discrimination between signal and background in the opposite-sign di-lepton+fat jet channel. The computed signal significance and LHC limits for different model scenarios are shown to be competitive and at least an order of magnitude better than existing limits. 

We also investigate the lepton flavor conserving modes in the flavor non-diagonal cases, for electron and muon flavors both in normal as well as inverted hierarchy. Such models are studied after utilizing extensive constrains coming from neutrino oscillation data, lepton flavor violation constraints and LEP considerations with general parametrization being constrained by non-unitarity. For $M_N=175$ GeV, a signal $\mu^{\pm} \mu^{\mp}+$fat jet with a significance more than $5$-$\sigma$ in the NH case can potentially be constrained in the near future at the $13$ TeV LHC, with a luminosity of $3000$ fb$^{-1}$.
\bigskip
\section*{Acknowledgement}
\label{sec:ack}
The work of PK and AB is partially supported by TDP project at Physical Research Laboratory (PRL), Department of Space, Government of India.  Primary part of the computation were performed using the Vikram-100 HPC resources at PRL. AT would like to thank S. Dube and A. Rane for discussions. AT would also like to acknowledge partial support from an SERB Early Career Research Award. PK and AB also gratefully acknowledge SUSY'17 and WHEPP'17 where parts of this work were presented and discussed.

\appendix
\section{Jet observable at LO+LL order}
 \begin{figure}[h!]
 	\centering
 	\includegraphics[width=0.22\textwidth]{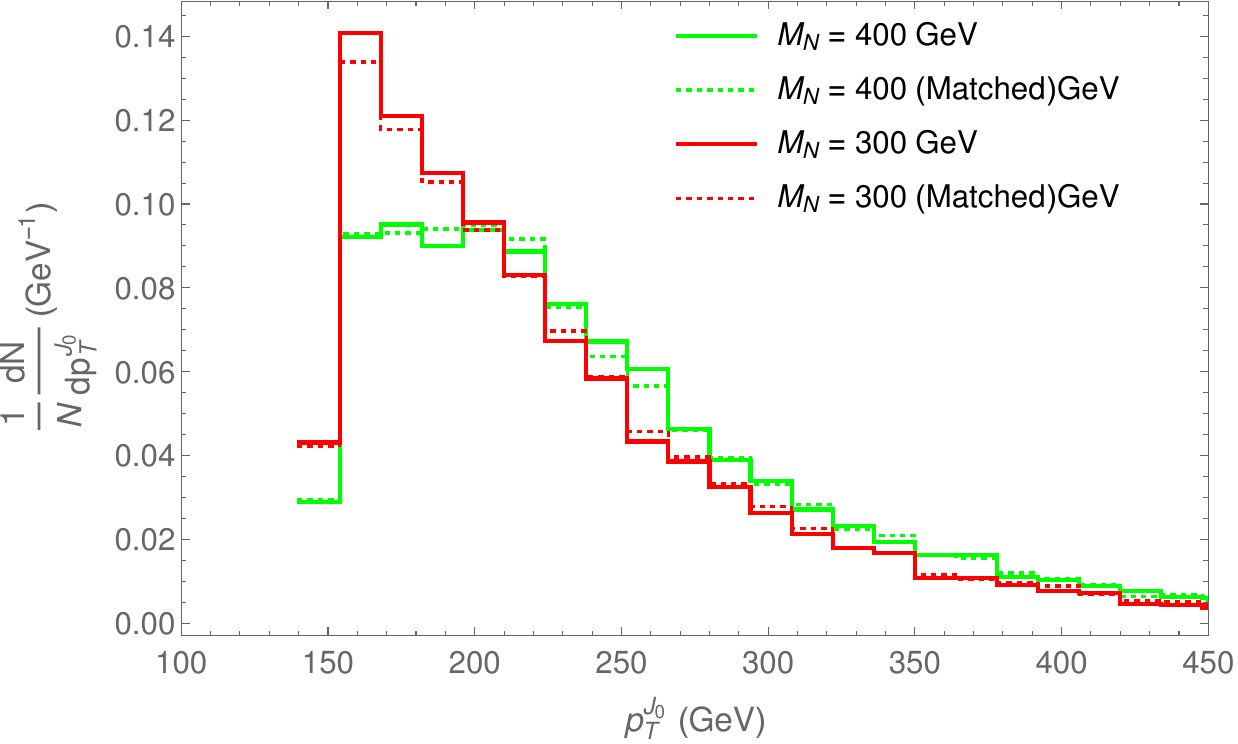}
 	\includegraphics[width=0.22\textwidth]{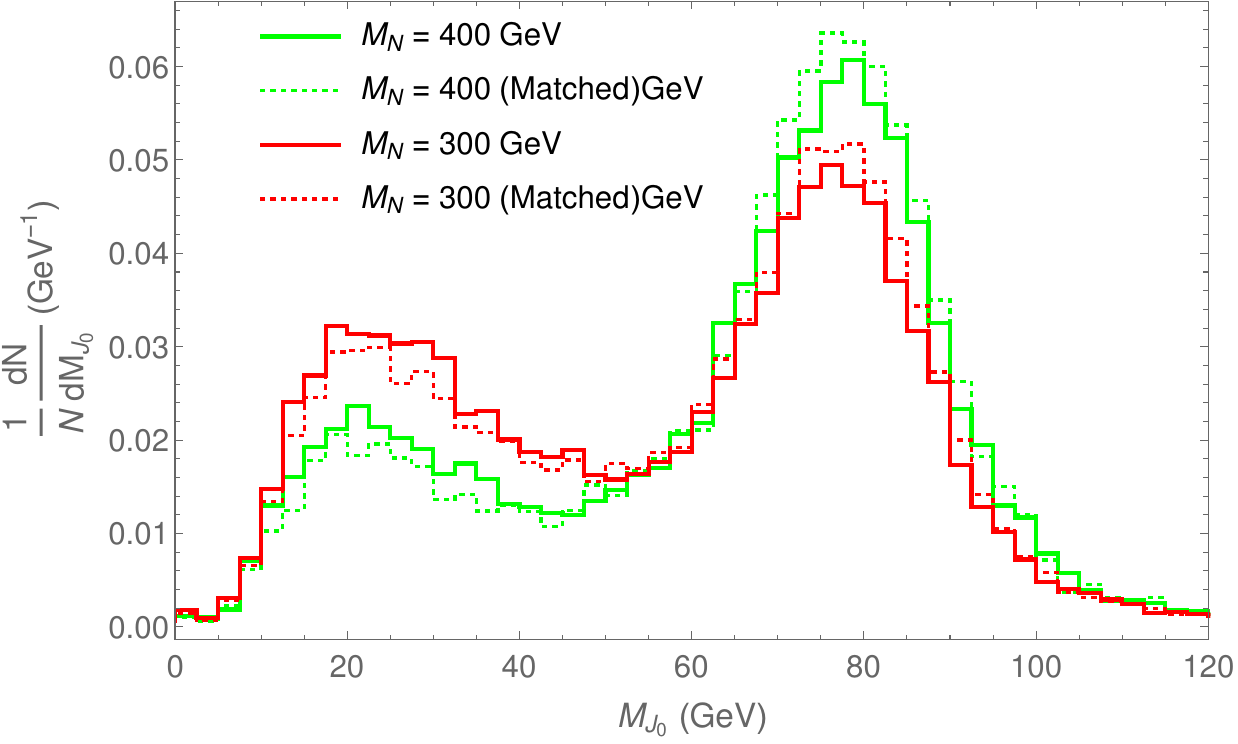}
 	\includegraphics[width=0.22\textwidth]{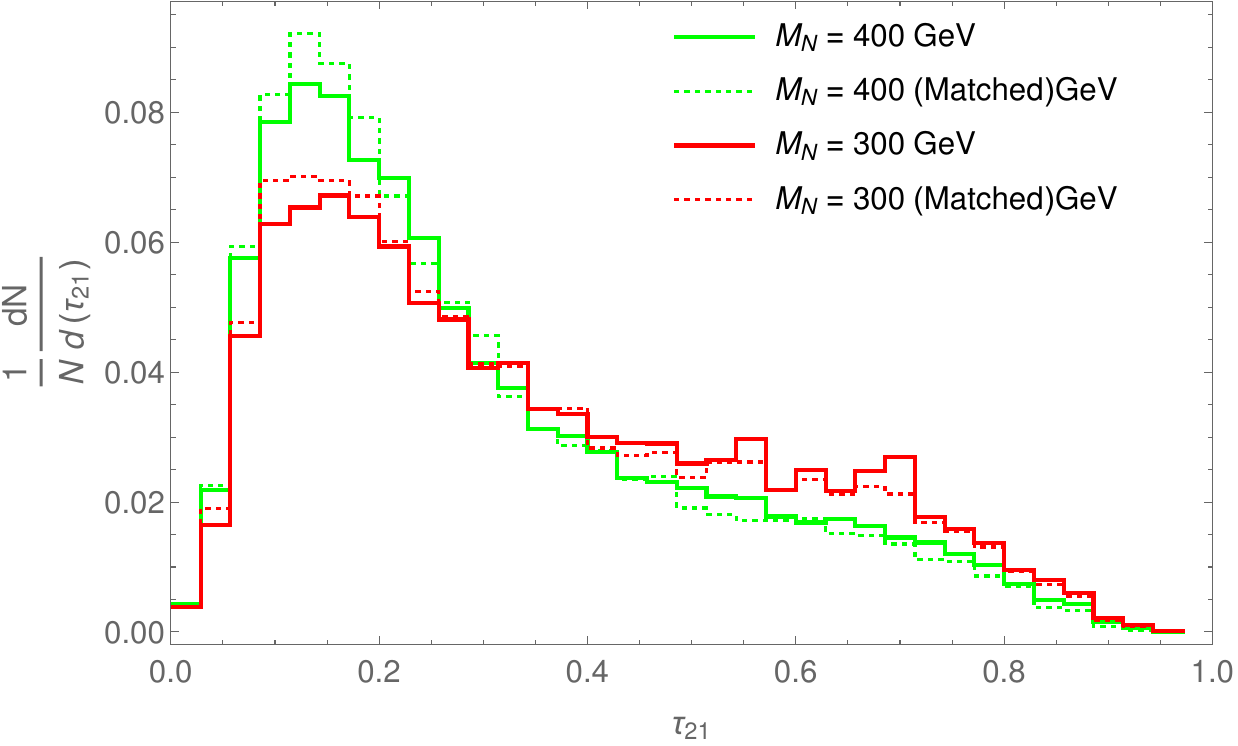}
 	\caption{Normalized differential distributions of transverse momentum $ p_T^{J_0}$ invariant mass $M^{J_0}$  and N-subjettiness ratio $\tau^{J_0}$  of the leading fat jet. These differential distributions are after the baseline selection cuts. The distribution of heavy neutrino benchmark points with $M_N$ = 300  and 400 GeV both for the unmatched (solid) and matched (dashed) samples.
 	}
 	\label{fig:fjpt}
 \end{figure}
We incorporate higher order corrections $(LO + LL)$  by performing MLM matching up to two jets. In Fig.~\ref{fig:fjpt} we compare the distributions of jet observable at LO and $LO + LL $ order. It is evident that the performance of the jet observables does not change drastically at $(LO + LL)$ order, Although it gives slightly better performance.


\bibliographystyle{apsrev4-1}
\bibliography{heavyN}

\begin{thebibliography}{101}%
\makeatletter
\providecommand \@ifxundefined [1]{%
 \@ifx{#1\undefined}
}%
\providecommand \@ifnum [1]{%
 \ifnum #1\expandafter \@firstoftwo
 \else \expandafter \@secondoftwo
 \fi
}%
\providecommand \@ifx [1]{%
 \ifx #1\expandafter \@firstoftwo
 \else \expandafter \@secondoftwo
 \fi
}%
\providecommand \natexlab [1]{#1}%
\providecommand \enquote  [1]{``#1''}%
\providecommand \bibnamefont  [1]{#1}%
\providecommand \bibfnamefont [1]{#1}%
\providecommand \citenamefont [1]{#1}%
\providecommand \href@noop [0]{\@secondoftwo}%
\providecommand \href [0]{\begingroup \@sanitize@url \@href}%
\providecommand \@href[1]{\@@startlink{#1}\@@href}%
\providecommand \@@href[1]{\endgroup#1\@@endlink}%
\providecommand \@sanitize@url [0]{\catcode `\\12\catcode `\$12\catcode
  `\&12\catcode `\#12\catcode `\^12\catcode `\_12\catcode `\%12\relax}%
\providecommand \@@startlink[1]{}%
\providecommand \@@endlink[0]{}%
\providecommand \url  [0]{\begingroup\@sanitize@url \@url }%
\providecommand \@url [1]{\endgroup\@href {#1}{\urlprefix }}%
\providecommand \urlprefix  [0]{URL }%
\providecommand \Eprint [0]{\href }%
\providecommand \doibase [0]{http://dx.doi.org/}%
\providecommand \selectlanguage [0]{\@gobble}%
\providecommand \bibinfo  [0]{\@secondoftwo}%
\providecommand \bibfield  [0]{\@secondoftwo}%
\providecommand \translation [1]{[#1]}%
\providecommand \BibitemOpen [0]{}%
\providecommand \bibitemStop [0]{}%
\providecommand \bibitemNoStop [0]{.\EOS\space}%
\providecommand \EOS [0]{\spacefactor3000\relax}%
\providecommand \BibitemShut  [1]{\csname bibitem#1\endcsname}%
\let\auto@bib@innerbib\@empty
\bibitem [{\citenamefont {Catanesi}(2013)}]{Catanesi:2013fxa}%
  \BibitemOpen
  \bibfield  {author} {\bibinfo {author} {\bibfnamefont {M.~G.}\ \bibnamefont
  {Catanesi}} (\bibinfo {collaboration} {T2K}),\ }\bibfield  {booktitle} {\emph
  {\bibinfo {booktitle} {{Proceedings, Neutrino Oscillation Workshop (NOW
  2012): Lecce, Italy, September 9-15, 2012}}},\ }\href {\doibase
  10.1016/j.nuclphysbps.2013.04.074} {\bibfield  {journal} {\bibinfo  {journal}
  {Nucl. Phys. Proc. Suppl.}\ }\textbf {\bibinfo {volume} {237-238}},\ \bibinfo
  {pages} {129} (\bibinfo {year} {2013})}\BibitemShut {NoStop}%
\bibitem [{\citenamefont {Adamson}\ \emph {et~al.}(2011)\citenamefont {Adamson}
  \emph {et~al.}}]{Adamson:2011qu}%
  \BibitemOpen
  \bibfield  {author} {\bibinfo {author} {\bibfnamefont {P.}~\bibnamefont
  {Adamson}} \emph {et~al.} (\bibinfo {collaboration} {MINOS}),\ }\href
  {\doibase 10.1103/PhysRevLett.107.181802} {\bibfield  {journal} {\bibinfo
  {journal} {Phys. Rev. Lett.}\ }\textbf {\bibinfo {volume} {107}},\ \bibinfo
  {pages} {181802} (\bibinfo {year} {2011})},\ \Eprint
  {http://arxiv.org/abs/1108.0015} {arXiv:1108.0015 [hep-ex]} \BibitemShut
  {NoStop}%
\bibitem [{\citenamefont {Abe}\ \emph {et~al.}(2012)\citenamefont {Abe} \emph
  {et~al.}}]{Abe:2011fz}%
  \BibitemOpen
  \bibfield  {author} {\bibinfo {author} {\bibfnamefont {Y.}~\bibnamefont
  {Abe}} \emph {et~al.} (\bibinfo {collaboration} {Double Chooz}),\ }\href
  {\doibase 10.1103/PhysRevLett.108.131801} {\bibfield  {journal} {\bibinfo
  {journal} {Phys. Rev. Lett.}\ }\textbf {\bibinfo {volume} {108}},\ \bibinfo
  {pages} {131801} (\bibinfo {year} {2012})},\ \Eprint
  {http://arxiv.org/abs/1112.6353} {arXiv:1112.6353 [hep-ex]} \BibitemShut
  {NoStop}%
\bibitem [{\citenamefont {An}\ \emph {et~al.}(2012)\citenamefont {An} \emph
  {et~al.}}]{An:2012eh}%
  \BibitemOpen
  \bibfield  {author} {\bibinfo {author} {\bibfnamefont {F.~P.}\ \bibnamefont
  {An}} \emph {et~al.} (\bibinfo {collaboration} {Daya Bay}),\ }\href {\doibase
  10.1103/PhysRevLett.108.171803} {\bibfield  {journal} {\bibinfo  {journal}
  {Phys. Rev. Lett.}\ }\textbf {\bibinfo {volume} {108}},\ \bibinfo {pages}
  {171803} (\bibinfo {year} {2012})},\ \Eprint {http://arxiv.org/abs/1203.1669}
  {arXiv:1203.1669 [hep-ex]} \BibitemShut {NoStop}%
\bibitem [{\citenamefont {Ahn}\ \emph {et~al.}(2012)\citenamefont {Ahn} \emph
  {et~al.}}]{Ahn:2012nd}%
  \BibitemOpen
  \bibfield  {author} {\bibinfo {author} {\bibfnamefont {J.~K.}\ \bibnamefont
  {Ahn}} \emph {et~al.} (\bibinfo {collaboration} {RENO}),\ }\href {\doibase
  10.1103/PhysRevLett.108.191802} {\bibfield  {journal} {\bibinfo  {journal}
  {Phys. Rev. Lett.}\ }\textbf {\bibinfo {volume} {108}},\ \bibinfo {pages}
  {191802} (\bibinfo {year} {2012})},\ \Eprint {http://arxiv.org/abs/1204.0626}
  {arXiv:1204.0626 [hep-ex]} \BibitemShut {NoStop}%
\bibitem [{\citenamefont {Minkowski}(1977)}]{Minkowski:1977sc}%
  \BibitemOpen
  \bibfield  {author} {\bibinfo {author} {\bibfnamefont {P.}~\bibnamefont
  {Minkowski}},\ }\href {\doibase 10.1016/0370-2693(77)90435-X} {\bibfield
  {journal} {\bibinfo  {journal} {Phys. Lett.}\ }\textbf {\bibinfo {volume}
  {67B}},\ \bibinfo {pages} {421} (\bibinfo {year} {1977})}\BibitemShut
  {NoStop}%
\bibitem [{\citenamefont {Yanagida}(1980)}]{Yanagida:1980xy}%
  \BibitemOpen
  \bibfield  {author} {\bibinfo {author} {\bibfnamefont {T.}~\bibnamefont
  {Yanagida}},\ }\href {\doibase 10.1143/PTP.64.1103} {\bibfield  {journal}
  {\bibinfo  {journal} {Prog. Theor. Phys.}\ }\textbf {\bibinfo {volume}
  {64}},\ \bibinfo {pages} {1103} (\bibinfo {year} {1980})}\BibitemShut
  {NoStop}%
\bibitem [{\citenamefont {Schechter}\ and\ \citenamefont
  {Valle}(1980)}]{Schechter:1980gr}%
  \BibitemOpen
  \bibfield  {author} {\bibinfo {author} {\bibfnamefont {J.}~\bibnamefont
  {Schechter}}\ and\ \bibinfo {author} {\bibfnamefont {J.~W.~F.}\ \bibnamefont
  {Valle}},\ }\href {\doibase 10.1103/PhysRevD.22.2227} {\bibfield  {journal}
  {\bibinfo  {journal} {Phys. Rev.}\ }\textbf {\bibinfo {volume} {D22}},\
  \bibinfo {pages} {2227} (\bibinfo {year} {1980})}\BibitemShut {NoStop}%
\bibitem [{\citenamefont {Sawada}\ and\ \citenamefont
  {Sugamoto}(1979)}]{Sawada:1979dis}%
  \BibitemOpen
  \bibinfo {editor} {\bibfnamefont {O.}~\bibnamefont {Sawada}}\ and\ \bibinfo
  {editor} {\bibfnamefont {A.}~\bibnamefont {Sugamoto}},\ eds.,\ \href@noop {}
  {\emph {\bibinfo {title} {{Proceedings: Workshop on the Unified Theories and
  the Baryon Number in the Universe}}}},\ \bibinfo {organization}
  {Natl.Lab.High Energy Phys.}\ (\bibinfo  {publisher} {Natl.Lab.High Energy
  Phys.},\ \bibinfo {address} {Tsukuba, Japan},\ \bibinfo {year}
  {1979})\BibitemShut {NoStop}%
\bibitem [{\citenamefont {Gell-Mann}\ \emph {et~al.}(1979)\citenamefont
  {Gell-Mann}, \citenamefont {Ramond},\ and\ \citenamefont
  {Slansky}}]{GellMann:1980vs}%
  \BibitemOpen
  \bibfield  {author} {\bibinfo {author} {\bibfnamefont {M.}~\bibnamefont
  {Gell-Mann}}, \bibinfo {author} {\bibfnamefont {P.}~\bibnamefont {Ramond}}, \
  and\ \bibinfo {author} {\bibfnamefont {R.}~\bibnamefont {Slansky}},\
  }\bibfield  {booktitle} {\emph {\bibinfo {booktitle} {{Supergravity Workshop
  Stony Brook, New York, September 27-28, 1979}}},\ }\href@noop {} {\bibfield
  {journal} {\bibinfo  {journal} {Conf. Proc.}\ }\textbf {\bibinfo {volume}
  {C790927}},\ \bibinfo {pages} {315} (\bibinfo {year} {1979})},\ \Eprint
  {http://arxiv.org/abs/1306.4669} {arXiv:1306.4669 [hep-th]} \BibitemShut
  {NoStop}%
\bibitem [{\citenamefont {Glashow}(1980)}]{Glashow:1979nm}%
  \BibitemOpen
  \bibfield  {author} {\bibinfo {author} {\bibfnamefont {S.~L.}\ \bibnamefont
  {Glashow}},\ }\bibfield  {booktitle} {\emph {\bibinfo {booktitle} {{Cargese
  Summer Institute: Quarks and Leptons Cargese, France, July 9-29, 1979}}},\
  }\href {\doibase 10.1007/978-1-4684-7197-7_15} {\bibfield  {journal}
  {\bibinfo  {journal} {NATO Sci. Ser. B}\ }\textbf {\bibinfo {volume} {61}},\
  \bibinfo {pages} {687} (\bibinfo {year} {1980})}\BibitemShut {NoStop}%
\bibitem [{\citenamefont {Mohapatra}\ and\ \citenamefont
  {Senjanovic}(1980)}]{Mohapatra:1979ia}%
  \BibitemOpen
  \bibfield  {author} {\bibinfo {author} {\bibfnamefont {R.~N.}\ \bibnamefont
  {Mohapatra}}\ and\ \bibinfo {author} {\bibfnamefont {G.}~\bibnamefont
  {Senjanovic}},\ }\href {\doibase 10.1103/PhysRevLett.44.912} {\bibfield
  {journal} {\bibinfo  {journal} {Phys. Rev. Lett.}\ }\textbf {\bibinfo
  {volume} {44}},\ \bibinfo {pages} {912} (\bibinfo {year} {1980})}\BibitemShut
  {NoStop}%
\bibitem [{\citenamefont {Weinberg}(1979)}]{Weinberg:1979sa}%
  \BibitemOpen
  \bibfield  {author} {\bibinfo {author} {\bibfnamefont {S.}~\bibnamefont
  {Weinberg}},\ }\href {\doibase 10.1103/PhysRevLett.43.1566} {\bibfield
  {journal} {\bibinfo  {journal} {Phys. Rev. Lett.}\ }\textbf {\bibinfo
  {volume} {43}},\ \bibinfo {pages} {1566} (\bibinfo {year}
  {1979})}\BibitemShut {NoStop}%
\bibitem [{\citenamefont {Mohapatra}(1986)}]{Mohapatra:1986aw}%
  \BibitemOpen
  \bibfield  {author} {\bibinfo {author} {\bibfnamefont {R.~N.}\ \bibnamefont
  {Mohapatra}},\ }\href {\doibase 10.1103/PhysRevLett.56.561} {\bibfield
  {journal} {\bibinfo  {journal} {Phys. Rev. Lett.}\ }\textbf {\bibinfo
  {volume} {56}},\ \bibinfo {pages} {561} (\bibinfo {year} {1986})}\BibitemShut
  {NoStop}%
\bibitem [{\citenamefont {Mohapatra}\ and\ \citenamefont
  {Valle}(1986)}]{Mohapatra:1986bd}%
  \BibitemOpen
  \bibfield  {author} {\bibinfo {author} {\bibfnamefont {R.~N.}\ \bibnamefont
  {Mohapatra}}\ and\ \bibinfo {author} {\bibfnamefont {J.~W.~F.}\ \bibnamefont
  {Valle}},\ }\bibfield  {booktitle} {\emph {\bibinfo {booktitle}
  {{Proceedings, 23RD International Conference on High Energy Physics, JULY
  16-23, 1986, Berkeley, CA}}},\ }\href {\doibase 10.1103/PhysRevD.34.1642}
  {\bibfield  {journal} {\bibinfo  {journal} {Phys. Rev.}\ }\textbf {\bibinfo
  {volume} {D34}},\ \bibinfo {pages} {1642} (\bibinfo {year}
  {1986})}\BibitemShut {NoStop}%
\bibitem [{\citenamefont {Das}\ \emph {et~al.}(2018{\natexlab{a}})\citenamefont
  {Das}, \citenamefont {Dev},\ and\ \citenamefont {Mohapatra}}]{Das:2017hmg}%
  \BibitemOpen
  \bibfield  {author} {\bibinfo {author} {\bibfnamefont {A.}~\bibnamefont
  {Das}}, \bibinfo {author} {\bibfnamefont {P.~S.~B.}\ \bibnamefont {Dev}}, \
  and\ \bibinfo {author} {\bibfnamefont {R.~N.}\ \bibnamefont {Mohapatra}},\
  }\href {\doibase 10.1103/PhysRevD.97.015018} {\bibfield  {journal} {\bibinfo
  {journal} {Phys. Rev. D}\ }\textbf {\bibinfo {volume} {97}},\ \bibinfo
  {pages} {015018} (\bibinfo {year} {2018}{\natexlab{a}})},\ \Eprint
  {http://arxiv.org/abs/1709.06553} {arXiv:1709.06553 [hep-ph]} \BibitemShut
  {NoStop}%
\bibitem [{\citenamefont {Das}\ \emph {et~al.}(2014)\citenamefont {Das},
  \citenamefont {Bhupal~Dev},\ and\ \citenamefont {Okada}}]{Das:2014jxa}%
  \BibitemOpen
  \bibfield  {author} {\bibinfo {author} {\bibfnamefont {A.}~\bibnamefont
  {Das}}, \bibinfo {author} {\bibfnamefont {P.~S.}\ \bibnamefont {Bhupal~Dev}},
  \ and\ \bibinfo {author} {\bibfnamefont {N.}~\bibnamefont {Okada}},\ }\href
  {\doibase 10.1016/j.physletb.2014.06.058} {\bibfield  {journal} {\bibinfo
  {journal} {Phys. Lett.}\ }\textbf {\bibinfo {volume} {B735}},\ \bibinfo
  {pages} {364} (\bibinfo {year} {2014})},\ \Eprint
  {http://arxiv.org/abs/1405.0177} {arXiv:1405.0177 [hep-ph]} \BibitemShut
  {NoStop}%
\bibitem [{\citenamefont {Das}\ and\ \citenamefont
  {Okada}(2016)}]{Das:2015toa}%
  \BibitemOpen
  \bibfield  {author} {\bibinfo {author} {\bibfnamefont {A.}~\bibnamefont
  {Das}}\ and\ \bibinfo {author} {\bibfnamefont {N.}~\bibnamefont {Okada}},\
  }\href {\doibase 10.1103/PhysRevD.93.033003} {\bibfield  {journal} {\bibinfo
  {journal} {Phys. Rev.}\ }\textbf {\bibinfo {volume} {D93}},\ \bibinfo {pages}
  {033003} (\bibinfo {year} {2016})},\ \Eprint
  {http://arxiv.org/abs/1510.04790} {arXiv:1510.04790 [hep-ph]} \BibitemShut
  {NoStop}%
\bibitem [{\citenamefont {'t~Hooft}(1980)}]{tHooft:1979rat}%
  \BibitemOpen
  \bibfield  {author} {\bibinfo {author} {\bibfnamefont {G.}~\bibnamefont
  {'t~Hooft}},\ }\bibfield  {booktitle} {\emph {\bibinfo {booktitle} {{Recent
  Developments in Gauge Theories. Proceedings, Nato Advanced Study Institute,
  Cargese, France, August 26 - September 8, 1979}}},\ }\href {\doibase
  10.1007/978-1-4684-7571-5_9} {\bibfield  {journal} {\bibinfo  {journal} {NATO
  Sci. Ser. B}\ }\textbf {\bibinfo {volume} {59}},\ \bibinfo {pages} {135}
  (\bibinfo {year} {1980})}\BibitemShut {NoStop}%
\bibitem [{\citenamefont {Seymour}(1994)}]{Seymour:1993mx}%
  \BibitemOpen
  \bibfield  {author} {\bibinfo {author} {\bibfnamefont {M.~H.}\ \bibnamefont
  {Seymour}},\ }\href {\doibase 10.1007/BF01559532} {\bibfield  {journal}
  {\bibinfo  {journal} {Z. Phys.}\ }\textbf {\bibinfo {volume} {C62}},\
  \bibinfo {pages} {127} (\bibinfo {year} {1994})}\BibitemShut {NoStop}%
\bibitem [{\citenamefont {Butterworth}\ \emph {et~al.}(2007)\citenamefont
  {Butterworth}, \citenamefont {Ellis},\ and\ \citenamefont
  {Raklev}}]{Butterworth:2007ke}%
  \BibitemOpen
  \bibfield  {author} {\bibinfo {author} {\bibfnamefont {J.~M.}\ \bibnamefont
  {Butterworth}}, \bibinfo {author} {\bibfnamefont {J.~R.}\ \bibnamefont
  {Ellis}}, \ and\ \bibinfo {author} {\bibfnamefont {A.~R.}\ \bibnamefont
  {Raklev}},\ }\href {\doibase 10.1088/1126-6708/2007/05/033} {\bibfield
  {journal} {\bibinfo  {journal} {JHEP}\ }\textbf {\bibinfo {volume} {05}},\
  \bibinfo {pages} {033} (\bibinfo {year} {2007})},\ \Eprint
  {http://arxiv.org/abs/hep-ph/0702150} {arXiv:hep-ph/0702150 [HEP-PH]}
  \BibitemShut {NoStop}%
\bibitem [{\citenamefont {Brooijmans}(2008)}]{Brooijmans:2008zza}%
  \BibitemOpen
  \bibfield  {author} {\bibinfo {author} {\bibfnamefont {G.}~\bibnamefont
  {Brooijmans}}\ }(\bibinfo {year} {2008})\BibitemShut {NoStop}%
\bibitem [{\citenamefont {Butterworth}\ \emph {et~al.}(2008)\citenamefont
  {Butterworth}, \citenamefont {Davison}, \citenamefont {Rubin},\ and\
  \citenamefont {Salam}}]{Butterworth:2008iy}%
  \BibitemOpen
  \bibfield  {author} {\bibinfo {author} {\bibfnamefont {J.~M.}\ \bibnamefont
  {Butterworth}}, \bibinfo {author} {\bibfnamefont {A.~R.}\ \bibnamefont
  {Davison}}, \bibinfo {author} {\bibfnamefont {M.}~\bibnamefont {Rubin}}, \
  and\ \bibinfo {author} {\bibfnamefont {G.~P.}\ \bibnamefont {Salam}},\ }\href
  {\doibase 10.1103/PhysRevLett.100.242001} {\bibfield  {journal} {\bibinfo
  {journal} {Phys. Rev. Lett.}\ }\textbf {\bibinfo {volume} {100}},\ \bibinfo
  {pages} {242001} (\bibinfo {year} {2008})},\ \Eprint
  {http://arxiv.org/abs/0802.2470} {arXiv:0802.2470 [hep-ph]} \BibitemShut
  {NoStop}%
\bibitem [{\citenamefont {Chakraborty}\ \emph {et~al.}(2018)\citenamefont
  {Chakraborty}, \citenamefont {Iyer},\ and\ \citenamefont
  {Roy}}]{Chakraborty:2017mbz}%
  \BibitemOpen
  \bibfield  {author} {\bibinfo {author} {\bibfnamefont {A.}~\bibnamefont
  {Chakraborty}}, \bibinfo {author} {\bibfnamefont {A.~M.}\ \bibnamefont
  {Iyer}}, \ and\ \bibinfo {author} {\bibfnamefont {T.~S.}\ \bibnamefont
  {Roy}},\ }\href {\doibase 10.1016/j.nuclphysb.2018.05.019} {\bibfield
  {journal} {\bibinfo  {journal} {Nucl. Phys. B}\ }\textbf {\bibinfo {volume}
  {932}},\ \bibinfo {pages} {439} (\bibinfo {year} {2018})},\ \Eprint
  {http://arxiv.org/abs/1707.07084} {arXiv:1707.07084 [hep-ph]} \BibitemShut
  {NoStop}%
\bibitem [{\citenamefont {Roy}\ and\ \citenamefont
  {Thalapillil}(2017)}]{Roy:2016qfv}%
  \BibitemOpen
  \bibfield  {author} {\bibinfo {author} {\bibfnamefont {T.~S.}\ \bibnamefont
  {Roy}}\ and\ \bibinfo {author} {\bibfnamefont {A.~M.}\ \bibnamefont
  {Thalapillil}},\ }\href {\doibase 10.1103/PhysRevD.95.075002} {\bibfield
  {journal} {\bibinfo  {journal} {Phys. Rev.}\ }\textbf {\bibinfo {volume}
  {D95}},\ \bibinfo {pages} {075002} (\bibinfo {year} {2017})},\ \Eprint
  {http://arxiv.org/abs/1609.04835} {arXiv:1609.04835 [hep-ph]} \BibitemShut
  {NoStop}%
\bibitem [{\citenamefont {Komiske}\ \emph {et~al.}(2017)\citenamefont
  {Komiske}, \citenamefont {Metodiev}, \citenamefont {Nachman},\ and\
  \citenamefont {Schwartz}}]{Komiske:2017ubm}%
  \BibitemOpen
  \bibfield  {author} {\bibinfo {author} {\bibfnamefont {P.~T.}\ \bibnamefont
  {Komiske}}, \bibinfo {author} {\bibfnamefont {E.~M.}\ \bibnamefont
  {Metodiev}}, \bibinfo {author} {\bibfnamefont {B.}~\bibnamefont {Nachman}}, \
  and\ \bibinfo {author} {\bibfnamefont {M.~D.}\ \bibnamefont {Schwartz}},\
  }\href {\doibase 10.1007/JHEP12(2017)051} {\bibfield  {journal} {\bibinfo
  {journal} {JHEP}\ }\textbf {\bibinfo {volume} {12}},\ \bibinfo {pages} {051}
  (\bibinfo {year} {2017})},\ \Eprint {http://arxiv.org/abs/1707.08600}
  {arXiv:1707.08600 [hep-ph]} \BibitemShut {NoStop}%
\bibitem [{\citenamefont {Adams}\ \emph {et~al.}(2015)\citenamefont {Adams}
  \emph {et~al.}}]{Adams:2015hiv}%
  \BibitemOpen
  \bibfield  {author} {\bibinfo {author} {\bibfnamefont {D.}~\bibnamefont
  {Adams}} \emph {et~al.},\ }\href {\doibase 10.1140/epjc/s10052-015-3587-2}
  {\bibfield  {journal} {\bibinfo  {journal} {Eur. Phys. J.}\ }\textbf
  {\bibinfo {volume} {C75}},\ \bibinfo {pages} {409} (\bibinfo {year}
  {2015})},\ \Eprint {http://arxiv.org/abs/1504.00679} {arXiv:1504.00679
  [hep-ph]} \BibitemShut {NoStop}%
\bibitem [{\citenamefont {Larkoski}\ \emph {et~al.}(2020)\citenamefont
  {Larkoski}, \citenamefont {Moult},\ and\ \citenamefont
  {Nachman}}]{Larkoski:2017jix}%
  \BibitemOpen
  \bibfield  {author} {\bibinfo {author} {\bibfnamefont {A.~J.}\ \bibnamefont
  {Larkoski}}, \bibinfo {author} {\bibfnamefont {I.}~\bibnamefont {Moult}}, \
  and\ \bibinfo {author} {\bibfnamefont {B.}~\bibnamefont {Nachman}},\ }\href
  {\doibase 10.1016/j.physrep.2019.11.001} {\bibfield  {journal} {\bibinfo
  {journal} {Phys. Rept.}\ }\textbf {\bibinfo {volume} {841}},\ \bibinfo
  {pages} {1} (\bibinfo {year} {2020})},\ \Eprint
  {http://arxiv.org/abs/1709.04464} {arXiv:1709.04464 [hep-ph]} \BibitemShut
  {NoStop}%
\bibitem [{\citenamefont {Izaguirre}\ and\ \citenamefont
  {Shuve}(2015)}]{Izaguirre:2015pga}%
  \BibitemOpen
  \bibfield  {author} {\bibinfo {author} {\bibfnamefont {E.}~\bibnamefont
  {Izaguirre}}\ and\ \bibinfo {author} {\bibfnamefont {B.}~\bibnamefont
  {Shuve}},\ }\href {\doibase 10.1103/PhysRevD.91.093010} {\bibfield  {journal}
  {\bibinfo  {journal} {Phys. Rev.}\ }\textbf {\bibinfo {volume} {D91}},\
  \bibinfo {pages} {093010} (\bibinfo {year} {2015})},\ \Eprint
  {http://arxiv.org/abs/1504.02470} {arXiv:1504.02470 [hep-ph]} \BibitemShut
  {NoStop}%
\bibitem [{\citenamefont {Antusch}\ \emph {et~al.}(2017)\citenamefont
  {Antusch}, \citenamefont {Cazzato},\ and\ \citenamefont
  {Fischer}}]{Antusch:2016ejd}%
  \BibitemOpen
  \bibfield  {author} {\bibinfo {author} {\bibfnamefont {S.}~\bibnamefont
  {Antusch}}, \bibinfo {author} {\bibfnamefont {E.}~\bibnamefont {Cazzato}}, \
  and\ \bibinfo {author} {\bibfnamefont {O.}~\bibnamefont {Fischer}},\ }\href
  {\doibase 10.1142/S0217751X17500786} {\bibfield  {journal} {\bibinfo
  {journal} {Int. J. Mod. Phys.}\ }\textbf {\bibinfo {volume} {A32}},\ \bibinfo
  {pages} {1750078} (\bibinfo {year} {2017})},\ \Eprint
  {http://arxiv.org/abs/1612.02728} {arXiv:1612.02728 [hep-ph]} \BibitemShut
  {NoStop}%
\bibitem [{\citenamefont {Mitra}\ \emph {et~al.}(2016)\citenamefont {Mitra},
  \citenamefont {Ruiz}, \citenamefont {Scott},\ and\ \citenamefont
  {Spannowsky}}]{Mitra:2016kov}%
  \BibitemOpen
  \bibfield  {author} {\bibinfo {author} {\bibfnamefont {M.}~\bibnamefont
  {Mitra}}, \bibinfo {author} {\bibfnamefont {R.}~\bibnamefont {Ruiz}},
  \bibinfo {author} {\bibfnamefont {D.~J.}\ \bibnamefont {Scott}}, \ and\
  \bibinfo {author} {\bibfnamefont {M.}~\bibnamefont {Spannowsky}},\ }\href
  {\doibase 10.1103/PhysRevD.94.095016} {\bibfield  {journal} {\bibinfo
  {journal} {Phys. Rev.}\ }\textbf {\bibinfo {volume} {D94}},\ \bibinfo {pages}
  {095016} (\bibinfo {year} {2016})},\ \Eprint
  {http://arxiv.org/abs/1607.03504} {arXiv:1607.03504 [hep-ph]} \BibitemShut
  {NoStop}%
\bibitem [{\citenamefont {Dube}\ \emph {et~al.}(2017)\citenamefont {Dube},
  \citenamefont {Gadkari},\ and\ \citenamefont {Thalapillil}}]{Dube:2017jgo}%
  \BibitemOpen
  \bibfield  {author} {\bibinfo {author} {\bibfnamefont {S.}~\bibnamefont
  {Dube}}, \bibinfo {author} {\bibfnamefont {D.}~\bibnamefont {Gadkari}}, \
  and\ \bibinfo {author} {\bibfnamefont {A.~M.}\ \bibnamefont {Thalapillil}},\
  }\href {\doibase 10.1103/PhysRevD.96.055031} {\bibfield  {journal} {\bibinfo
  {journal} {Phys. Rev.}\ }\textbf {\bibinfo {volume} {D96}},\ \bibinfo {pages}
  {055031} (\bibinfo {year} {2017})},\ \Eprint
  {http://arxiv.org/abs/1707.00008} {arXiv:1707.00008 [hep-ph]} \BibitemShut
  {NoStop}%
\bibitem [{\citenamefont {Cox}\ \emph {et~al.}(2018)\citenamefont {Cox},
  \citenamefont {Han},\ and\ \citenamefont {Yanagida}}]{Cox:2017eme}%
  \BibitemOpen
  \bibfield  {author} {\bibinfo {author} {\bibfnamefont {P.}~\bibnamefont
  {Cox}}, \bibinfo {author} {\bibfnamefont {C.}~\bibnamefont {Han}}, \ and\
  \bibinfo {author} {\bibfnamefont {T.~T.}\ \bibnamefont {Yanagida}},\ }\href
  {\doibase 10.1007/JHEP01(2018)037} {\bibfield  {journal} {\bibinfo  {journal}
  {JHEP}\ }\textbf {\bibinfo {volume} {01}},\ \bibinfo {pages} {037} (\bibinfo
  {year} {2018})},\ \Eprint {http://arxiv.org/abs/1707.04532} {arXiv:1707.04532
  [hep-ph]} \BibitemShut {NoStop}%
\bibitem [{\citenamefont {Leonardi}\ \emph {et~al.}(2016)\citenamefont
  {Leonardi}, \citenamefont {Alunni}, \citenamefont {Romeo}, \citenamefont
  {Fanò},\ and\ \citenamefont {Panella}}]{Leonardi:2015qna}%
  \BibitemOpen
  \bibfield  {author} {\bibinfo {author} {\bibfnamefont {R.}~\bibnamefont
  {Leonardi}}, \bibinfo {author} {\bibfnamefont {L.}~\bibnamefont {Alunni}},
  \bibinfo {author} {\bibfnamefont {F.}~\bibnamefont {Romeo}}, \bibinfo
  {author} {\bibfnamefont {L.}~\bibnamefont {Fanò}}, \ and\ \bibinfo {author}
  {\bibfnamefont {O.}~\bibnamefont {Panella}},\ }\href {\doibase
  10.1140/epjc/s10052-016-4396-y} {\bibfield  {journal} {\bibinfo  {journal}
  {Eur. Phys. J.}\ }\textbf {\bibinfo {volume} {C76}},\ \bibinfo {pages} {593}
  (\bibinfo {year} {2016})},\ \Eprint {http://arxiv.org/abs/1510.07988}
  {arXiv:1510.07988 [hep-ph]} \BibitemShut {NoStop}%
\bibitem [{\citenamefont {Sirunyan}\ \emph {et~al.}(2017)\citenamefont
  {Sirunyan} \emph {et~al.}}]{Sirunyan:2017xnz}%
  \BibitemOpen
  \bibfield  {author} {\bibinfo {author} {\bibfnamefont {A.}~\bibnamefont
  {Sirunyan}} \emph {et~al.} (\bibinfo {collaboration} {CMS}),\ }\href
  {\doibase 10.1016/j.physletb.2017.11.001} {\bibfield  {journal} {\bibinfo
  {journal} {Phys. Lett. B}\ }\textbf {\bibinfo {volume} {775}},\ \bibinfo
  {pages} {315} (\bibinfo {year} {2017})},\ \Eprint
  {http://arxiv.org/abs/1706.08578} {arXiv:1706.08578 [hep-ex]} \BibitemShut
  {NoStop}%
\bibitem [{\citenamefont {Tang}(2018)}]{Tang:2017plx}%
  \BibitemOpen
  \bibfield  {author} {\bibinfo {author} {\bibfnamefont {Y.-L.}\ \bibnamefont
  {Tang}},\ }\href {\doibase 10.1103/PhysRevD.98.035043} {\bibfield  {journal}
  {\bibinfo  {journal} {Phys. Rev. D}\ }\textbf {\bibinfo {volume} {98}},\
  \bibinfo {pages} {035043} (\bibinfo {year} {2018})},\ \Eprint
  {http://arxiv.org/abs/1712.10108} {arXiv:1712.10108 [hep-ph]} \BibitemShut
  {NoStop}%
\bibitem [{\citenamefont {Dev}\ \emph {et~al.}(2016)\citenamefont {Dev},
  \citenamefont {Kim},\ and\ \citenamefont {Mohapatra}}]{Dev:2015kca}%
  \BibitemOpen
  \bibfield  {author} {\bibinfo {author} {\bibfnamefont {P.~S.~B.}\
  \bibnamefont {Dev}}, \bibinfo {author} {\bibfnamefont {D.}~\bibnamefont
  {Kim}}, \ and\ \bibinfo {author} {\bibfnamefont {R.~N.}\ \bibnamefont
  {Mohapatra}},\ }\href {\doibase 10.1007/JHEP01(2016)118} {\bibfield
  {journal} {\bibinfo  {journal} {JHEP}\ }\textbf {\bibinfo {volume} {01}},\
  \bibinfo {pages} {118} (\bibinfo {year} {2016})},\ \Eprint
  {http://arxiv.org/abs/1510.04328} {arXiv:1510.04328 [hep-ph]} \BibitemShut
  {NoStop}%
\bibitem [{\citenamefont {Das}\ \emph {et~al.}(2018{\natexlab{b}})\citenamefont
  {Das}, \citenamefont {Konar},\ and\ \citenamefont
  {Thalapillil}}]{Das:2017gke}%
  \BibitemOpen
  \bibfield  {author} {\bibinfo {author} {\bibfnamefont {A.}~\bibnamefont
  {Das}}, \bibinfo {author} {\bibfnamefont {P.}~\bibnamefont {Konar}}, \ and\
  \bibinfo {author} {\bibfnamefont {A.}~\bibnamefont {Thalapillil}},\ }\href
  {\doibase 10.1007/JHEP02(2018)083} {\bibfield  {journal} {\bibinfo  {journal}
  {JHEP}\ }\textbf {\bibinfo {volume} {02}},\ \bibinfo {pages} {083} (\bibinfo
  {year} {2018}{\natexlab{b}})},\ \Eprint {http://arxiv.org/abs/1709.09712}
  {arXiv:1709.09712 [hep-ph]} \BibitemShut {NoStop}%
\bibitem [{\citenamefont {del Aguila}\ and\ \citenamefont
  {Aguilar-Saavedra}(2009)}]{delAguila:2008cj}%
  \BibitemOpen
  \bibfield  {author} {\bibinfo {author} {\bibfnamefont {F.}~\bibnamefont {del
  Aguila}}\ and\ \bibinfo {author} {\bibfnamefont {J.~A.}\ \bibnamefont
  {Aguilar-Saavedra}},\ }\href {\doibase 10.1016/j.nuclphysb.2008.12.029}
  {\bibfield  {journal} {\bibinfo  {journal} {Nucl. Phys.}\ }\textbf {\bibinfo
  {volume} {B813}},\ \bibinfo {pages} {22} (\bibinfo {year} {2009})},\ \Eprint
  {http://arxiv.org/abs/0808.2468} {arXiv:0808.2468 [hep-ph]} \BibitemShut
  {NoStop}%
\bibitem [{\citenamefont {Malinsky}\ \emph {et~al.}(2009)\citenamefont
  {Malinsky}, \citenamefont {Ohlsson}, \citenamefont {Xing},\ and\
  \citenamefont {Zhang}}]{Malinsky:2009df}%
  \BibitemOpen
  \bibfield  {author} {\bibinfo {author} {\bibfnamefont {M.}~\bibnamefont
  {Malinsky}}, \bibinfo {author} {\bibfnamefont {T.}~\bibnamefont {Ohlsson}},
  \bibinfo {author} {\bibfnamefont {Z.-z.}\ \bibnamefont {Xing}}, \ and\
  \bibinfo {author} {\bibfnamefont {H.}~\bibnamefont {Zhang}},\ }\href
  {\doibase 10.1016/j.physletb.2009.07.038} {\bibfield  {journal} {\bibinfo
  {journal} {Phys. Lett.}\ }\textbf {\bibinfo {volume} {B679}},\ \bibinfo
  {pages} {242} (\bibinfo {year} {2009})},\ \Eprint
  {http://arxiv.org/abs/0905.2889} {arXiv:0905.2889 [hep-ph]} \BibitemShut
  {NoStop}%
\bibitem [{\citenamefont {Garg}\ \emph {et~al.}(2017)\citenamefont {Garg},
  \citenamefont {Goswami}, \citenamefont {Vishnudath},\ and\ \citenamefont
  {Khan}}]{Garg:2017iva}%
  \BibitemOpen
  \bibfield  {author} {\bibinfo {author} {\bibfnamefont {I.}~\bibnamefont
  {Garg}}, \bibinfo {author} {\bibfnamefont {S.}~\bibnamefont {Goswami}},
  \bibinfo {author} {\bibfnamefont {K.~N.}\ \bibnamefont {Vishnudath}}, \ and\
  \bibinfo {author} {\bibfnamefont {N.}~\bibnamefont {Khan}},\ }\href {\doibase
  10.1103/PhysRevD.96.055020} {\bibfield  {journal} {\bibinfo  {journal} {Phys.
  Rev.}\ }\textbf {\bibinfo {volume} {D96}},\ \bibinfo {pages} {055020}
  (\bibinfo {year} {2017})},\ \Eprint {http://arxiv.org/abs/1706.08851}
  {arXiv:1706.08851 [hep-ph]} \BibitemShut {NoStop}%
\bibitem [{\citenamefont {Abada}\ and\ \citenamefont
  {Lucente}(2014)}]{Abada:2014vea}%
  \BibitemOpen
  \bibfield  {author} {\bibinfo {author} {\bibfnamefont {A.}~\bibnamefont
  {Abada}}\ and\ \bibinfo {author} {\bibfnamefont {M.}~\bibnamefont
  {Lucente}},\ }\href {\doibase 10.1016/j.nuclphysb.2014.06.003} {\bibfield
  {journal} {\bibinfo  {journal} {Nucl. Phys.}\ }\textbf {\bibinfo {volume}
  {B885}},\ \bibinfo {pages} {651} (\bibinfo {year} {2014})},\ \Eprint
  {http://arxiv.org/abs/1401.1507} {arXiv:1401.1507 [hep-ph]} \BibitemShut
  {NoStop}%
\bibitem [{\citenamefont {Das}\ and\ \citenamefont {Okada}(2013)}]{Das:2012ze}%
  \BibitemOpen
  \bibfield  {author} {\bibinfo {author} {\bibfnamefont {A.}~\bibnamefont
  {Das}}\ and\ \bibinfo {author} {\bibfnamefont {N.}~\bibnamefont {Okada}},\
  }\href {\doibase 10.1103/PhysRevD.88.113001} {\bibfield  {journal} {\bibinfo
  {journal} {Phys. Rev.}\ }\textbf {\bibinfo {volume} {D88}},\ \bibinfo {pages}
  {113001} (\bibinfo {year} {2013})},\ \Eprint {http://arxiv.org/abs/1207.3734}
  {arXiv:1207.3734 [hep-ph]} \BibitemShut {NoStop}%
\bibitem [{\citenamefont {Das}\ and\ \citenamefont
  {Okada}(2017)}]{Das:2017nvm}%
  \BibitemOpen
  \bibfield  {author} {\bibinfo {author} {\bibfnamefont {A.}~\bibnamefont
  {Das}}\ and\ \bibinfo {author} {\bibfnamefont {N.}~\bibnamefont {Okada}},\
  }\href {\doibase 10.1016/j.physletb.2017.09.042} {\bibfield  {journal}
  {\bibinfo  {journal} {Phys. Lett.}\ }\textbf {\bibinfo {volume} {B774}},\
  \bibinfo {pages} {32} (\bibinfo {year} {2017})},\ \Eprint
  {http://arxiv.org/abs/1702.04668} {arXiv:1702.04668 [hep-ph]} \BibitemShut
  {NoStop}%
\bibitem [{\citenamefont {Casas}\ and\ \citenamefont
  {Ibarra}(2001)}]{Casas:2001sr}%
  \BibitemOpen
  \bibfield  {author} {\bibinfo {author} {\bibfnamefont {J.~A.}\ \bibnamefont
  {Casas}}\ and\ \bibinfo {author} {\bibfnamefont {A.}~\bibnamefont {Ibarra}},\
  }\href {\doibase 10.1016/S0550-3213(01)00475-8} {\bibfield  {journal}
  {\bibinfo  {journal} {Nucl. Phys.}\ }\textbf {\bibinfo {volume} {B618}},\
  \bibinfo {pages} {171} (\bibinfo {year} {2001})},\ \Eprint
  {http://arxiv.org/abs/hep-ph/0103065} {arXiv:hep-ph/0103065 [hep-ph]}
  \BibitemShut {NoStop}%
\bibitem [{\citenamefont {Das}\ \emph {et~al.}(2016{\natexlab{a}})\citenamefont
  {Das}, \citenamefont {Nagata},\ and\ \citenamefont {Okada}}]{Das:2016akd}%
  \BibitemOpen
  \bibfield  {author} {\bibinfo {author} {\bibfnamefont {A.}~\bibnamefont
  {Das}}, \bibinfo {author} {\bibfnamefont {N.}~\bibnamefont {Nagata}}, \ and\
  \bibinfo {author} {\bibfnamefont {N.}~\bibnamefont {Okada}},\ }\href
  {\doibase 10.1007/JHEP03(2016)049} {\bibfield  {journal} {\bibinfo  {journal}
  {JHEP}\ }\textbf {\bibinfo {volume} {03}},\ \bibinfo {pages} {049} (\bibinfo
  {year} {2016}{\natexlab{a}})},\ \Eprint {http://arxiv.org/abs/1601.05079}
  {arXiv:1601.05079 [hep-ph]} \BibitemShut {NoStop}%
\bibitem [{\citenamefont {Franceschini}\ \emph {et~al.}(2008)\citenamefont
  {Franceschini}, \citenamefont {Hambye},\ and\ \citenamefont
  {Strumia}}]{Franceschini:2008pz}%
  \BibitemOpen
  \bibfield  {author} {\bibinfo {author} {\bibfnamefont {R.}~\bibnamefont
  {Franceschini}}, \bibinfo {author} {\bibfnamefont {T.}~\bibnamefont
  {Hambye}}, \ and\ \bibinfo {author} {\bibfnamefont {A.}~\bibnamefont
  {Strumia}},\ }\href {\doibase 10.1103/PhysRevD.78.033002} {\bibfield
  {journal} {\bibinfo  {journal} {Phys. Rev.}\ }\textbf {\bibinfo {volume}
  {D78}},\ \bibinfo {pages} {033002} (\bibinfo {year} {2008})},\ \Eprint
  {http://arxiv.org/abs/0805.1613} {arXiv:0805.1613 [hep-ph]} \BibitemShut
  {NoStop}%
\bibitem [{\citenamefont {Bambhaniya}\ \emph
  {et~al.}(2015{\natexlab{a}})\citenamefont {Bambhaniya}, \citenamefont
  {Goswami}, \citenamefont {Khan}, \citenamefont {Konar},\ and\ \citenamefont
  {Mondal}}]{Bambhaniya:2014kga}%
  \BibitemOpen
  \bibfield  {author} {\bibinfo {author} {\bibfnamefont {G.}~\bibnamefont
  {Bambhaniya}}, \bibinfo {author} {\bibfnamefont {S.}~\bibnamefont {Goswami}},
  \bibinfo {author} {\bibfnamefont {S.}~\bibnamefont {Khan}}, \bibinfo {author}
  {\bibfnamefont {P.}~\bibnamefont {Konar}}, \ and\ \bibinfo {author}
  {\bibfnamefont {T.}~\bibnamefont {Mondal}},\ }\href {\doibase
  10.1103/PhysRevD.91.075007} {\bibfield  {journal} {\bibinfo  {journal} {Phys.
  Rev.}\ }\textbf {\bibinfo {volume} {D91}},\ \bibinfo {pages} {075007}
  (\bibinfo {year} {2015}{\natexlab{a}})},\ \Eprint
  {http://arxiv.org/abs/1410.5687} {arXiv:1410.5687 [hep-ph]} \BibitemShut
  {NoStop}%
\bibitem [{\citenamefont {Bambhaniya}\ \emph
  {et~al.}(2015{\natexlab{b}})\citenamefont {Bambhaniya}, \citenamefont {Khan},
  \citenamefont {Konar},\ and\ \citenamefont {Mondal}}]{Bambhaniya:2014hla}%
  \BibitemOpen
  \bibfield  {author} {\bibinfo {author} {\bibfnamefont {G.}~\bibnamefont
  {Bambhaniya}}, \bibinfo {author} {\bibfnamefont {S.}~\bibnamefont {Khan}},
  \bibinfo {author} {\bibfnamefont {P.}~\bibnamefont {Konar}}, \ and\ \bibinfo
  {author} {\bibfnamefont {T.}~\bibnamefont {Mondal}},\ }\href {\doibase
  10.1103/PhysRevD.91.095007} {\bibfield  {journal} {\bibinfo  {journal} {Phys.
  Rev.}\ }\textbf {\bibinfo {volume} {D91}},\ \bibinfo {pages} {095007}
  (\bibinfo {year} {2015}{\natexlab{b}})},\ \Eprint
  {http://arxiv.org/abs/1411.6866} {arXiv:1411.6866 [hep-ph]} \BibitemShut
  {NoStop}%
\bibitem [{\citenamefont {Patrignani}\ \emph {et~al.}(2016)\citenamefont
  {Patrignani} \emph {et~al.}}]{Patrignani:2016xqp}%
  \BibitemOpen
  \bibfield  {author} {\bibinfo {author} {\bibfnamefont {C.}~\bibnamefont
  {Patrignani}} \emph {et~al.} (\bibinfo {collaboration} {Particle Data
  Group}),\ }\href {\doibase 10.1088/1674-1137/40/10/100001} {\bibfield
  {journal} {\bibinfo  {journal} {Chin. Phys.}\ }\textbf {\bibinfo {volume}
  {C40}},\ \bibinfo {pages} {100001} (\bibinfo {year} {2016})}\BibitemShut
  {NoStop}%
\bibitem [{\citenamefont {Antusch}\ \emph {et~al.}(2006)\citenamefont
  {Antusch}, \citenamefont {Biggio}, \citenamefont {Fernandez-Martinez},
  \citenamefont {Gavela},\ and\ \citenamefont {Lopez-Pavon}}]{Antusch:2006vwa}%
  \BibitemOpen
  \bibfield  {author} {\bibinfo {author} {\bibfnamefont {S.}~\bibnamefont
  {Antusch}}, \bibinfo {author} {\bibfnamefont {C.}~\bibnamefont {Biggio}},
  \bibinfo {author} {\bibfnamefont {E.}~\bibnamefont {Fernandez-Martinez}},
  \bibinfo {author} {\bibfnamefont {M.~B.}\ \bibnamefont {Gavela}}, \ and\
  \bibinfo {author} {\bibfnamefont {J.}~\bibnamefont {Lopez-Pavon}},\ }\href
  {\doibase 10.1088/1126-6708/2006/10/084} {\bibfield  {journal} {\bibinfo
  {journal} {JHEP}\ }\textbf {\bibinfo {volume} {10}},\ \bibinfo {pages} {084}
  (\bibinfo {year} {2006})},\ \Eprint {http://arxiv.org/abs/hep-ph/0607020}
  {arXiv:hep-ph/0607020 [hep-ph]} \BibitemShut {NoStop}%
\bibitem [{\citenamefont {Abada}\ \emph {et~al.}(2007)\citenamefont {Abada},
  \citenamefont {Biggio}, \citenamefont {Bonnet}, \citenamefont {Gavela},\ and\
  \citenamefont {Hambye}}]{Abada:2007ux}%
  \BibitemOpen
  \bibfield  {author} {\bibinfo {author} {\bibfnamefont {A.}~\bibnamefont
  {Abada}}, \bibinfo {author} {\bibfnamefont {C.}~\bibnamefont {Biggio}},
  \bibinfo {author} {\bibfnamefont {F.}~\bibnamefont {Bonnet}}, \bibinfo
  {author} {\bibfnamefont {M.~B.}\ \bibnamefont {Gavela}}, \ and\ \bibinfo
  {author} {\bibfnamefont {T.}~\bibnamefont {Hambye}},\ }\href {\doibase
  10.1088/1126-6708/2007/12/061} {\bibfield  {journal} {\bibinfo  {journal}
  {JHEP}\ }\textbf {\bibinfo {volume} {12}},\ \bibinfo {pages} {061} (\bibinfo
  {year} {2007})},\ \Eprint {http://arxiv.org/abs/0707.4058} {arXiv:0707.4058
  [hep-ph]} \BibitemShut {NoStop}%
\bibitem [{\citenamefont {Ibarra}\ \emph {et~al.}(2010)\citenamefont {Ibarra},
  \citenamefont {Molinaro},\ and\ \citenamefont {Petcov}}]{Ibarra:2010xw}%
  \BibitemOpen
  \bibfield  {author} {\bibinfo {author} {\bibfnamefont {A.}~\bibnamefont
  {Ibarra}}, \bibinfo {author} {\bibfnamefont {E.}~\bibnamefont {Molinaro}}, \
  and\ \bibinfo {author} {\bibfnamefont {S.~T.}\ \bibnamefont {Petcov}},\
  }\href {\doibase 10.1007/JHEP09(2010)108} {\bibfield  {journal} {\bibinfo
  {journal} {JHEP}\ }\textbf {\bibinfo {volume} {09}},\ \bibinfo {pages} {108}
  (\bibinfo {year} {2010})},\ \Eprint {http://arxiv.org/abs/1007.2378}
  {arXiv:1007.2378 [hep-ph]} \BibitemShut {NoStop}%
\bibitem [{\citenamefont {Ibarra}\ \emph {et~al.}(2011)\citenamefont {Ibarra},
  \citenamefont {Molinaro},\ and\ \citenamefont {Petcov}}]{Ibarra:2011xn}%
  \BibitemOpen
  \bibfield  {author} {\bibinfo {author} {\bibfnamefont {A.}~\bibnamefont
  {Ibarra}}, \bibinfo {author} {\bibfnamefont {E.}~\bibnamefont {Molinaro}}, \
  and\ \bibinfo {author} {\bibfnamefont {S.~T.}\ \bibnamefont {Petcov}},\
  }\href {\doibase 10.1103/PhysRevD.84.013005} {\bibfield  {journal} {\bibinfo
  {journal} {Phys. Rev.}\ }\textbf {\bibinfo {volume} {D84}},\ \bibinfo {pages}
  {013005} (\bibinfo {year} {2011})},\ \Eprint {http://arxiv.org/abs/1103.6217}
  {arXiv:1103.6217 [hep-ph]} \BibitemShut {NoStop}%
\bibitem [{\citenamefont {Dinh}\ \emph {et~al.}(2012)\citenamefont {Dinh},
  \citenamefont {Ibarra}, \citenamefont {Molinaro},\ and\ \citenamefont
  {Petcov}}]{Dinh:2012bp}%
  \BibitemOpen
  \bibfield  {author} {\bibinfo {author} {\bibfnamefont {D.~N.}\ \bibnamefont
  {Dinh}}, \bibinfo {author} {\bibfnamefont {A.}~\bibnamefont {Ibarra}},
  \bibinfo {author} {\bibfnamefont {E.}~\bibnamefont {Molinaro}}, \ and\
  \bibinfo {author} {\bibfnamefont {S.~T.}\ \bibnamefont {Petcov}},\ }\href
  {\doibase 10.1007/JHEP09(2013)023, 10.1007/JHEP08(2012)125} {\bibfield
  {journal} {\bibinfo  {journal} {JHEP}\ }\textbf {\bibinfo {volume} {08}},\
  \bibinfo {pages} {125} (\bibinfo {year} {2012})},\ \bibinfo {note} {[Erratum:
  JHEP09,023(2013)]},\ \Eprint {http://arxiv.org/abs/1205.4671}
  {arXiv:1205.4671 [hep-ph]} \BibitemShut {NoStop}%
\bibitem [{\citenamefont {Adam}\ \emph {et~al.}(2011)\citenamefont {Adam} \emph
  {et~al.}}]{Adam:2011ch}%
  \BibitemOpen
  \bibfield  {author} {\bibinfo {author} {\bibfnamefont {J.}~\bibnamefont
  {Adam}} \emph {et~al.} (\bibinfo {collaboration} {MEG}),\ }\href {\doibase
  10.1103/PhysRevLett.107.171801} {\bibfield  {journal} {\bibinfo  {journal}
  {Phys. Rev. Lett.}\ }\textbf {\bibinfo {volume} {107}},\ \bibinfo {pages}
  {171801} (\bibinfo {year} {2011})},\ \Eprint {http://arxiv.org/abs/1107.5547}
  {arXiv:1107.5547 [hep-ex]} \BibitemShut {NoStop}%
\bibitem [{\citenamefont {Aubert}\ \emph {et~al.}(2010)\citenamefont {Aubert}
  \emph {et~al.}}]{Aubert:2009ag}%
  \BibitemOpen
  \bibfield  {author} {\bibinfo {author} {\bibfnamefont {B.}~\bibnamefont
  {Aubert}} \emph {et~al.} (\bibinfo {collaboration} {BaBar}),\ }\href
  {\doibase 10.1103/PhysRevLett.104.021802} {\bibfield  {journal} {\bibinfo
  {journal} {Phys. Rev. Lett.}\ }\textbf {\bibinfo {volume} {104}},\ \bibinfo
  {pages} {021802} (\bibinfo {year} {2010})},\ \Eprint
  {http://arxiv.org/abs/0908.2381} {arXiv:0908.2381 [hep-ex]} \BibitemShut
  {NoStop}%
\bibitem [{\citenamefont {O'Leary}\ \emph {et~al.}(2010)\citenamefont {O'Leary}
  \emph {et~al.}}]{OLeary:2010hau}%
  \BibitemOpen
  \bibfield  {author} {\bibinfo {author} {\bibfnamefont {B.}~\bibnamefont
  {O'Leary}} \emph {et~al.} (\bibinfo {collaboration} {SuperB}),\ }\href@noop
  {} {\  (\bibinfo {year} {2010})},\ \Eprint {http://arxiv.org/abs/1008.1541}
  {arXiv:1008.1541 [hep-ex]} \BibitemShut {NoStop}%
\bibitem [{\citenamefont {Baldini}\ \emph {et~al.}(2016)\citenamefont {Baldini}
  \emph {et~al.}}]{TheMEG:2016wtm}%
  \BibitemOpen
  \bibfield  {author} {\bibinfo {author} {\bibfnamefont {A.~M.}\ \bibnamefont
  {Baldini}} \emph {et~al.} (\bibinfo {collaboration} {MEG}),\ }\href {\doibase
  10.1140/epjc/s10052-016-4271-x} {\bibfield  {journal} {\bibinfo  {journal}
  {Eur. Phys. J.}\ }\textbf {\bibinfo {volume} {C76}},\ \bibinfo {pages} {434}
  (\bibinfo {year} {2016})},\ \Eprint {http://arxiv.org/abs/1605.05081}
  {arXiv:1605.05081 [hep-ex]} \BibitemShut {NoStop}%
\bibitem [{\citenamefont {Drewes}\ \emph {et~al.}(2017)\citenamefont {Drewes}
  \emph {et~al.}}]{Adhikari:2016bei}%
  \BibitemOpen
  \bibfield  {author} {\bibinfo {author} {\bibfnamefont {M.}~\bibnamefont
  {Drewes}} \emph {et~al.},\ }\href {\doibase 10.1088/1475-7516/2017/01/025}
  {\bibfield  {journal} {\bibinfo  {journal} {JCAP}\ }\textbf {\bibinfo
  {volume} {1701}},\ \bibinfo {pages} {025} (\bibinfo {year} {2017})},\ \Eprint
  {http://arxiv.org/abs/1602.04816} {arXiv:1602.04816 [hep-ph]} \BibitemShut
  {NoStop}%
\bibitem [{\citenamefont {Fernandez-Martinez}\ \emph
  {et~al.}(2016)\citenamefont {Fernandez-Martinez}, \citenamefont
  {Hernandez-Garcia},\ and\ \citenamefont
  {Lopez-Pavon}}]{Fernandez-Martinez:2016lgt}%
  \BibitemOpen
  \bibfield  {author} {\bibinfo {author} {\bibfnamefont {E.}~\bibnamefont
  {Fernandez-Martinez}}, \bibinfo {author} {\bibfnamefont {J.}~\bibnamefont
  {Hernandez-Garcia}}, \ and\ \bibinfo {author} {\bibfnamefont
  {J.}~\bibnamefont {Lopez-Pavon}},\ }\href {\doibase 10.1007/JHEP08(2016)033}
  {\bibfield  {journal} {\bibinfo  {journal} {JHEP}\ }\textbf {\bibinfo
  {volume} {08}},\ \bibinfo {pages} {033} (\bibinfo {year} {2016})},\ \Eprint
  {http://arxiv.org/abs/1605.08774} {arXiv:1605.08774 [hep-ph]} \BibitemShut
  {NoStop}%
\bibitem [{\citenamefont {Antusch}\ \emph {et~al.}(2018)\citenamefont
  {Antusch}, \citenamefont {Cazzato}, \citenamefont {Fischer}, \citenamefont
  {Hammad},\ and\ \citenamefont {Wang}}]{Antusch:2018bgr}%
  \BibitemOpen
  \bibfield  {author} {\bibinfo {author} {\bibfnamefont {S.}~\bibnamefont
  {Antusch}}, \bibinfo {author} {\bibfnamefont {E.}~\bibnamefont {Cazzato}},
  \bibinfo {author} {\bibfnamefont {O.}~\bibnamefont {Fischer}}, \bibinfo
  {author} {\bibfnamefont {A.}~\bibnamefont {Hammad}}, \ and\ \bibinfo {author}
  {\bibfnamefont {K.}~\bibnamefont {Wang}},\ }\href {\doibase
  10.1007/JHEP10(2018)067} {\bibfield  {journal} {\bibinfo  {journal} {JHEP}\
  }\textbf {\bibinfo {volume} {10}},\ \bibinfo {pages} {067} (\bibinfo {year}
  {2018})},\ \Eprint {http://arxiv.org/abs/1805.11400} {arXiv:1805.11400
  [hep-ph]} \BibitemShut {NoStop}%
\bibitem [{\citenamefont {Alwall}\ \emph {et~al.}(2011)\citenamefont {Alwall},
  \citenamefont {Herquet}, \citenamefont {Maltoni}, \citenamefont {Mattelaer},\
  and\ \citenamefont {Stelzer}}]{Alwall:2011uj}%
  \BibitemOpen
  \bibfield  {author} {\bibinfo {author} {\bibfnamefont {J.}~\bibnamefont
  {Alwall}}, \bibinfo {author} {\bibfnamefont {M.}~\bibnamefont {Herquet}},
  \bibinfo {author} {\bibfnamefont {F.}~\bibnamefont {Maltoni}}, \bibinfo
  {author} {\bibfnamefont {O.}~\bibnamefont {Mattelaer}}, \ and\ \bibinfo
  {author} {\bibfnamefont {T.}~\bibnamefont {Stelzer}},\ }\href {\doibase
  10.1007/JHEP06(2011)128} {\bibfield  {journal} {\bibinfo  {journal} {JHEP}\
  }\textbf {\bibinfo {volume} {06}},\ \bibinfo {pages} {128} (\bibinfo {year}
  {2011})},\ \Eprint {http://arxiv.org/abs/1106.0522} {arXiv:1106.0522
  [hep-ph]} \BibitemShut {NoStop}%
\bibitem [{\citenamefont {Alwall}\ \emph {et~al.}(2014)\citenamefont {Alwall},
  \citenamefont {Frederix}, \citenamefont {Frixione}, \citenamefont {Hirschi},
  \citenamefont {Maltoni}, \citenamefont {Mattelaer}, \citenamefont {Shao},
  \citenamefont {Stelzer}, \citenamefont {Torrielli},\ and\ \citenamefont
  {Zaro}}]{Alwall:2014hca}%
  \BibitemOpen
  \bibfield  {author} {\bibinfo {author} {\bibfnamefont {J.}~\bibnamefont
  {Alwall}}, \bibinfo {author} {\bibfnamefont {R.}~\bibnamefont {Frederix}},
  \bibinfo {author} {\bibfnamefont {S.}~\bibnamefont {Frixione}}, \bibinfo
  {author} {\bibfnamefont {V.}~\bibnamefont {Hirschi}}, \bibinfo {author}
  {\bibfnamefont {F.}~\bibnamefont {Maltoni}}, \bibinfo {author} {\bibfnamefont
  {O.}~\bibnamefont {Mattelaer}}, \bibinfo {author} {\bibfnamefont {H.~S.}\
  \bibnamefont {Shao}}, \bibinfo {author} {\bibfnamefont {T.}~\bibnamefont
  {Stelzer}}, \bibinfo {author} {\bibfnamefont {P.}~\bibnamefont {Torrielli}},
  \ and\ \bibinfo {author} {\bibfnamefont {M.}~\bibnamefont {Zaro}},\ }\href
  {\doibase 10.1007/JHEP07(2014)079} {\bibfield  {journal} {\bibinfo  {journal}
  {JHEP}\ }\textbf {\bibinfo {volume} {07}},\ \bibinfo {pages} {079} (\bibinfo
  {year} {2014})},\ \Eprint {http://arxiv.org/abs/1405.0301} {arXiv:1405.0301
  [hep-ph]} \BibitemShut {NoStop}%
\bibitem [{\citenamefont {Sjostrand}\ \emph {et~al.}(2006)\citenamefont
  {Sjostrand}, \citenamefont {Mrenna},\ and\ \citenamefont
  {Skands}}]{Sjostrand:2006za}%
  \BibitemOpen
  \bibfield  {author} {\bibinfo {author} {\bibfnamefont {T.}~\bibnamefont
  {Sjostrand}}, \bibinfo {author} {\bibfnamefont {S.}~\bibnamefont {Mrenna}}, \
  and\ \bibinfo {author} {\bibfnamefont {P.~Z.}\ \bibnamefont {Skands}},\
  }\href {\doibase 10.1088/1126-6708/2006/05/026} {\bibfield  {journal}
  {\bibinfo  {journal} {JHEP}\ }\textbf {\bibinfo {volume} {05}},\ \bibinfo
  {pages} {026} (\bibinfo {year} {2006})},\ \Eprint
  {http://arxiv.org/abs/hep-ph/0603175} {arXiv:hep-ph/0603175 [hep-ph]}
  \BibitemShut {NoStop}%
\bibitem [{\citenamefont {Ball}\ \emph {et~al.}(2015)\citenamefont {Ball} \emph
  {et~al.}}]{Ball:2014uwa}%
  \BibitemOpen
  \bibfield  {author} {\bibinfo {author} {\bibfnamefont {R.~D.}\ \bibnamefont
  {Ball}} \emph {et~al.} (\bibinfo {collaboration} {NNPDF}),\ }\href {\doibase
  10.1007/JHEP04(2015)040} {\bibfield  {journal} {\bibinfo  {journal} {JHEP}\
  }\textbf {\bibinfo {volume} {04}},\ \bibinfo {pages} {040} (\bibinfo {year}
  {2015})},\ \Eprint {http://arxiv.org/abs/1410.8849} {arXiv:1410.8849
  [hep-ph]} \BibitemShut {NoStop}%
\bibitem [{\citenamefont
  {{\url{https://cp3.irmp.ucl.ac.be/projects/madgraph/wiki/FAQ-General-13}}}()}]{madgraph_scale}%
  \BibitemOpen
  \bibfield  {author} {\bibinfo {author} {\bibnamefont
  {{\url{https://cp3.irmp.ucl.ac.be/projects/madgraph/wiki/FAQ-General-13}}}},\
  }\href@noop {} {\ }\BibitemShut {NoStop}%
\bibitem [{\citenamefont {Mangano}\ \emph {et~al.}(2007)\citenamefont
  {Mangano}, \citenamefont {Moretti}, \citenamefont {Piccinini},\ and\
  \citenamefont {Treccani}}]{Mangano:2006rw}%
  \BibitemOpen
  \bibfield  {author} {\bibinfo {author} {\bibfnamefont {M.~L.}\ \bibnamefont
  {Mangano}}, \bibinfo {author} {\bibfnamefont {M.}~\bibnamefont {Moretti}},
  \bibinfo {author} {\bibfnamefont {F.}~\bibnamefont {Piccinini}}, \ and\
  \bibinfo {author} {\bibfnamefont {M.}~\bibnamefont {Treccani}},\ }\href
  {\doibase 10.1088/1126-6708/2007/01/013} {\bibfield  {journal} {\bibinfo
  {journal} {JHEP}\ }\textbf {\bibinfo {volume} {01}},\ \bibinfo {pages} {013}
  (\bibinfo {year} {2007})},\ \Eprint {http://arxiv.org/abs/hep-ph/0611129}
  {arXiv:hep-ph/0611129 [hep-ph]} \BibitemShut {NoStop}%
\bibitem [{\citenamefont {Hoeche}\ \emph {et~al.}(2006)\citenamefont {Hoeche},
  \citenamefont {Krauss}, \citenamefont {Lavesson}, \citenamefont {Lonnblad},
  \citenamefont {Mangano}, \citenamefont {Schalicke},\ and\ \citenamefont
  {Schumann}}]{Hoche:2006ph}%
  \BibitemOpen
  \bibfield  {author} {\bibinfo {author} {\bibfnamefont {S.}~\bibnamefont
  {Hoeche}}, \bibinfo {author} {\bibfnamefont {F.}~\bibnamefont {Krauss}},
  \bibinfo {author} {\bibfnamefont {N.}~\bibnamefont {Lavesson}}, \bibinfo
  {author} {\bibfnamefont {L.}~\bibnamefont {Lonnblad}}, \bibinfo {author}
  {\bibfnamefont {M.}~\bibnamefont {Mangano}}, \bibinfo {author} {\bibfnamefont
  {A.}~\bibnamefont {Schalicke}}, \ and\ \bibinfo {author} {\bibfnamefont
  {S.}~\bibnamefont {Schumann}},\ }in\ \href
  {http://inspirehep.net/record/709818/files/arXiv:hep-ph_0602031.pdf} {\emph
  {\bibinfo {booktitle} {{HERA and the LHC: A Workshop on the implications of
  HERA for LHC physics: Proceedings Part A}}}}\ (\bibinfo {year} {2006})\
  \Eprint {http://arxiv.org/abs/hep-ph/0602031} {arXiv:hep-ph/0602031 [hep-ph]}
  \BibitemShut {NoStop}%
\bibitem [{\citenamefont {Alwall}\ \emph {et~al.}(2008)\citenamefont {Alwall}
  \emph {et~al.}}]{Alwall:2007fs}%
  \BibitemOpen
  \bibfield  {author} {\bibinfo {author} {\bibfnamefont {J.}~\bibnamefont
  {Alwall}} \emph {et~al.},\ }\href {\doibase 10.1140/epjc/s10052-007-0490-5}
  {\bibfield  {journal} {\bibinfo  {journal} {Eur. Phys. J.}\ }\textbf
  {\bibinfo {volume} {C53}},\ \bibinfo {pages} {473} (\bibinfo {year}
  {2008})},\ \Eprint {http://arxiv.org/abs/0706.2569} {arXiv:0706.2569
  [hep-ph]} \BibitemShut {NoStop}%
\bibitem [{\citenamefont {de~Favereau}\ \emph {et~al.}(2014)\citenamefont
  {de~Favereau}, \citenamefont {Delaere}, \citenamefont {Demin}, \citenamefont
  {Giammanco}, \citenamefont {Lemaître}, \citenamefont {Mertens},\ and\
  \citenamefont {Selvaggi}}]{deFavereau:2013fsa}%
  \BibitemOpen
  \bibfield  {author} {\bibinfo {author} {\bibfnamefont {J.}~\bibnamefont
  {de~Favereau}}, \bibinfo {author} {\bibfnamefont {C.}~\bibnamefont
  {Delaere}}, \bibinfo {author} {\bibfnamefont {P.}~\bibnamefont {Demin}},
  \bibinfo {author} {\bibfnamefont {A.}~\bibnamefont {Giammanco}}, \bibinfo
  {author} {\bibfnamefont {V.}~\bibnamefont {Lemaître}}, \bibinfo {author}
  {\bibfnamefont {A.}~\bibnamefont {Mertens}}, \ and\ \bibinfo {author}
  {\bibfnamefont {M.}~\bibnamefont {Selvaggi}} (\bibinfo {collaboration}
  {DELPHES 3}),\ }\href {\doibase 10.1007/JHEP02(2014)057} {\bibfield
  {journal} {\bibinfo  {journal} {JHEP}\ }\textbf {\bibinfo {volume} {02}},\
  \bibinfo {pages} {057} (\bibinfo {year} {2014})},\ \Eprint
  {http://arxiv.org/abs/1307.6346} {arXiv:1307.6346 [hep-ex]} \BibitemShut
  {NoStop}%
\bibitem [{\citenamefont {Cacciari}\ \emph {et~al.}(2012)\citenamefont
  {Cacciari}, \citenamefont {Salam},\ and\ \citenamefont
  {Soyez}}]{Cacciari:2011ma}%
  \BibitemOpen
  \bibfield  {author} {\bibinfo {author} {\bibfnamefont {M.}~\bibnamefont
  {Cacciari}}, \bibinfo {author} {\bibfnamefont {G.~P.}\ \bibnamefont {Salam}},
  \ and\ \bibinfo {author} {\bibfnamefont {G.}~\bibnamefont {Soyez}},\ }\href
  {\doibase 10.1140/epjc/s10052-012-1896-2} {\bibfield  {journal} {\bibinfo
  {journal} {Eur. Phys. J.}\ }\textbf {\bibinfo {volume} {C72}},\ \bibinfo
  {pages} {1896} (\bibinfo {year} {2012})},\ \Eprint
  {http://arxiv.org/abs/1111.6097} {arXiv:1111.6097 [hep-ph]} \BibitemShut
  {NoStop}%
\bibitem [{\citenamefont {Dokshitzer}\ \emph {et~al.}(1997)\citenamefont
  {Dokshitzer}, \citenamefont {Leder}, \citenamefont {Moretti},\ and\
  \citenamefont {Webber}}]{Dokshitzer:1997in}%
  \BibitemOpen
  \bibfield  {author} {\bibinfo {author} {\bibfnamefont {Y.~L.}\ \bibnamefont
  {Dokshitzer}}, \bibinfo {author} {\bibfnamefont {G.~D.}\ \bibnamefont
  {Leder}}, \bibinfo {author} {\bibfnamefont {S.}~\bibnamefont {Moretti}}, \
  and\ \bibinfo {author} {\bibfnamefont {B.~R.}\ \bibnamefont {Webber}},\
  }\href {\doibase 10.1088/1126-6708/1997/08/001} {\bibfield  {journal}
  {\bibinfo  {journal} {JHEP}\ }\textbf {\bibinfo {volume} {08}},\ \bibinfo
  {pages} {001} (\bibinfo {year} {1997})},\ \Eprint
  {http://arxiv.org/abs/hep-ph/9707323} {arXiv:hep-ph/9707323 [hep-ph]}
  \BibitemShut {NoStop}%
\bibitem [{\citenamefont {Chatrchyan}\ \emph {et~al.}(2013)\citenamefont
  {Chatrchyan} \emph {et~al.}}]{Chatrchyan:2012jua}%
  \BibitemOpen
  \bibfield  {author} {\bibinfo {author} {\bibfnamefont {S.}~\bibnamefont
  {Chatrchyan}} \emph {et~al.} (\bibinfo {collaboration} {CMS}),\ }\href
  {\doibase 10.1088/1748-0221/8/04/P04013} {\bibfield  {journal} {\bibinfo
  {journal} {JINST}\ }\textbf {\bibinfo {volume} {8}},\ \bibinfo {pages}
  {P04013} (\bibinfo {year} {2013})},\ \Eprint {http://arxiv.org/abs/1211.4462}
  {arXiv:1211.4462 [hep-ex]} \BibitemShut {NoStop}%
\bibitem [{\citenamefont {Catani}\ \emph {et~al.}(2009)\citenamefont {Catani},
  \citenamefont {Cieri}, \citenamefont {Ferrera}, \citenamefont {de~Florian},\
  and\ \citenamefont {Grazzini}}]{Catani:2009sm}%
  \BibitemOpen
  \bibfield  {author} {\bibinfo {author} {\bibfnamefont {S.}~\bibnamefont
  {Catani}}, \bibinfo {author} {\bibfnamefont {L.}~\bibnamefont {Cieri}},
  \bibinfo {author} {\bibfnamefont {G.}~\bibnamefont {Ferrera}}, \bibinfo
  {author} {\bibfnamefont {D.}~\bibnamefont {de~Florian}}, \ and\ \bibinfo
  {author} {\bibfnamefont {M.}~\bibnamefont {Grazzini}},\ }\href {\doibase
  10.1103/PhysRevLett.103.082001} {\bibfield  {journal} {\bibinfo  {journal}
  {Phys. Rev. Lett.}\ }\textbf {\bibinfo {volume} {103}},\ \bibinfo {pages}
  {082001} (\bibinfo {year} {2009})},\ \Eprint {http://arxiv.org/abs/0903.2120}
  {arXiv:0903.2120 [hep-ph]} \BibitemShut {NoStop}%
\bibitem [{\citenamefont {Grazzini}\ \emph {et~al.}(2016)\citenamefont
  {Grazzini}, \citenamefont {Kallweit}, \citenamefont {Rathlev},\ and\
  \citenamefont {Wiesemann}}]{Grazzini:2016swo}%
  \BibitemOpen
  \bibfield  {author} {\bibinfo {author} {\bibfnamefont {M.}~\bibnamefont
  {Grazzini}}, \bibinfo {author} {\bibfnamefont {S.}~\bibnamefont {Kallweit}},
  \bibinfo {author} {\bibfnamefont {D.}~\bibnamefont {Rathlev}}, \ and\
  \bibinfo {author} {\bibfnamefont {M.}~\bibnamefont {Wiesemann}},\ }\href
  {\doibase 10.1016/j.physletb.2016.08.017} {\bibfield  {journal} {\bibinfo
  {journal} {Phys. Lett.}\ }\textbf {\bibinfo {volume} {B761}},\ \bibinfo
  {pages} {179} (\bibinfo {year} {2016})},\ \Eprint
  {http://arxiv.org/abs/1604.08576} {arXiv:1604.08576 [hep-ph]} \BibitemShut
  {NoStop}%
\bibitem [{\citenamefont {Campbell}\ \emph {et~al.}(2011)\citenamefont
  {Campbell}, \citenamefont {Ellis},\ and\ \citenamefont
  {Williams}}]{Campbell:2011bn}%
  \BibitemOpen
  \bibfield  {author} {\bibinfo {author} {\bibfnamefont {J.~M.}\ \bibnamefont
  {Campbell}}, \bibinfo {author} {\bibfnamefont {R.~K.}\ \bibnamefont {Ellis}},
  \ and\ \bibinfo {author} {\bibfnamefont {C.}~\bibnamefont {Williams}},\
  }\href {\doibase 10.1007/JHEP07(2011)018} {\bibfield  {journal} {\bibinfo
  {journal} {JHEP}\ }\textbf {\bibinfo {volume} {07}},\ \bibinfo {pages} {018}
  (\bibinfo {year} {2011})},\ \Eprint {http://arxiv.org/abs/1105.0020}
  {arXiv:1105.0020 [hep-ph]} \BibitemShut {NoStop}%
\bibitem [{\citenamefont {Nhung}\ \emph {et~al.}(2013)\citenamefont {Nhung},
  \citenamefont {Ninh},\ and\ \citenamefont {Weber}}]{Nhung:2013jta}%
  \BibitemOpen
  \bibfield  {author} {\bibinfo {author} {\bibfnamefont {D.~T.}\ \bibnamefont
  {Nhung}}, \bibinfo {author} {\bibfnamefont {L.~D.}\ \bibnamefont {Ninh}}, \
  and\ \bibinfo {author} {\bibfnamefont {M.~M.}\ \bibnamefont {Weber}},\ }\href
  {\doibase 10.1007/JHEP12(2013)096} {\bibfield  {journal} {\bibinfo  {journal}
  {JHEP}\ }\textbf {\bibinfo {volume} {12}},\ \bibinfo {pages} {096} (\bibinfo
  {year} {2013})},\ \Eprint {http://arxiv.org/abs/1307.7403} {arXiv:1307.7403
  [hep-ph]} \BibitemShut {NoStop}%
\bibitem [{\citenamefont {Muselli}\ \emph {et~al.}(2015)\citenamefont
  {Muselli}, \citenamefont {Bonvini}, \citenamefont {Forte}, \citenamefont
  {Marzani},\ and\ \citenamefont {Ridolfi}}]{Muselli:2015kba}%
  \BibitemOpen
  \bibfield  {author} {\bibinfo {author} {\bibfnamefont {C.}~\bibnamefont
  {Muselli}}, \bibinfo {author} {\bibfnamefont {M.}~\bibnamefont {Bonvini}},
  \bibinfo {author} {\bibfnamefont {S.}~\bibnamefont {Forte}}, \bibinfo
  {author} {\bibfnamefont {S.}~\bibnamefont {Marzani}}, \ and\ \bibinfo
  {author} {\bibfnamefont {G.}~\bibnamefont {Ridolfi}},\ }\href {\doibase
  10.1007/JHEP08(2015)076} {\bibfield  {journal} {\bibinfo  {journal} {JHEP}\
  }\textbf {\bibinfo {volume} {08}},\ \bibinfo {pages} {076} (\bibinfo {year}
  {2015})},\ \Eprint {http://arxiv.org/abs/1505.02006} {arXiv:1505.02006
  [hep-ph]} \BibitemShut {NoStop}%
\bibitem [{\citenamefont {Das}\ \emph {et~al.}(2016{\natexlab{b}})\citenamefont
  {Das}, \citenamefont {Konar},\ and\ \citenamefont {Majhi}}]{Das:2016hof}%
  \BibitemOpen
  \bibfield  {author} {\bibinfo {author} {\bibfnamefont {A.}~\bibnamefont
  {Das}}, \bibinfo {author} {\bibfnamefont {P.}~\bibnamefont {Konar}}, \ and\
  \bibinfo {author} {\bibfnamefont {S.}~\bibnamefont {Majhi}},\ }\href
  {\doibase 10.1007/JHEP06(2016)019} {\bibfield  {journal} {\bibinfo  {journal}
  {JHEP}\ }\textbf {\bibinfo {volume} {06}},\ \bibinfo {pages} {019} (\bibinfo
  {year} {2016}{\natexlab{b}})},\ \Eprint {http://arxiv.org/abs/1604.00608}
  {arXiv:1604.00608 [hep-ph]} \BibitemShut {NoStop}%
\bibitem [{\citenamefont {Degrande}\ \emph {et~al.}(2016)\citenamefont
  {Degrande}, \citenamefont {Mattelaer}, \citenamefont {Ruiz},\ and\
  \citenamefont {Turner}}]{Degrande:2016aje}%
  \BibitemOpen
  \bibfield  {author} {\bibinfo {author} {\bibfnamefont {C.}~\bibnamefont
  {Degrande}}, \bibinfo {author} {\bibfnamefont {O.}~\bibnamefont {Mattelaer}},
  \bibinfo {author} {\bibfnamefont {R.}~\bibnamefont {Ruiz}}, \ and\ \bibinfo
  {author} {\bibfnamefont {J.}~\bibnamefont {Turner}},\ }\href {\doibase
  10.1103/PhysRevD.94.053002} {\bibfield  {journal} {\bibinfo  {journal} {Phys.
  Rev.}\ }\textbf {\bibinfo {volume} {D94}},\ \bibinfo {pages} {053002}
  (\bibinfo {year} {2016})},\ \Eprint {http://arxiv.org/abs/1602.06957}
  {arXiv:1602.06957 [hep-ph]} \BibitemShut {NoStop}%
\bibitem [{\citenamefont {Thaler}\ and\ \citenamefont
  {Van~Tilburg}(2011)}]{Thaler:2010tr}%
  \BibitemOpen
  \bibfield  {author} {\bibinfo {author} {\bibfnamefont {J.}~\bibnamefont
  {Thaler}}\ and\ \bibinfo {author} {\bibfnamefont {K.}~\bibnamefont
  {Van~Tilburg}},\ }\href {\doibase 10.1007/JHEP03(2011)015} {\bibfield
  {journal} {\bibinfo  {journal} {JHEP}\ }\textbf {\bibinfo {volume} {03}},\
  \bibinfo {pages} {015} (\bibinfo {year} {2011})},\ \Eprint
  {http://arxiv.org/abs/1011.2268} {arXiv:1011.2268 [hep-ph]} \BibitemShut
  {NoStop}%
\bibitem [{\citenamefont {Thaler}\ and\ \citenamefont
  {Van~Tilburg}(2012)}]{Thaler:2011gf}%
  \BibitemOpen
  \bibfield  {author} {\bibinfo {author} {\bibfnamefont {J.}~\bibnamefont
  {Thaler}}\ and\ \bibinfo {author} {\bibfnamefont {K.}~\bibnamefont
  {Van~Tilburg}},\ }\href {\doibase 10.1007/JHEP02(2012)093} {\bibfield
  {journal} {\bibinfo  {journal} {JHEP}\ }\textbf {\bibinfo {volume} {02}},\
  \bibinfo {pages} {093} (\bibinfo {year} {2012})},\ \Eprint
  {http://arxiv.org/abs/1108.2701} {arXiv:1108.2701 [hep-ph]} \BibitemShut
  {NoStop}%
\bibitem [{\citenamefont {Cacciari}\ and\ \citenamefont
  {Salam}(2006)}]{Cacciari:2005hq}%
  \BibitemOpen
  \bibfield  {author} {\bibinfo {author} {\bibfnamefont {M.}~\bibnamefont
  {Cacciari}}\ and\ \bibinfo {author} {\bibfnamefont {G.~P.}\ \bibnamefont
  {Salam}},\ }\href {\doibase 10.1016/j.physletb.2006.08.037} {\bibfield
  {journal} {\bibinfo  {journal} {Phys. Lett.}\ }\textbf {\bibinfo {volume}
  {B641}},\ \bibinfo {pages} {57} (\bibinfo {year} {2006})},\ \Eprint
  {http://arxiv.org/abs/hep-ph/0512210} {arXiv:hep-ph/0512210 [hep-ph]}
  \BibitemShut {NoStop}%
\bibitem [{\citenamefont {Khachatryan}\ \emph {et~al.}(2014)\citenamefont
  {Khachatryan} \emph {et~al.}}]{Khachatryan:2014vla}%
  \BibitemOpen
  \bibfield  {author} {\bibinfo {author} {\bibfnamefont {V.}~\bibnamefont
  {Khachatryan}} \emph {et~al.} (\bibinfo {collaboration} {CMS}),\ }\href
  {\doibase 10.1007/JHEP12(2014)017} {\bibfield  {journal} {\bibinfo  {journal}
  {JHEP}\ }\textbf {\bibinfo {volume} {12}},\ \bibinfo {pages} {017} (\bibinfo
  {year} {2014})},\ \Eprint {http://arxiv.org/abs/1410.4227} {arXiv:1410.4227
  [hep-ex]} \BibitemShut {NoStop}%
\bibitem [{\citenamefont {Wobisch}\ and\ \citenamefont
  {Wengler}(1998)}]{Wobisch:1998wt}%
  \BibitemOpen
  \bibfield  {author} {\bibinfo {author} {\bibfnamefont {M.}~\bibnamefont
  {Wobisch}}\ and\ \bibinfo {author} {\bibfnamefont {T.}~\bibnamefont
  {Wengler}},\ }in\ \href
  {http://inspirehep.net/record/484872/files/arXiv:hep-ph_9907280.pdf} {\emph
  {\bibinfo {booktitle} {{Monte Carlo generators for HERA physics. Proceedings,
  Workshop, Hamburg, Germany, 1998-1999}}}}\ (\bibinfo {year} {1998})\ pp.\
  \bibinfo {pages} {270--279},\ \Eprint {http://arxiv.org/abs/hep-ph/9907280}
  {arXiv:hep-ph/9907280 [hep-ph]} \BibitemShut {NoStop}%
\bibitem [{\citenamefont {Lester}\ and\ \citenamefont
  {Summers}(1999)}]{Lester:1999tx}%
  \BibitemOpen
  \bibfield  {author} {\bibinfo {author} {\bibfnamefont {C.~G.}\ \bibnamefont
  {Lester}}\ and\ \bibinfo {author} {\bibfnamefont {D.~J.}\ \bibnamefont
  {Summers}},\ }\href {\doibase 10.1016/S0370-2693(99)00945-4} {\bibfield
  {journal} {\bibinfo  {journal} {Phys. Lett.}\ }\textbf {\bibinfo {volume}
  {B463}},\ \bibinfo {pages} {99} (\bibinfo {year} {1999})},\ \Eprint
  {http://arxiv.org/abs/hep-ph/9906349} {arXiv:hep-ph/9906349 [hep-ph]}
  \BibitemShut {NoStop}%
\bibitem [{\citenamefont {Cho}\ \emph {et~al.}(2008)\citenamefont {Cho},
  \citenamefont {Choi}, \citenamefont {Kim},\ and\ \citenamefont
  {Park}}]{Cho:2007qv}%
  \BibitemOpen
  \bibfield  {author} {\bibinfo {author} {\bibfnamefont {W.~S.}\ \bibnamefont
  {Cho}}, \bibinfo {author} {\bibfnamefont {K.}~\bibnamefont {Choi}}, \bibinfo
  {author} {\bibfnamefont {Y.~G.}\ \bibnamefont {Kim}}, \ and\ \bibinfo
  {author} {\bibfnamefont {C.~B.}\ \bibnamefont {Park}},\ }\href {\doibase
  10.1103/PhysRevLett.100.171801} {\bibfield  {journal} {\bibinfo  {journal}
  {Phys. Rev. Lett.}\ }\textbf {\bibinfo {volume} {100}},\ \bibinfo {pages}
  {171801} (\bibinfo {year} {2008})},\ \Eprint {http://arxiv.org/abs/0709.0288}
  {arXiv:0709.0288 [hep-ph]} \BibitemShut {NoStop}%
\bibitem [{\citenamefont {Barr}\ \emph {et~al.}(2011)\citenamefont {Barr},
  \citenamefont {Khoo}, \citenamefont {Konar}, \citenamefont {Kong},
  \citenamefont {Lester}, \citenamefont {Matchev},\ and\ \citenamefont
  {Park}}]{Barr:2011xt}%
  \BibitemOpen
  \bibfield  {author} {\bibinfo {author} {\bibfnamefont {A.~J.}\ \bibnamefont
  {Barr}}, \bibinfo {author} {\bibfnamefont {T.~J.}\ \bibnamefont {Khoo}},
  \bibinfo {author} {\bibfnamefont {P.}~\bibnamefont {Konar}}, \bibinfo
  {author} {\bibfnamefont {K.}~\bibnamefont {Kong}}, \bibinfo {author}
  {\bibfnamefont {C.~G.}\ \bibnamefont {Lester}}, \bibinfo {author}
  {\bibfnamefont {K.~T.}\ \bibnamefont {Matchev}}, \ and\ \bibinfo {author}
  {\bibfnamefont {M.}~\bibnamefont {Park}},\ }\href {\doibase
  10.1103/PhysRevD.84.095031} {\bibfield  {journal} {\bibinfo  {journal} {Phys.
  Rev.}\ }\textbf {\bibinfo {volume} {D84}},\ \bibinfo {pages} {095031}
  (\bibinfo {year} {2011})},\ \Eprint {http://arxiv.org/abs/1105.2977}
  {arXiv:1105.2977 [hep-ph]} \BibitemShut {NoStop}%
\bibitem [{\citenamefont {Burns}\ \emph {et~al.}(2009)\citenamefont {Burns},
  \citenamefont {Kong}, \citenamefont {Matchev},\ and\ \citenamefont
  {Park}}]{Burns:2008va}%
  \BibitemOpen
  \bibfield  {author} {\bibinfo {author} {\bibfnamefont {M.}~\bibnamefont
  {Burns}}, \bibinfo {author} {\bibfnamefont {K.}~\bibnamefont {Kong}},
  \bibinfo {author} {\bibfnamefont {K.~T.}\ \bibnamefont {Matchev}}, \ and\
  \bibinfo {author} {\bibfnamefont {M.}~\bibnamefont {Park}},\ }\href {\doibase
  10.1088/1126-6708/2009/03/143} {\bibfield  {journal} {\bibinfo  {journal}
  {JHEP}\ }\textbf {\bibinfo {volume} {03}},\ \bibinfo {pages} {143} (\bibinfo
  {year} {2009})},\ \Eprint {http://arxiv.org/abs/0810.5576} {arXiv:0810.5576
  [hep-ph]} \BibitemShut {NoStop}%
\bibitem [{\citenamefont {Konar}\ \emph {et~al.}(2010)\citenamefont {Konar},
  \citenamefont {Kong}, \citenamefont {Matchev},\ and\ \citenamefont
  {Park}}]{Konar:2009qr}%
  \BibitemOpen
  \bibfield  {author} {\bibinfo {author} {\bibfnamefont {P.}~\bibnamefont
  {Konar}}, \bibinfo {author} {\bibfnamefont {K.}~\bibnamefont {Kong}},
  \bibinfo {author} {\bibfnamefont {K.~T.}\ \bibnamefont {Matchev}}, \ and\
  \bibinfo {author} {\bibfnamefont {M.}~\bibnamefont {Park}},\ }\href {\doibase
  10.1007/JHEP04(2010)086} {\bibfield  {journal} {\bibinfo  {journal} {JHEP}\
  }\textbf {\bibinfo {volume} {04}},\ \bibinfo {pages} {086} (\bibinfo {year}
  {2010})},\ \Eprint {http://arxiv.org/abs/0911.4126} {arXiv:0911.4126
  [hep-ph]} \BibitemShut {NoStop}%
\bibitem [{\citenamefont {Atre}\ \emph {et~al.}(2009)\citenamefont {Atre},
  \citenamefont {Han}, \citenamefont {Pascoli},\ and\ \citenamefont
  {Zhang}}]{Atre:2009rg}%
  \BibitemOpen
  \bibfield  {author} {\bibinfo {author} {\bibfnamefont {A.}~\bibnamefont
  {Atre}}, \bibinfo {author} {\bibfnamefont {T.}~\bibnamefont {Han}}, \bibinfo
  {author} {\bibfnamefont {S.}~\bibnamefont {Pascoli}}, \ and\ \bibinfo
  {author} {\bibfnamefont {B.}~\bibnamefont {Zhang}},\ }\href {\doibase
  10.1088/1126-6708/2009/05/030} {\bibfield  {journal} {\bibinfo  {journal}
  {JHEP}\ }\textbf {\bibinfo {volume} {05}},\ \bibinfo {pages} {030} (\bibinfo
  {year} {2009})},\ \Eprint {http://arxiv.org/abs/0901.3589} {arXiv:0901.3589
  [hep-ph]} \BibitemShut {NoStop}%
\bibitem [{\citenamefont {Nardi}\ \emph {et~al.}(1994)\citenamefont {Nardi},
  \citenamefont {Roulet},\ and\ \citenamefont {Tommasini}}]{Nardi:1994iv}%
  \BibitemOpen
  \bibfield  {author} {\bibinfo {author} {\bibfnamefont {E.}~\bibnamefont
  {Nardi}}, \bibinfo {author} {\bibfnamefont {E.}~\bibnamefont {Roulet}}, \
  and\ \bibinfo {author} {\bibfnamefont {D.}~\bibnamefont {Tommasini}},\ }\href
  {\doibase 10.1016/0370-2693(94)90736-6} {\bibfield  {journal} {\bibinfo
  {journal} {Phys. Lett.}\ }\textbf {\bibinfo {volume} {B327}},\ \bibinfo
  {pages} {319} (\bibinfo {year} {1994})},\ \Eprint
  {http://arxiv.org/abs/hep-ph/9402224} {arXiv:hep-ph/9402224 [hep-ph]}
  \BibitemShut {NoStop}%
\bibitem [{\citenamefont {Nardi}\ \emph {et~al.}(1995)\citenamefont {Nardi},
  \citenamefont {Roulet},\ and\ \citenamefont {Tommasini}}]{Nardi:1994nw}%
  \BibitemOpen
  \bibfield  {author} {\bibinfo {author} {\bibfnamefont {E.}~\bibnamefont
  {Nardi}}, \bibinfo {author} {\bibfnamefont {E.}~\bibnamefont {Roulet}}, \
  and\ \bibinfo {author} {\bibfnamefont {D.}~\bibnamefont {Tommasini}},\ }\href
  {\doibase 10.1016/0370-2693(95)91542-M} {\bibfield  {journal} {\bibinfo
  {journal} {Phys. Lett.}\ }\textbf {\bibinfo {volume} {B344}},\ \bibinfo
  {pages} {225} (\bibinfo {year} {1995})},\ \Eprint
  {http://arxiv.org/abs/hep-ph/9409310} {arXiv:hep-ph/9409310 [hep-ph]}
  \BibitemShut {NoStop}%
\bibitem [{\citenamefont {Kim}\ \emph {et~al.}(2019)\citenamefont {Kim},
  \citenamefont {Kong}, \citenamefont {Matchev},\ and\ \citenamefont
  {Park}}]{Kim:2018cxf}%
  \BibitemOpen
  \bibfield  {author} {\bibinfo {author} {\bibfnamefont {J.~H.}\ \bibnamefont
  {Kim}}, \bibinfo {author} {\bibfnamefont {K.}~\bibnamefont {Kong}}, \bibinfo
  {author} {\bibfnamefont {K.~T.}\ \bibnamefont {Matchev}}, \ and\ \bibinfo
  {author} {\bibfnamefont {M.}~\bibnamefont {Park}},\ }\href {\doibase
  10.1103/PhysRevLett.122.091801} {\bibfield  {journal} {\bibinfo  {journal}
  {Phys. Rev. Lett.}\ }\textbf {\bibinfo {volume} {122}},\ \bibinfo {pages}
  {091801} (\bibinfo {year} {2019})},\ \Eprint
  {http://arxiv.org/abs/1807.11498} {arXiv:1807.11498 [hep-ph]} \BibitemShut
  {NoStop}%
\bibitem [{\citenamefont
  {{\url{https://twiki.cern.ch/twiki/pub/Main/ABCDMethod/ABCDGuide_draft18Oct18.pdf}}}()}]{ABCD}%
  \BibitemOpen
  \bibfield  {author} {\bibinfo {author} {\bibnamefont
  {{\url{https://twiki.cern.ch/twiki/pub/Main/ABCDMethod/ABCDGuide_draft18Oct18.pdf}}}},\
  }\href@noop {} {\ }\BibitemShut {NoStop}%
\bibitem [{\citenamefont {Dev}\ \emph {et~al.}(2014)\citenamefont {Dev},
  \citenamefont {Pilaftsis},\ and\ \citenamefont {Yang}}]{Dev:2013wba}%
  \BibitemOpen
  \bibfield  {author} {\bibinfo {author} {\bibfnamefont {P.~S.~B.}\
  \bibnamefont {Dev}}, \bibinfo {author} {\bibfnamefont {A.}~\bibnamefont
  {Pilaftsis}}, \ and\ \bibinfo {author} {\bibfnamefont {U.-k.}\ \bibnamefont
  {Yang}},\ }\href {\doibase 10.1103/PhysRevLett.112.081801} {\bibfield
  {journal} {\bibinfo  {journal} {Phys. Rev. Lett.}\ }\textbf {\bibinfo
  {volume} {112}},\ \bibinfo {pages} {081801} (\bibinfo {year} {2014})},\
  \Eprint {http://arxiv.org/abs/1308.2209} {arXiv:1308.2209 [hep-ph]}
  \BibitemShut {NoStop}%
\bibitem [{\citenamefont {Alva}\ \emph {et~al.}(2015)\citenamefont {Alva},
  \citenamefont {Han},\ and\ \citenamefont {Ruiz}}]{Alva:2014gxa}%
  \BibitemOpen
  \bibfield  {author} {\bibinfo {author} {\bibfnamefont {D.}~\bibnamefont
  {Alva}}, \bibinfo {author} {\bibfnamefont {T.}~\bibnamefont {Han}}, \ and\
  \bibinfo {author} {\bibfnamefont {R.}~\bibnamefont {Ruiz}},\ }\href {\doibase
  10.1007/JHEP02(2015)072} {\bibfield  {journal} {\bibinfo  {journal} {JHEP}\
  }\textbf {\bibinfo {volume} {02}},\ \bibinfo {pages} {072} (\bibinfo {year}
  {2015})},\ \Eprint {http://arxiv.org/abs/1411.7305} {arXiv:1411.7305
  [hep-ph]} \BibitemShut {NoStop}%
\bibitem [{\citenamefont {de~Blas}(2013)}]{deBlas:2013gla}%
  \BibitemOpen
  \bibfield  {author} {\bibinfo {author} {\bibfnamefont {J.}~\bibnamefont
  {de~Blas}},\ }\bibfield  {booktitle} {\emph {\bibinfo {booktitle}
  {{Proceedings, 1st Large Hadron Collider Physics Conference (LHCP 2013):
  Barcelona, Spain, May 13-18, 2013}}},\ }\href {\doibase
  10.1051/epjconf/20136019008} {\bibfield  {journal} {\bibinfo  {journal} {EPJ
  Web Conf.}\ }\textbf {\bibinfo {volume} {60}},\ \bibinfo {pages} {19008}
  (\bibinfo {year} {2013})},\ \Eprint {http://arxiv.org/abs/1307.6173}
  {arXiv:1307.6173 [hep-ph]} \BibitemShut {NoStop}%
\bibitem [{\citenamefont {del Aguila}\ \emph {et~al.}(2008)\citenamefont {del
  Aguila}, \citenamefont {de~Blas},\ and\ \citenamefont
  {Perez-Victoria}}]{delAguila:2008pw}%
  \BibitemOpen
  \bibfield  {author} {\bibinfo {author} {\bibfnamefont {F.}~\bibnamefont {del
  Aguila}}, \bibinfo {author} {\bibfnamefont {J.}~\bibnamefont {de~Blas}}, \
  and\ \bibinfo {author} {\bibfnamefont {M.}~\bibnamefont {Perez-Victoria}},\
  }\href {\doibase 10.1103/PhysRevD.78.013010} {\bibfield  {journal} {\bibinfo
  {journal} {Phys. Rev.}\ }\textbf {\bibinfo {volume} {D78}},\ \bibinfo {pages}
  {013010} (\bibinfo {year} {2008})},\ \Eprint {http://arxiv.org/abs/0803.4008}
  {arXiv:0803.4008 [hep-ph]} \BibitemShut {NoStop}%
\bibitem [{\citenamefont {Akhmedov}\ \emph {et~al.}(2013)\citenamefont
  {Akhmedov}, \citenamefont {Kartavtsev}, \citenamefont {Lindner},
  \citenamefont {Michaels},\ and\ \citenamefont {Smirnov}}]{Akhmedov:2013hec}%
  \BibitemOpen
  \bibfield  {author} {\bibinfo {author} {\bibfnamefont {E.}~\bibnamefont
  {Akhmedov}}, \bibinfo {author} {\bibfnamefont {A.}~\bibnamefont
  {Kartavtsev}}, \bibinfo {author} {\bibfnamefont {M.}~\bibnamefont {Lindner}},
  \bibinfo {author} {\bibfnamefont {L.}~\bibnamefont {Michaels}}, \ and\
  \bibinfo {author} {\bibfnamefont {J.}~\bibnamefont {Smirnov}},\ }\href
  {\doibase 10.1007/JHEP05(2013)081} {\bibfield  {journal} {\bibinfo  {journal}
  {JHEP}\ }\textbf {\bibinfo {volume} {05}},\ \bibinfo {pages} {081} (\bibinfo
  {year} {2013})},\ \Eprint {http://arxiv.org/abs/1302.1872} {arXiv:1302.1872
  [hep-ph]} \BibitemShut {NoStop}%
\end{thebibliography}%

\end{document}